\documentclass[12pt,pdflatex]{sn-jnl}

\usepackage{lineno}

\usepackage{graphicx}%
\usepackage{multirow}%
\usepackage{amsmath,amssymb,amsfonts}%
\usepackage{amsthm}%
\usepackage{mathrsfs}%
\usepackage[title]{appendix}%
\usepackage{xcolor}%
\usepackage{textcomp}%
\usepackage{manyfoot}%
\usepackage{booktabs}%
\usepackage{algorithm}%
\usepackage{algorithmicx}%
\usepackage{algpseudocode}%
\usepackage{listings}%
\usepackage[section]{placeins}

\usepackage[round]{natbib}
\begin{document}

\title{Conditional diffusion models for downscaling and bias correction of Earth system model precipitation

\thanks{Contact: Michael Aich, michael.aich@tum.de} 
}

\author*[1,2]{ \fnm{Michael} \sur{Aich}}\email{michael.aich@tum.de}
\author[1,2]{\fnm{Philipp} \sur{Hess}}
\author[3]{\fnm{Baoxiang} \sur{Pan}}
\author[1,2]{ \fnm{Sebastian} \sur{Bathiany}}
\author[1,2]{ \fnm{Yu} \sur{Huang}}
\author[1,2,4]{ \fnm{Niklas} \sur{Boers}}

\affil[1]{\small Technical University of Munich, Germany; Munich Climate Center; TUM School of Engineering and Design, Department of Aerospace and Geodesy, Earth System Modelling Group}
\affil[2]{\small Potsdam Institute for Climate Impact Research, Potsdam, Germany}
\affil[3]{\small Institute of Atmospheric Physics, Chinese Academy of Sciences, Beijing, China}
\affil[4]{\small Global Systems Institute and Department of Mathematics, University of Exeter, Exeter, UK}

\abstract{Climate change exacerbates extreme weather events like heavy rainfall and flooding. As these events cause severe socioeconomic damage, accurate high-resolution simulation of precipitation is imperative. However, existing Earth System Models (ESMs) struggle to resolve small-scale dynamics and suffer from biases. Traditional statistical bias correction and downscaling methods fall short in improving spatial structure, while recent deep learning methods lack controllability and suffer from unstable training. Here, we propose a machine learning framework for simultaneous bias correction and downscaling. We first map observational and ESM data to a shared embedding space, where both are unbiased towards each other, and then train a conditional diffusion model to reverse the mapping. Only observational data is used for the training, so that the diffusion model can be employed to correct and downscale any ESM field without need for retraining. Our approach ensures statistical fidelity and preserves spatial patterns larger than a chosen spatial correction scale. We demonstrate that our approach outperforms existing statistical and deep learning methods especially regarding extreme events.}

\maketitle

\noindent{\bfseries Short Summary:} Accurately simulating rainfall is essential to understand the impacts of climate change, especially extreme events such as floods and droughts. Climate models simulate the atmosphere at a coarse resolution and often misrepresent precipitation, leading to biased and overly smooth fields. We improve climate model precipitation using a generative machine learning model that is data-efficient, preserves key climate signals such as trends and variability, and significantly improves the representation of extreme events. 

\clearpage

\section{Introduction}
With global warming, we anticipate more intense rainfall events and associated natural hazards, e.g., in terms of floods and landslides, in many regions of the world \cite{IPCC_AR6_SYR_SPM_2023}. 
Understanding and accurately simulating precipitation is particularly important for adaptation planning and, hence, for mitigating damages and reducing risks associated with climate change. Earth System Models (ESMs) play a crucial role in simulating precipitation patterns for both historical and future scenarios. However, these simulations are computationally extremely demanding, primarily because they require solving complex partial differential equations. To manage the computational load, ESMs resort to approximate solutions on discretized grids with coarse spatial resolution (typically around 100~km). The consequence is that these models do not resolve small-scale dynamics, such as many of the processes relevant to precipitation generation. This leads to considerable biases in the ESM fields compared to observations. Moreover, the coarse spatial resolution prevents accurate projections of localized precipitation extremes. Therefore, precipitation fields simulated by ESMs cannot be used directly for impact assessments \cite{zelinka2020causes} and especially tasks such as water resource and flood management, which require precise spatial data at high resolution \cite{gutmann2014intercomparison}. 

Statistical bias correction methods can be used as a post-processing to adjust statistical biases. Quantile mapping (QM) is the most common method for improving the statistics of ESM precipitation fields \cite{tong2021bias, gudmundsson2012downscaling, cannon2015bias,miao2019improving}. QM reduces the bias using a mapping that, locally at each grid cell, aligns the estimated cumulative distribution of the model output with the observed precipitation patterns over a reference time period. Although QM is effective in correcting distributions of single grid cells, it falls short in improving the spatial structure and patterns of precipitation simulations \cite{hess2022physically}. A visual inspection shows that ESM precipitation remains too smooth compared to the observational data after applying quantile mapping.\newline 

To address these problems, deep learning methods have recently been introduced \cite{pan2019improving,li2022convolutional, hess2023deep, pan2021learning, franccois2021adjusting, hess2022physically}. In these approaches, the statistical relationships between model simulations and observational data are learned implicitly.
A general constraint when using machine learning methods for bias correction is that individual samples of observational and Earth System Model data are always unpaired. In this context, a sample is a specific weather situation at a specific point in time. The reason for this lack of pairs is that simulations, even with very similar initial conditions, diverge already after a short period of time due to the chaotic nature of the underlying atmospheric dynamics. Currently, one can, therefore, not utilize the wide range of supervised machine learning (ML) techniques that have shown great success in various disciplines in recent years and the available options are consequently restricted to self- and unsupervised machine learning methods. Recent studies \cite{hess2023deep, pan2021learning, franccois2021adjusting, hess2022physically} applied generative adversarial networks (GANs \cite{goodfellow2020generative}) and specifically cycleGANs \cite{Zhu_2017_ICCV} to improve upon existing bias correction techniques. A major limitation of GAN-based approaches is that the stability and convergence of the training process are difficult to control and that it is challenging to find metrics that indicate training convergence. In addition, GANs often suffer from mode collapse, where only a part of the target probability distribution is approximated by the GAN.\newline

As noted above, the low spatial resolution of ESM fields prevents local risk and impact assessment, necessitating the additional use of downscaling methods. In line with the climate literature, we refer to increasing the spatial resolution as downscaling throughout our manuscript, although we are aware that, especially in the machine learning literature, the term upsampling is more prevalent. We use the term downscaling only when we want to increase the information content in an image as well as the number of pixels. When we refer to upsampling (downsampling), we only mean an increase (decrease) in the number of pixels. 
Statistical downscaling aims to learn a transformation from the low-resolution ESM fields to high-resolution observations. Recent developments lean towards using machine learning methods for this task \cite{rampal2022high, hobeichi2023using, rampal2024downscaling}.
The potential for machine learning-based downscaling methods was already shown in \cite{vandal2017deepsd, van2023deep, doury2023regional, doury2024suitability, rampal2025reliable}.

Recently, Hess et al. \cite{hess2025fast} used an unconditional consistency model (CM) for downscaling 3° $\times$ 3.75° precipitation data to 0.75°$\times$ 0.9375°. Our work addresses the more challenging task of downscaling from 1°$\times$ 1.25° to 0.25°$\times$ 0.25° resolution, a scale essential for regional impact assessments. We show that the consistency model applied to our higher resolution setting with limited amounts of training data struggles to approximate the distribution, highlighting an advantage of our conditional training approach. The analysis is further extended to out-of-distribution scenarios, particularly those involving extreme precipitation and future emission projections.

\begin{figure*}[!htb]
    \centering
    \includegraphics[width=0.88\textwidth]{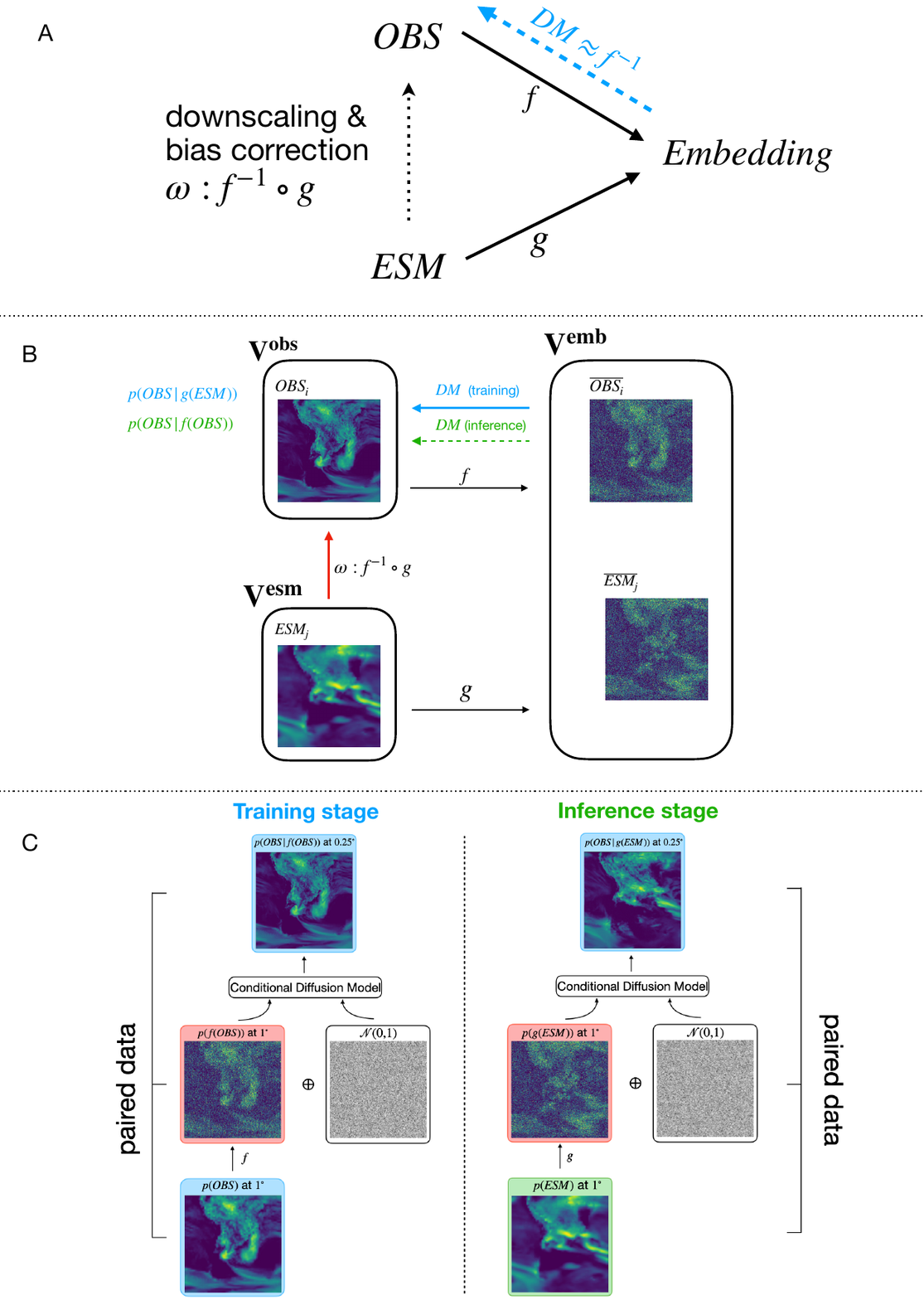}
    \caption{ \textbf{Schematic overview of our approach}. \textbf{(A)} Bias correction and downscaling can be formulated as a mapping $\omega$ from the ESM data space to the data space of observations (OBS) used for training. We first map both datasets to a shared embedding space and then learn the inverse of the mapping $f$ with a DM. We achieve a correction of the ESM data by applying $DM \circ g$. 
    \textbf{(B)} Our framework allows to train a single model for bias correction and downscaling in a supervised way despite the unpaired nature of OBS and ESM fields. We construct functions $f,g$ that map $OBS \in \mathbf{V^{obs}}$ and $ESM \in \mathbf{V^{esm}}$ fields to a shared embedding space $\mathbf{V^{emb}}$. Note that this embedding space does not enforce pairing between individual fields, but a similar distribution between the embedded fields. By inverting $f$, we can rewrite $\omega$ as $\omega = f^{-1} \circ g$. We learn the inverse $f^{-1}$ with a conditional diffusion model. This model is trained (blue arrow) on pairs of observational data to approximate the map from $f(OBS)$ to OBS. Because $f(OBS)$ and $g(ESM)$ share the embedding space (and are identically distributed by construction), we can evaluate (green arrow) the $DM$ on the embedded ESM data $g(ESM)$ and thereby approximate the bias correction and downscaling function $\omega = f^{-1} \circ g \approx DM \circ g$, without the need of paired data between OBS and ESM.
    The indices $i,j$ highlight that the two exemplary fields $ESM_j$ and $ESM_i$ are not paired. \newline
    \textbf{(C)}  Left: Training process of the conditional DM $DM\approx f^{-1}$. Note that the individual samples of the input OBS and their embeddings $f(OBS)$, as well as the embeddings $f(OBS)$ and the output of $DM\approx f^{-1}$ are paired, respectively. Right: Inference process of $DM\approx f^{-1}$. In this case, the individual samples of the input ESM, their embeddings $g(ESM)$, and the output of $DM\approx f^{-1}$ are paired, respectively. It is not necessary for the training embedding samples to be paired with the inference embedding samples. See fig.~\ref*{fig:diffusion_model} for details.}
    \label{fig:main_fig}
\end{figure*}

Diffusion models (DMs) have recently emerged as the state-of-the-art ML approach for conditional image generation \cite{saharia2022photorealistic, rombach2022high, saharia2022image} and image-to-image translation \cite{saharia2022palette}, mostly outperforming GANs across different tasks. Diffusion models (Fig. \ref{fig:main_fig} and fig.~\ref*{fig:diffusion_model}) avoid the common issues present with GANs in exchange for slower inference speed. A diffusion model consists of a forward and a backward process. During the forward process, noise is added to an image in subsequent steps to gradually remove its content. The amount of noise added follows a predefined equation. During the backward process, a neural network is trained to reverse each of these individual noising steps to recover the original image. The trained diffusion model can generate an image of the training data distribution, given a noise image as input. 
Recent work \cite{wan2024debias} introduced a framework for downscaling and bias correction, combining a diffusion model that is responsible for downscaling and a model based on optimal transport responsible for bias correction. Optimal transport \cite{cuturi2013sinkhorn} learns a map between two data distributions in an unsupervised setting. However, this framework is computationally expensive and has so far only been demonstrated on synthetic datasets, without evaluation on real-world observational or ESM fields. In contrast, our approach is computationally efficient by combining computationally efficient QM for large-scale bias correction with a conditional diffusion model that performs both small-scale bias correction and downscaling by generating matching small-scale patterns. We demonstrate its effectiveness for precipitation data, highlighting its ability to correct biases, downscale accurately, and capture extremes, uncertainties, and trends. A major advantage is that our conditional training allows us to use a relatively small dataset for training and still capture the distribution accurately. In contrast, unconditional models often need considerably more data to capture the full data variability, as we also show in our comparison to \cite{hess2025fast} (see fig.~\ref*{fig:CM_comp}).

Existing work leveraging state-of-the-art ML methods for bias correction and downscaling does not systematically investigate out-of-distribution scenarios like future emission scenarios and especially the representation of extreme events of the generative models in detail. Understanding the generalization performance of the models under these conditions is, however, crucial for impact modelers who rely on these outputs for risk assessments under future climate conditions. We will therefore present a detailed analysis of the generalization capability of our approach, both in terms of its performance in preserving climate change trends, as well as in capturing extreme events and their trends. 

A major challenge in bias correction and downscaling of ESMs is that the whole class of state-of-the-art supervised machine learning methods is not applicable in this setting. This is due to two fundamental issues. First, due to the chaotic nature of atmosphere and ocean dynamics, ESM simulations and observational data are inherently unpaired. This means that the weather on a specific day in an ESM simulation does not correspond to the observed weather on the same day, which prevents directly training a supervised ML method on the task. Second, training a ML model on observational data and applying it to ESM data is unreliable due to the substantial distribution shift between both datasets caused by systematic biases in the ESM. This violation of the assumption of independently and identically distributed (i.i.d.) data leads to poor generalization. Our proposed framework directly addresses both challenges. We reformulate the problem in a novel way, which allows us to train arbitrary ML models in a conditional setup without the need for explicit ESM-observation pairs, while at the same time resolving the distribution shift. 

We present a novel framework based on state-of-the-art conditional diffusion models that allows us to perform both bias correction and downscaling with one single neural network, which only takes precipitation as input and output. We use a conditional diffusion model (Fig.~\ref{fig:main_fig} and fig. ~\ref*{fig:diffusion_model}) to correct low-resolution (LR) ESM fields toward high-resolution (HR) observational data (OBS). The supervised formulation of the task allows us to train a conditional diffusion model that is more data efficient (requiring less training data) than its unconditional counterpart because it is trained to only learn the small-scale precipitation patterns, given the large-scale patterns. The model then learns to copy the correct large-scale information from the condition channel. An unconditional model that learns to approximate the full distribution of precipitation at all scales is unnecessarily complex for the task.
In general, our task of bias correction and downscaling can be seen as taking a field from a distribution $p(\text{ESM})$ and transforming it into a field from a conditional distribution $p(\text{OBS}|\text{ESM})$. 

A key idea of our framework is to reformulate the problem in a way that yields a clear training objective. A key part of it is a statistical mapping to an embedding space, which ensures that training and inference data are identically distributed. We achieve this by introducing transformations $f$ and $g$ that map observational (OBS) and ESM data to a shared embedding space (see Methods and Fig. \ref{fig:main_fig}A). This space is explicitly designed to solve the two fundamental issues mentioned above: it creates a valid supervised objective by providing paired samples of observational data and their perturbed embeddings (OBS, f(OBS)), and it ensures the training and inference distributions match by making the distributions of the embeddings $f(OBS)$ and $g(ESM)$ similar. On this shared embedding space, we can train a conditional diffusion model to approximate the inverse of $f$ (Fig. \ref{fig:main_fig}B and Fig. \ref{fig:main_fig}C). The neural network is trained to predict the clean OBS data given the embedded OBS data, thereby only relying on pairs between OBS and $f(OBS)$. For inference, the ESM data is mapped into the same embedding space using the transformation $g$. The statistical similarity of the resulting embeddings $f(OBS)$ and $g(ESM)$ enables the diffusion model, which was trained exclusively on observational data, to generalize effectively to downscale and bias-correct the ESM fields. The diffusion model will map the embedded ESM data towards the distribution of observational data, resulting in bias-corrected and downscaled ESM fields.

\begin{figure*}[htb]
    \centering
    \includegraphics[width=\textwidth]{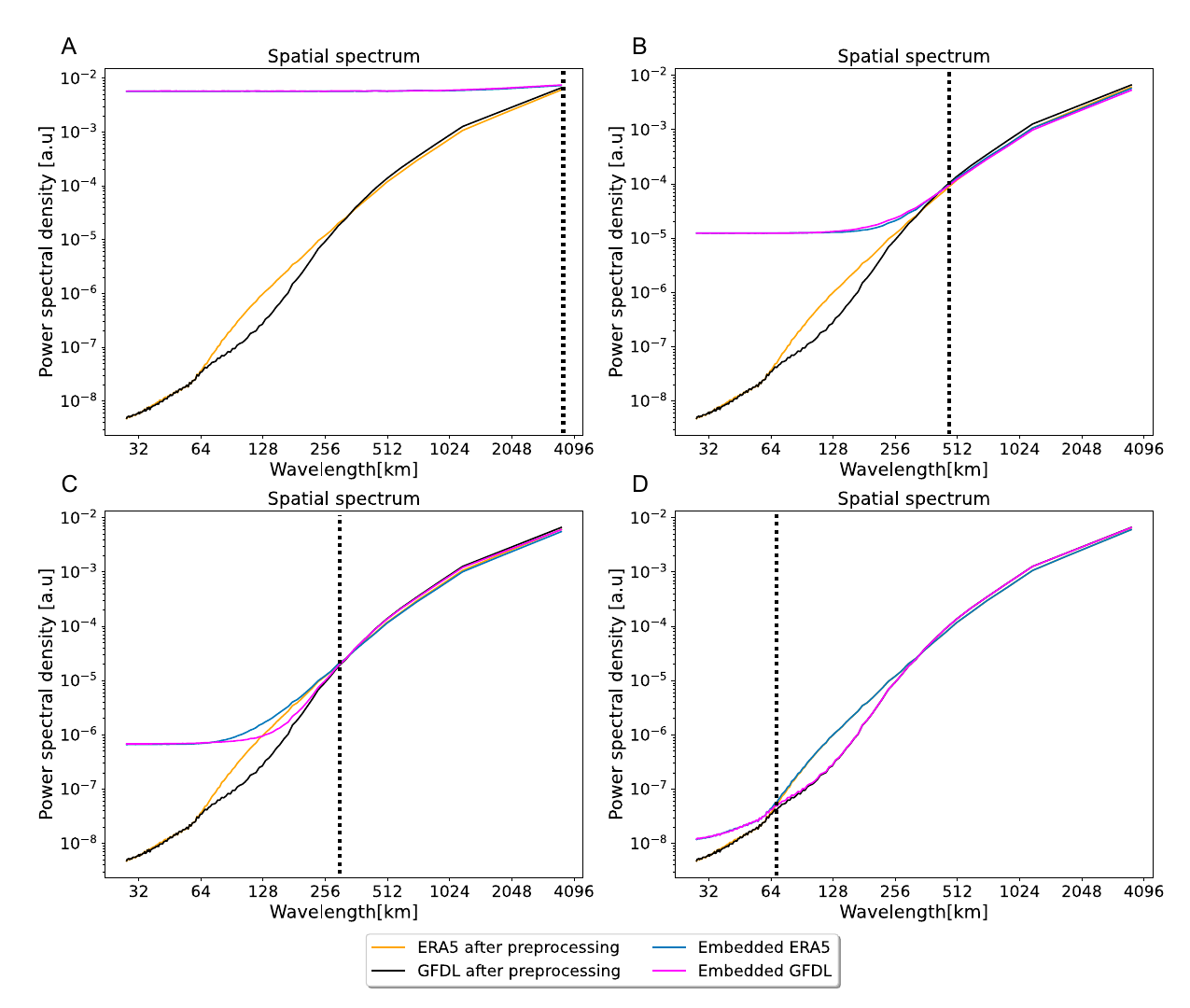}
    \caption{\textbf{ Power spectral densities (PSDs) for different choices noising scale of the diffusion model}. The noising scale $s$ (dashed line) is a hyperparameter that can be chosen depending on the ESM and observational datasets, as well as on the specific task. For the maximal choice of $s$ \textbf{(A)} all information in the observations (ERA5) and model simulations (GFDL) is noised and thereby destroyed. Conditioning on pure noise makes the task equivalent to unconditional image generation. The diffusion model will learn to generate observational fields with no relation to the ESM fields. When $s$ is chosen to be minimal, there will be no noising and the conditional generation will directly replicate the condition, i.e. the ESM field.
    In \textbf{(B)} we chose $s$ as the point where the PSDs of the observational and simulated datasets intersect. We then apply sufficiently many forward diffusion noising steps to both datasets, destroying small-scale structure until they agree in the PSD. We call scales smaller than $s$ small scales and scales larger than $s$ large scales. In \textbf{(C)} and \textbf{(D)}, the effects of choosing a smaller noise scale $s$ are shown. Prior knowledge about the ESM or its accuracy can also guide the choice of $s$.
    }
    \label{fig:noise_level}
\end{figure*}

This framework offers great flexibility as it can be applied to any ESM, with minimal adjustments in the embedding pipeline. The embedding transformation for the ESM has two key components. First, we use quantile mapping (QM) as a fast and effective method to correct large-scale biases in the ESM. Second, we introduce noise to remove small-scale information in the precipitation fields. We define large scales as those spatial scales that are effectively corrected using QM alone, while smaller spatial scales, which require additional correction, are referred to as small scales (Fig. \ref{fig:noise_level}). This noise selectively targets small-scale patterns, leaving intact large-scale patterns. In our approach, quantile mapping addresses large-scale biases, while the small-scale biases and downscaling are handled by our diffusion model. The task of our model is then to perform downscaling and bias correction by regenerating these small-scale features, in a way that ensures consistency with the preserved large-scale patterns. When applying our framework on a different region or ESM, it is computationally inexpensive to recompute the quantile mapping (QM) for the embedding transformation.

\section{Results}\label{chapter2}

The ability of the diffusion model $DM$ to approximate $f^{-1}$ and the effectiveness of the transformations $f,g$ will determine the overall performance of the downscaling and bias correction model $\omega = DM \circ g $. 
Therefore, we first investigate the effectiveness of the embedding transformations $f$ and $g$, followed by an analysis of the downscaling and bias correction performance of the diffusion model $DM$, on the observational dataset. Once we have shown that both work as expected, we investigate the performance of the diffusion model in bias correction and downscaling of the ESM precipitation fields.
Without loss of generality, we chose the 0.25° ERA5 reanalysis \cite{hersbach_era5_2020} data as observational data and the state-of-the-art GFDL-ESM4 \cite{dunne_gfdl_2020} at 1° as our ESM. 
\newline

\subsection{Embedding evaluation} 

Transformations $f,g$ are chosen so that they map observational (OBS) and model (ESM) data to a common embedding space $\mathbf{V^{emb}}$, where all samples are identically distributed. For constructing $f$ and $g$ we need $f(ERA5)$ and $g(GFDL)$ to be unbiased with respect to each other. The transformations need to be chosen such that the embedded data share the same distribution and the same power spectral density (PSD). We assess if they are statistically unbiased towards each other by analyzing their histograms and latitude / longitude profiles, as well as their spatial PSDs (after applying pre-processing transformations). Figure \ref*{fig:eval_emb} shows that $f(ERA5)$ and $g(GFDL)$ have the same spatial distribution (fig.~\ref*{fig:eval_emb}A) with minor differences in temporal statistics shown by the histogram (fig.~\ref*{fig:eval_emb}B) and latitudinal/ longitudinal profiles (fig.~\ref*{fig:eval_emb}C and fig.~\ref*{fig:eval_emb}D). \newline

The individual operations that make up the transformations $f$ and $g$ do not change the large-scale patterns of their respective inputs, as desired for a valid bias correction. The goal of downscaling and bias correction $\omega$ (Fig. \ref{fig:main_fig}) is to rely on the unbiased large-scale patterns of the ESM and correct statistics, as well as small-scale patterns. The transformation $g$ preserves the unbiased information from the ESM by construction. Therefore, we want the diffusion model, approximating $f^{-1}$, to also preserve unbiased information.

The extreme case of erasing all detail with large amounts of noise (Fig. \ref{fig:noise_level}A) leads to learning the unconditional distribution $p(ERA5)$, which is then not a correction of $GFDL$ but a generative emulation of the ERA5 reanalysis data. We tested this by adding the same amount of noise to the output of our diffusion model that was added to create $g(GFDL)$. This ensures that both the downscaled and bias-corrected fields, as well as the original GFDL fields, lack the small-scale details up to the same point. 

To verify that large-scale patterns are preserved by the diffusion model, we compute image similarity metrics between the low pass filtered version of the input of the diffusion model (embedded ERA5 data $f(ERA5)$) and the low pass filtered output of the diffusion model $DM(f(ERA5))$. The output of the low pass filter leaves the large-scale features unchanged. The comparison yields an average structural similarity index (SSIM \cite{SSIM}) value of 0.85 and a Pearson correlation coefficient of 0.95 for the validation dataset. This verifies that large-scale patterns are well preserved by the diffusion model. \newline

Our diffusion model is able to reconstruct high-resolution fields following the ERA5 distribution from embedded ERA5 fields $f(ERA5)$, with only minor discrepancies in small-scale patterns (fig. ~\ref*{fig:era5_bc_eval}A). A comparison between the mean absolute spatial-temporal difference between the first downsampled and then bilinearly upsampled ERA5 and the ground truth ERA5 fields at 0.25° yields a mean bias of 0.27 mm d$^{-1}$. The downscaling of our diffusion model reduces this bias to 0.21 mm d$^{-1}$ (at 0.25°). Our diffusion model thus approximates $f^{-1}$ well, and we successfully created a shared embedding space in which $f(ERA5)$ and $g(GFDL)$ are identically distributed.

\subsection{Evaluation of downscaling and bias correction performance} \label{bc_performance}

\begin{figure*}[!htb]
    \centering
    \includegraphics[width=0.7\textwidth]{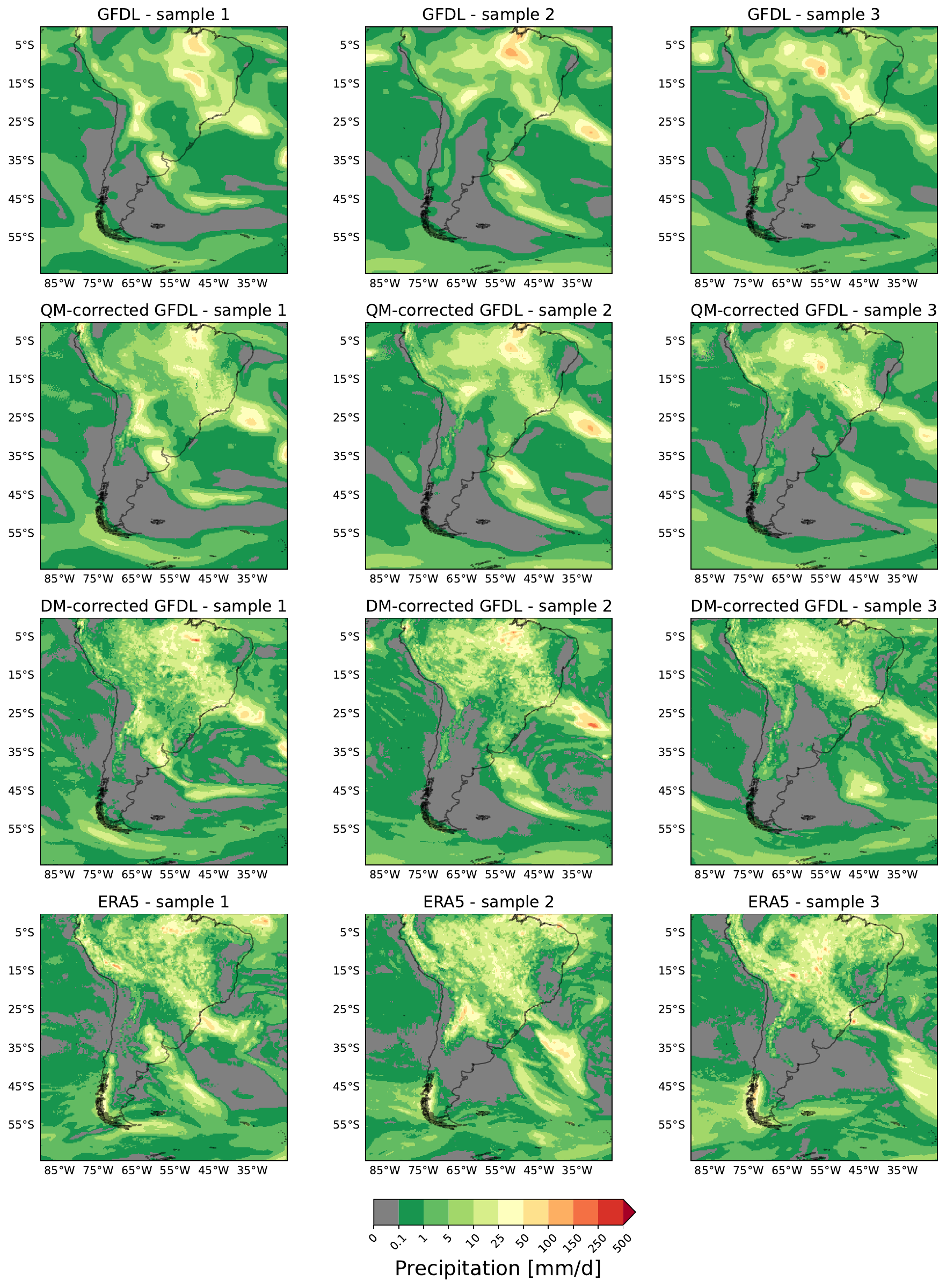}
    \caption{\textbf{Comparative visualization of individual randomly selected samples}. Each row presents three samples of the same dataset. The top row shows GFDL ESM4 data, bilinearly upsampled to 0.25° to match the other fields. The second row shows QM-corrected and the third row diffusion model-corrected GFDL fields. The bottom row shows samples of the original ERA5 data, which are unpaired to the GFDL fields above. Visual inspection shows that the diffusion model correction greatly improves upon the QM correction in terms of producing realistic spatial patterns, since the QM-corrected fields remain way too blurry compared to the HR ERA5 data. The overall large-scale patterns are preserved by the DM. There is no visual difference between the details and sharpness of diffusion model-corrected GFDL fields compared to ERA5.}
    \label{fig:individual_samples}
\end{figure*}

\begin{figure*}[!htb]
    \centering
    \includegraphics[width=1\textwidth]{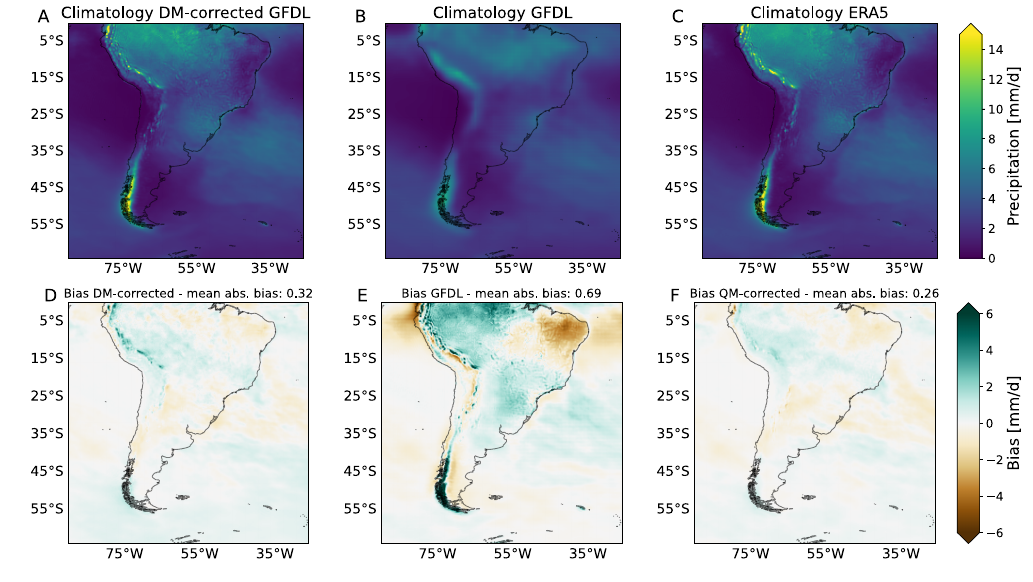}
    \caption{\textbf{Comparison of climatologies and model biases}. The first row shows the climatology of \textbf{(A)} the diffusion model-corrected GFDL at 0.25°, \textbf{(B)} the GFDL ESM4 model, upsampled to 0.25° and \textbf{(C)} the 0.25° ERA5 data. The second row shows the bias of the GFDL and the QM- and diffusion model-corrections, defined as the difference between long-term temporal averages of all validation samples. Specifically, the temporally averaged bias fields with respect to ERA5 are shown for \textbf{(D)} the diffusion model correction, \textbf{(E)} the uncorrected GFDL and \textbf{(F)} the QM correction. Results indicate a substantial improvement of our diffusion model (A) and the benchmark (C) over just upsampling GFDL to 0.25°. The absolute bias on top of each panel is given by the mean absolute value of the differences over the spatial and temporal dimension with respect to ERA5.
    }
    \label{fig:absolute_bias}
\end{figure*}

\begin{figure*}[!htb]
    \centering
    \includegraphics[width=1.0\textwidth]{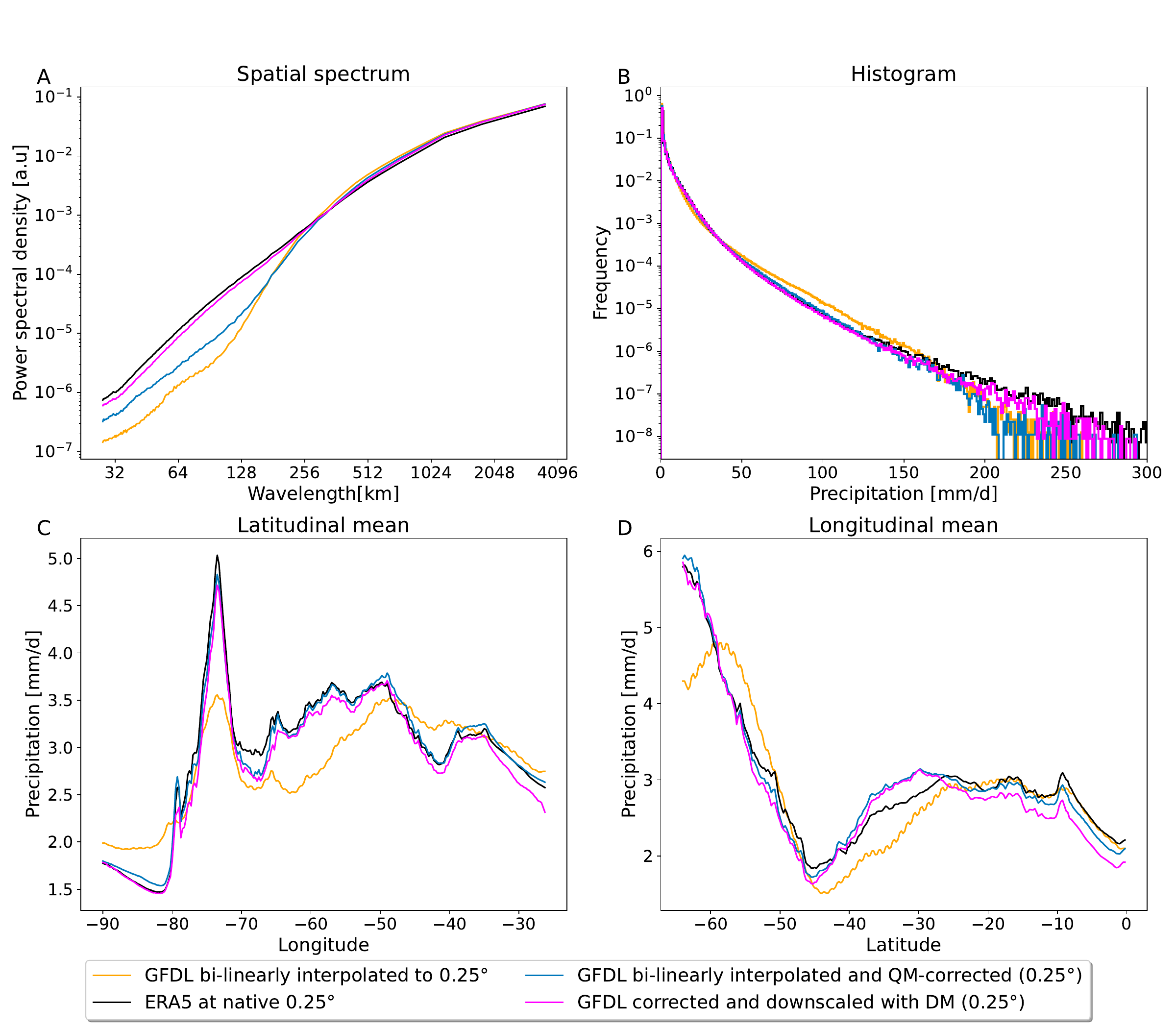}
    \caption{\textbf{Evaluation of our diffusion model's performance for downscaling and bias correction}. Comparison of GFDL (bilinearly upsampled to 0.25°) (orange) 
    and ERA5 (black) to diffusion model-corrected GFDL (magenta) and QM-corrected GFDL fields (blue) as our benchmark. 
    The Power spectral density (PSD) plot \textbf{(A)} shows that the diffusion model corrects the small-scale spatial details far better than our benchmark. The spectrum aligns very well with the high resolution ERA5 target data. The histograms \textbf{(B)} as well as the latitude \textbf{(C)} and longitude \textbf{(D)} profiles show substantial improvements compared to the uncorrected GFDL data.
    }
    \label{fig:gfdl_bc_combined_bias_map}
\end{figure*}

We investigate the inference performance of our diffusion model on embedded GFDL data $g(GFDL)$. We compare the downscaling and bias correction performance of our diffusion model to a benchmark consisting of first applying bilinear upsampling followed by QM for bias correction.

Figure \ref{fig:individual_samples} presents a qualitative comparison between the different individual precipitation fields. The upsampled $GFDL$ fields, as well as our benchmark are visually too smooth. They therefore appear blurry compared to the ERA5 precipitation fields despite having the same spatial resolution of 0.25°. Our diffusion model produces high-resolution detailed outputs that are visually indistinguishable from the $ERA5$ reanalysis that we treat as the ground truth. We also compared our diffusion model to a different state-of-the-art diffusion model implementation, EDM \cite{karras2022elucidating}. The EDM model was trained for the same number of epochs, while taking twice as long for one. The EDM almost perfectly corrects the spectrum (fig.~\ref*{fig:edm}A). However in both the histogram (fig.~\ref*{fig:edm}B) as well as in latitudinal and longitudinal profiles (fig.~\ref*{fig:edm}C and fig.~\ref*{fig:edm}D) the EDM model is inferior to our proposed diffusion model. We also compared our method against a VQ-VAE-based generative model, finding that our model outperforms it across these metrics (for details, see SI Sec. S\ref*{VAE} and fig.~\ref*{fig:vqvae_eval}).

To further validate our choice of architecture, we also compare the diffusion model's performance against two other state-of-the-art deep learning models, a UNet and a Transformer, using the same experimental setup. The results (fig.~\ref*{fig:compare_DM_transformer_CNN}) show a significant advantage for the DM in reproducing small-scale spatial patterns, by aligning better with the ERA5 reference spectrum (fig.~\ref*{fig:compare_DM_transformer_CNN}A). In contrast, all three models perform comparably well in correcting the overall precipitation distribution and the latitudinal/longitudinal mean profiles (fig.~\ref*{fig:compare_DM_transformer_CNN}B-D). The generative process of the diffusion model is particularly well-suited for correcting the high-frequency spatial details. Another advantage over both deterministic models is the DM's stochasticity, which allows for the generation of ensembles to quantify uncertainty.

The analysis of temporally averaged precipitation fields shows that the climatology of the diffusion model-corrected GFDL data (Fig. \ref{fig:absolute_bias}A) and the high-resolution ERA5 data (Fig. \ref{fig:absolute_bias}C) is more accurate and less smooth than the climatology of the GFDL data (Fig. \ref{fig:absolute_bias}B). A comparison between the absolute temporally and absolute spatial-temporally averaged diffusion model corrected GFDL and ERA5 fields (Fig. \ref{fig:absolute_bias}D) yields a bias of 0.32 mm d$^{-1}$. This is a substantial improvement over the original GFDL dataset, which yields a bias of 0.69 mm d$^{-1}$ (Fig. \ref{fig:absolute_bias}E). Our diffusion model performs comparably with the state-of-the-art bias correction performance of our benchmark, which is by design optimal for this task, at 0.26 mm d$^{-1}$ (Fig. \ref{fig:absolute_bias}F). For a quantitative comparison including Root Mean Square Error (RMSE) and Pearson correlation for these climatologies, see Table ~\ref*{tab:metrics}.

There are large differences between the GFDL and ERA5 data in small-scale patterns (Fig. \ref{fig:gfdl_bc_combined_bias_map}A). The histogram of precipitation intensities (Fig. \ref{fig:gfdl_bc_combined_bias_map}B) also confirms that the ESM is only really accurate for precipitation events up to 40 mm d$^{-1}$, after which the respective frequencies diverge. The latitudinal and longitudinal mean profiles (Fig. \ref{fig:gfdl_bc_combined_bias_map}C and Fig. \ref{fig:gfdl_bc_combined_bias_map}D) indicate the presence of regional biases.

Our framework demonstrates comparable skill to the QM-based benchmark in correcting the latitude and longitude profiles, for which QM is near optimal by construction (Fig. \ref{fig:gfdl_bc_combined_bias_map}C and Fig. \ref{fig:gfdl_bc_combined_bias_map}D). Comparing the histograms (Fig. \ref{fig:gfdl_bc_combined_bias_map}B and fig.~\ref*{fig:absolute_errors}) shows that our diffusion model is superior compared to the benchmark, strongly outperforming it for extreme values, in particular.

For the spatial patterns and especially the small-scale spatial features, the QM benchmark shows only slight improvements over the original GFDL data (Fig. \ref{fig:gfdl_bc_combined_bias_map}A). The diffusion model is vastly superior in correcting these small-scale spatial patterns (Fig. \ref{fig:gfdl_bc_combined_bias_map}A and Fig. \ref{fig:individual_samples}) and almost completely removes the small-scale biases, as seen in the spatial PSD. 

To verify that large-scale patterns are preserved by the diffusion model, we compute image similarity metrics between the low-pass-filtered embedded GFDL data and the low-pass filtered output of the diffusion model. The comparison yields an average structural similarity index value (SSIM \cite{SSIM}) of 0.77 and a Pearson correlation coefficient of 0.90, verifying that large-scale patterns are well preserved by the diffusion model. 

We also assess our model's performance on extreme precipitation events. For this, we use the R95p metric, which is defined as the total annual precipitation from wet days (PR $>$ 1 mm d$^{-1}$) that exceed the 95th percentile of our reference period. The difference between the R95p values for the ERA5 and DM corrected GFDL (fig.~\ref*{fig:r95p}A), the ERA5 and QM corrected GFDL (fig.~\ref*{fig:r95p}B) and ERA5 and GFDL (fig.~\ref*{fig:r95p}C), demonstrate that the diffusion model effectively corrects the bias in extreme precipitation events, performing at least as well as the quantile mapping correction. 
To further test the model's performance on correcting characteristics of rainfall events in the tail of the distribution, we conduct a return-level analysis for extreme rainfall events (fig.~\ref*{fig:return_intervals}). We calculated the average return periods for both moderately extreme ($>$50 mm d$^{-1}$) and very extreme ($>$80 mm d$^{-1}$) events. The raw GFDL model has a significant wet bias, substantially underestimating the return periods (3.33 years and 4.60 years) compared to the ERA5 reference (4.11 years and 7.38 years). Our DM successfully mitigates this bias, yielding more realistic return periods of 4.18 and 7.98 years. 

We show that the spatial correlation between the climatologies is improved through our method by computing the Pearson correlation between the temporally averaged fields. The Pearson correlation between ERA5 and GFDL climatology is 0.83, while the correlation between ERA5 and DM-corrected GFDL is 0.98, which is the same as that for the QM-corrected GFDL data. We also investigate how our DM captures the statistics of consecutive dry days (CDD) and consecutive wet days (CWD) compared to the QM benchmark and the raw GFDL (fig.~\ref*{fig:dryday}). Our diffusion model produces superior CDD (fig.~\ref*{fig:dryday_diff}A and fig.~\ref*{fig:dryday_diff}B) and CWD (fig.~\ref*{fig:dryday_diff}D and fig.~\ref*{fig:dryday_diff}E) statistics compared to our QM benchmark and GFDL, as shown in the difference plots of CDD / CWD.

Our method therefore accurately preserves the large-scale precipitation content, while successfully correcting small-scale structure of the precipitation fields, as well as statistical biases in histograms and latitude / longitude profiles (Fig. \ref{fig:gfdl_bc_combined_bias_map}). Finally, we confirmed the temporal consistency of our model by analyzing autocorrelation (fig.~\ref*{fig:AC}) and seasonal spell duration (fig.~\ref*{fig:seasonal_scatter}). We further validated the robustness of our metrics over an extended validation period (1995-2014) (fig.~\ref*{fig:extended_metrics}).

We also test our framework on a different region of similar size over South Asia. We choose the same GFDL dataset and keep the experimental setup and evaluation identical to the South American region. The setup for quantile mapping the South Asia GFDL data and creating the benchmark data is also the same. We retrained our DM on mapping embedded ERA5 data (over South Asia) to the original ERA5 data. The noising scale in this experiment is the same as for South America, as the PSDs for both regions diverge around the same spatial scale. The evaluation (fig.~\ref*{fig:asia_metrics}) confirms that our DM successfully corrects precipitation biases in this new region and most notably outperforms the QM baseline in representing small-scale spatial features.

To further assess our framework's robustness, we conducted an additional experiment using a different ESM. We replaced the GFDL dataset with the MPI-ESM-HR model while keeping the experimental setup and evaluation protocol identical. The MPI and GFDL data diverge at a similar spatial scale in the PSD over the South American domain, allowing us to use the same noising scale hyperparameter s. Quantile delta mapping was applied in the same way as for the GFDL data. Consequently, our diffusion model did not require retraining and could be applied directly to the embedded MPI data at inference. Evaluation on our main metrics (fig.~\ref*{fig:MPI_metrics}) demonstrates our framework's ability to generalize to different ESMs. Our DM not only restores spatial variability across all scales significantly better than the QM benchmark (fig.~\ref*{fig:MPI_metrics}A), but also shows superior ability to reproduce the frequency of extreme precipitation events (fig.~\ref*{fig:MPI_metrics}B).

\subsection{Evaluation of ensemble spread}

One of the key strengths of our method lies in its capability to generate a diverse ensemble of downscaled and bias-corrected fields from a single condition. We therefore evaluate the ability of our diffusion model to represent and produce accurate estimates of uncertainty, a critical aspect for robust climate modeling and decision-making. We generate a 50-member DM ensemble by running the model 50 times, each conditioned on the same low-resolution ERA5 year, producing one-year trajectories. The corresponding high-resolution year serves as the ground truth. Our results demonstrate that the DM ensemble effectively reproduces the correct precipitation patterns, as shown by the close alignment between the ensemble mean and the high resolution ground truth of ERA5 over the annual cycle (fig.~\ref*{fig:ensemble_mean}). Probabilistic performance, evaluated using CRPS, highlights that the DM significantly outperforms a bilinear baseline, with lower mean CRPS values (0.76 mm d$^{-1}$ vs 0.90 mm d$^{-1}$), as well as better temporally and spatially averaged CRPS (fig.~\ref*{fig:crps}). Furthermore, we confirm that the DM ensemble produces well-calibrated uncertainty estimates with a spread-skill plot. Our model achieves near-perfect alignment with the 1:1 line, indicating an accurate representation of uncertainty (fig.~\ref*{fig:spread_skill_plot}). For more details see SI Sec. S\ref*{ensembleSI}.

\subsection{Evaluation on future climate scenarios}

\begin{figure}[htb]
    \centering
    \includegraphics[width=0.8\textwidth]{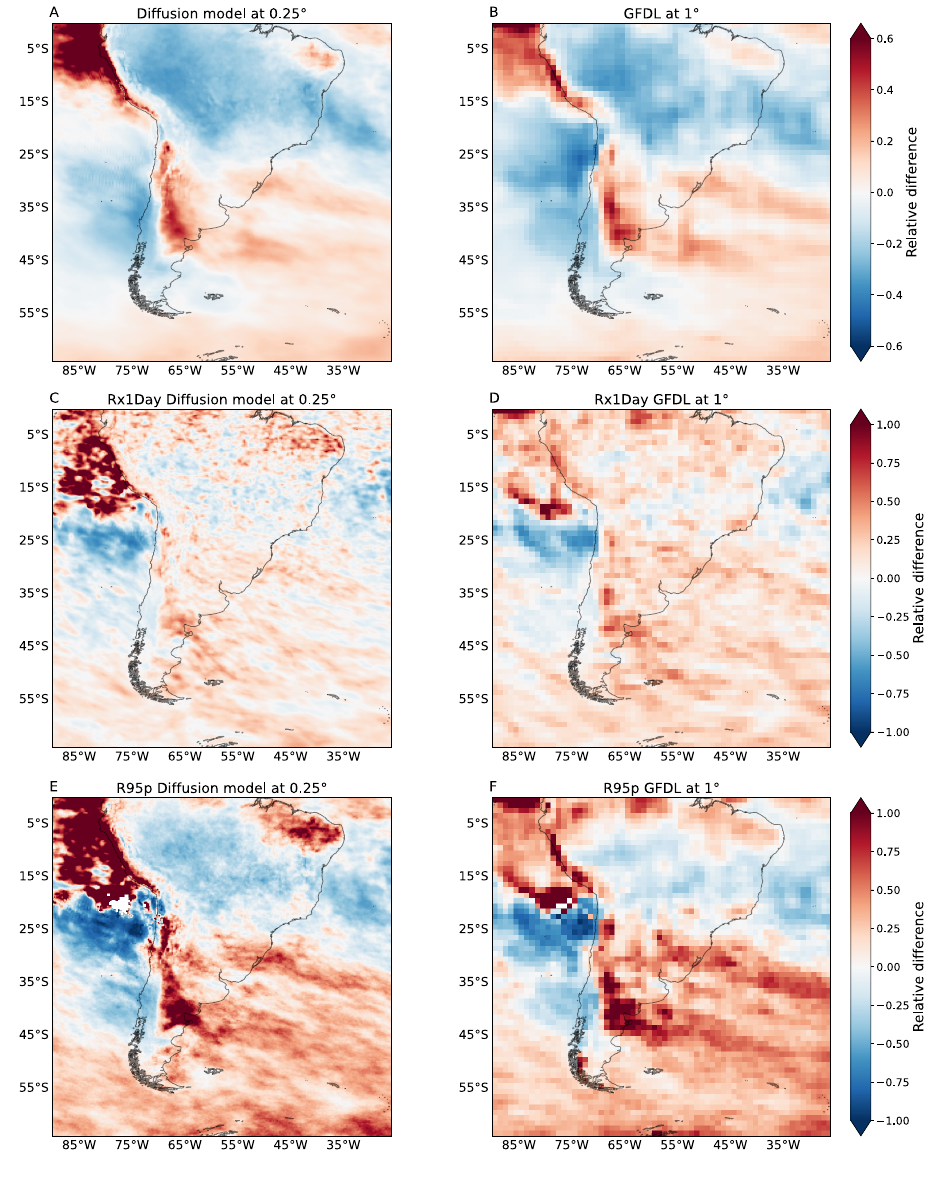}
    \caption{\textbf{Comparison of relative climate change signals}. We compute the relative climate change signal between the late 21st century (2081-2100) under the GFDL SSP5-8.5 scenario and the historical GFDL period (1995–2014). In \textbf{(A)} and \textbf{(B)}, we show that our diffusion model successfully preserves the mean precipitation climate change signal in the downscaled 0.25° GFDL fields, matching the change of the original 1° GFDL data. Positive values (red) indicate an intensification of precipitation, while negative values (blue) indicate reductions. In \textbf{(C-D)} and \textbf{(E-F)} we evaluate how well the DM-correction preserves the climate change signal for extreme events in historical and future scenarios. Both the Rx1Day \textbf{(C-D)} as well as R95p in \textbf{(E-F)} show that the DM-downscaling does preserve the climate change signal for extreme events. There are only slight differences over the north western part of our fields, where the DM-correction predicts more slightly more extremes for the SSP5-8.5 scenario. This is in line with the bias correction capabilities of the DM, correcting the under-representation of extreme precipitation in the original GFDL data. }
    \phantomsection\label{fig:climate_change_signal}
\end{figure}

Evaluating the performance of downscaling models is crucial for their application in climate impact studies under future climate scenarios. We assess our diffusion model's ability to preserve climate change signals in the underlying ESM simulations by applying it to a high-emission future scenario (SSP5-8.5). Figure \ref{fig:climate_change_signal} compares the relative climate change signal between the late 21st century (2081-2100) and the historical period (1995–2014) for annual mean and annual extreme precipitation. We find that our downscaled 0.25° fields successfully capture the mean precipitation change, closely matching the pattern and magnitude shown in the original 1° GFDL data (Fig. \ref{fig:climate_change_signal}A and Fig. \ref{fig:climate_change_signal}B). The diffusion model also robustly preserves the climate change signal for extreme precipitation indices, including Rx1Day (wettest day for each year) and R95p (Fig. \ref{fig:climate_change_signal}C - Fig. \ref{fig:climate_change_signal}F). The spatial patterns of change for the extremes are well-reproduced in the DM-corrected output compared to the original model data. Notably, slight differences are observed in the northwestern domain (Fig. \ref{fig:climate_change_signal}C and Fig. \ref{fig:climate_change_signal}E), where the DM-correction projects a slightly stronger increase in extreme events under SSP5-8.5. A slight increase in extremes aligns with the diffusion model's bias correction capabilities, reflecting its role in addressing the known under-representation of extreme precipitation in the original GFDL simulations.

Furthermore, we demonstrate that our conditionally trained diffusion model generalizes robustly to unseen future emission scenarios by accurately preserving regional precipitation trends without requiring retraining. We analyze the full annual mean precipitation timeseries from 2015 to 2100 over two representative regions, one exhibiting a strong negative trend and one with a pronounced positive trend (fig.~\ref*{fig:trends_ssp_regional}). For each region, we compare the annual mean precipitation from the original GFDL SSP5-8.5 data at 1° with the DM-corrected output at 0.25° resolution. The diffusion model consistently preserves the direction and magnitude of the trends found in the original GFDL data across the entire timeseries, for both the negative (fig.~\ref*{fig:trends_ssp_regional} blue) and positive trend (fig.~\ref*{fig:trends_ssp_regional} red) regions. This demonstrates the model’s ability to maintain physically meaningful long-term changes in precipitation, further supporting its generalization capability to future scenarios. Note that the absolute values do not have to coincide, as our model corrects the bias and hence the numerical values. Our model can generalize to unseen climates, preserving the trends, since there is no decrease in performance during inference on GFDL SSP5-8.5 data. Note that our set-up generalized to unseen climate scenarios without any external constraints. The reason why our model preserves trends well is likely given by the fact that the trend is dominated by the large-scale patterns and our model learned to rely on the large-scale patterns of the condition and only generates small-scale patterns.

\section{Discussion}\label{chapter3}

We introduced a framework based on generative machine learning that allows both bias correction and downscaling of Earth system model fields with a single diffusion model. We achieve this by first mapping observational fields and ESM data to a shared embedding space and then applying the learned inverse of the observation embedding transformation to the embedded ESM fields. We learn the inverse transformation with a conditional diffusion model. Although the underlying observational and ESM fields are unpaired, our framework allows for training on paired data (between observations and embedded observations, see above) and therefore any supervised machine learning method can be adopted to the task, which allows for more flexibility. Supervised methods are often superior in performance and more natural for the downscaling application. The diffusion model is trained on individual samples and has successfully learned to reproduce the statistics of observational data. For the observational ground truth, we chose the ERA5 reanalysis, and for the ESM data to be corrected and downscaled, we chose fields from GFDL-ESM4. 

We demonstrated our framework's robustness and generalizability in two additional experiments (Sec. \ref{bc_performance}). When applying the model to a new geographical region in South Asia with the same ESM, the DM requires retraining to adapt to the new regional characteristics. In contrast, when applying the framework to an entirely different ESM (MPI-ESM) over the South American region, the core DM did not need to be retrained since the same noising scale hyperparameter could be used. For different ESMs a new noising scale hyperparameter could be necessary, requiring retraining of our DM with a different noising scale; however this depends on the choice of the spatial scale below which bias correction is desired, and for comparable outputs, we recommend to keep the noising scale $s$ fixed for different ESMs. For example, to correct multiple ESMs at once, one can use the most heavily biased model to select the noising scale. A single diffusion model can then be trained to correct all ESMs at once, saving significant computational resources during inference. In general, we expect that many ESMs (like the MPI and GFDL model we use) will have similar spatial scales up to which they can capture realistic spatial precipitation features, because they have a similar resolution and have similar limitations from parameterization schemes. In all cases, readjusting the computationally inexpensive Quantile Delta Mapping (QDM) is a required step in the embedding process. The results will also depend on the specific quantile mapping scheme, QDM is chosen to preserve trends.
\newline

Our diffusion model corrects small-scale biases of the ESM fields, while completely preserving the large-scale structures, which is key for impact assessments, especially with regard to extremes and local impacts in terms of floods or landslides. The diffusion model performs particularly well for extreme events where traditional methods struggle. The method improves the temporal precipitation distribution at the grid cell level and surpasses the state-of-the-art approach (quantile mapping) in correcting spatial patterns. The downscaling performance has also been shown to be excellent. The diffusion model manages to generate small-scale details for the low resolution ESM data, that match those of high resolution observations. Our model preserves relevant information from the large scales, such as trends and extremes, and generates bias corrected and downscaled precipitation fields with adequate uncertainties. \newline

We show that our method is robust in the out-of-distribution setting of downscaling and bias-correcting the SSP5-8.5 future emission scenario. It is critical for impact assessments that our model is able to accurately preserve the climate change signal of the original SSP5-8.5 data. \newline

A key innovation of our approach is the embedding strategy, which makes the training process independent of the source ESM (apart from a single data-dependent hyperparameter setting the spatial scale below which the fields are corrected), which not only allows the framework to be flexibly applied to downscale and bias-correct a wide range of ESMs but also allows it to be used with different state-of-the-art machine learning backbone models.
Another key advantage of our framework is its data efficiency. In our conditional approach the model only needs to learn how to generate small-scale features given the large-scale ones. The task is considerably less demanding than that of unconditional models (e.g., Hess et al. \cite{hess2025fast}), which must learn the entire data distribution from scratch during training. This data efficiency makes our method applicable to datasets with shorter record lengths than ERA5, such as newer observational products.

Indeed, comparing results for generated climatologies between our conditional DM and the unconditional consistency model (CM) by Hess et al. \cite{hess2025fast}, it becomes apparent that the CM struggles to learn the target distribution accurately, leading to blurring (fig.~\ref*{fig:CM_comp}) that would hinder applications for impact assessments.\newline

Our method is not specific to ERA5 and GFDL because the training of the diffusion model does not directly depend on the ESM choice. A specific ESM choice will only modify a hyperparameter in the embedding transformations $f$ and $g$. This, however, requires almost no fine-tuning, as the temporal frequencies can always be matched with quantile mapping. The only parameter that might change for different datasets and use cases is the amount of noise that is added to the observational and ESM datasets. We choose the amount of noise such that the PSDs of the observational ground truth and the ESM fields align beyond a certain scale. This means that we have complete flexibility in deciding which patterns we want to preserve and which we want to correct. This is a major advantage over existing GAN based approaches.\newline

We can decrease the level of detail that is preserved by the diffusion model through increasing the amount of noise added in the transformations $f$ and $g$. The amount of noise added is directly proportional to the freedom the diffusion model has in generating diverse outputs and inversely proportional to the model's ability to preserve large-scale patterns. \newline

The downscaled and bias corrected fields will automatically inherit time consistency between different samples up to the noising scale. This means that ESM fields showing two successive days will still look like two successive days after the correction. Future work could build a video diffusion model that inputs and outputs full time series instead of single frames, in order to guarantee time consistency across all scales. \newline

We focused on precipitation data over the South American continent, because of its heavily tailed distribution and the pronounced spatial intermittency. Especially at small scales, precipitation data is extremely challenging to model and therefore serves as a reasonable choice to show the framework's capabilities in a particularly difficult setting. Regional data is chosen due to computational constraints, yet the diverse terrain of our study region, encompassing land, sea, and a wide range of altitudes, enables robust testing of the downscaling and bias correction performance, also given the substantial biases of the GFDL model in this region. We also conducted additional experiments for another region over South Asia, and using another ESM, namely the MPI-ESM-HR, in order to confirm the generality of our approach. The extension to global scales is straightforward and requires no major changes in the architecture. We intend to include more variables in a consistent manner on a global scale in future research. Optimizing the inference strategy, with speedup techniques such as distillation \cite{luhman_knowledge_2021}, to decrease the sampling time will prove helpful in this context.  \newline

As for any ML model, the ability to generate the rarest extremes is limited by their frequency in the training data. Our conditional approach helps mitigate this to some extent by inheriting the large-scale patterns for these events directly from the ESM. \newline

It is straightforward to extend our methodology to downscaling and bias correction of numerical or data-driven weather predictions on short- to medium-range or even seasonal temporal scales. This would not require any fundamental changes to the architecture. This would, however, require a target dataset with sufficiently high resolution. The ability of the diffusion model to not disturb the temporal consistency between samples can be useful in this scenario. Future work could then focus on extending this model to a multivariate setting, which would be essential for weather prediction and for assessing physical consistency between variables.

\section{Materials and Methods}\label{chapter4}
\subsection{Data} 
For the study region, we focus on the South American continent and the surrounding oceans. Specifically, the targeted area spans from latitude 0°N to 63°S and from longitude -90°W to -27°E. For the ablation study of the South Asian region, we selected an area from 0.75°N to 64.5°N latitude and from 42°E to 105.75°E longitude. The training period comprises ERA5 data from 1992-01-01 to 2011-01-01. The range of years included for the evaluation on ERA5 and GFDL spans from 2011-01-02 to 2014-12-01. Additionally, an extended 20-year window (1995–2014) is used for analyses requiring greater statistical robustness.

\subsubsection*{ERA5}
ERA5 \cite{hersbach_era5_2020} is a state-of-the-art atmospheric reanalysis dataset provided by the European Center for Medium-Range Weather Forecasting (ECMWF). Reanalysis refers to the process of combining observations from various sources, such as weather stations, satellites, and other instruments, with a numerical weather model to create a continuous and comprehensive representation of the Earth's atmosphere. We use the daily total precipitation data at 0.25° horizontal resolution as the target for the diffusion model.

\subsubsection*{GFDL}
The climate model output is taken from a state-of-the-art ESM from Phase 6 of the Coupled Model Intercomparison Project (CMIP6), namely GFDL-ESM4 \cite{dunne_gfdl_2020}. We abbreviate the model with GFDL throughout the paper. The dataset contains daily precipitation data of the first ensemble member (r1i1p1f1) of the historical simulation (esm-hist). The data is available from 1850 to 2014, at 1° latitudinal and 1.25° longitudinal resolution and a daily temporal resolution. 

GFDL-ESM4 \cite{dunne_gfdl_2020} SSP5-8.5 represents a high-emission future pathway. We use daily-resolution data from the CMIP6 archive, provided at 1° latitude and 1.25° longitude spatial resolution, covering the period from 2015 to 2100.

\subsubsection*{MPI}
For our ablation study, we repeat our experiments for the MPI-ESM HR model \cite{gutjahr2019max}. We abbreviate MPI-ESM-HR with MPI in the paper. The data has 0.9375°×0.9375° spatial resolution. We use daily data from 1992 to 2014 using data from 1992-2011 for training and 2011 to 2014 for inference.

\subsection*{Benchmark dataset}

In order to benchmark our method, we first apply bilinear interpolation to increase the resolution of the GFDL fields from 1° to 0.25°. After that, we apply quantile delta mapping \cite{cannon2015bias} to fit the upsampled GFDL data to the original 0.25° ERA5 data. QM is fitted on past observations and can then be used to correct the statistics of any (past/present) ESM field towards that reference period. We use quantile delta mapping (QDM) and chose the ERA5 training period from 1992-01-01 to 2011-01-01 as the reference period to fit the GFDL to ERA5. The benchmark dataset to evaluate our approach is then constructed by applying QM to the GFDL validation period (2011-01-02 to 2014-12-01). Some analyses required a longer evaluation period (1995-2014). To create a fair benchmark for these specific cases QDM was also recalibrated, it was both fitted and applied using data exclusively from this 1995-2014 window. For the SSP5-8.5 data, we use the 1995 to 2014 period of ERA5 as reference data and the historical GFDL data as the model input to fit the QDM. We then apply this mapping to the full time period of the GFDL SSP5-8.5 data (2015–2100).

\subsection*{Data pre-processing}

The units of the GFDL data and MPI data are $\text{kg m}^{-2} \text{s}^{-1}$, and for ERA5 $\text{m} \text{h}^{-1}$. For consistency, both are transformed to $\text{mm d$^{-1}$}$.

Our pre-processing pipeline consists of: 

\begin{itemize}
    \item Only GFDL: rescaling the original 1°$\times$ 1.25° GFDL data to 1$\times$1° (64$\times$64 pixel). 
    \item Only MPI: rescaling the original 0.9375°$\times$ 0.9375° GFDL data to 1$\times$1° (64$\times$64 pixel). 
    \item Add +1 mm d$^{-1}$ precipitation to each value in order to be able to apply a log-transformation to the data.
    \item Apply the logarithm with base 10 in order to compress the range of values.
    \item Standardize the data, i.e. subtract the mean and divide by the standard deviation to facilitate training convergence.
    \item Transform the data to the range [-1,1] to facilitate the convergence of the training.

\end{itemize}

An ablation study (fig.~\ref*{fig:compare_preprocess}) confirms the choice of our precipitation pre-processing pipeline, showing that omitting the log-transformation or the final range scaling leads to spectral discrepancies or distributional biases. As part of the transformation $g$, the 1° GFDL data is bilinearly upsampled. This and the downsampling and upsampling of ERA5 data, which is part of $f$, are already done during pre-processing.
The downsampling of 0.25° ERA5 data (256$\times$256 pixel) to 1° (64$\times$64 pixel) is done by only keeping every fourth pixel in each field. For the just mentioned upsampling, we apply bilinear interpolation to increase the resolution from 1° to 0.25°. Note that bilinear interpolation to 0.25° does not increase the amount of information in the images compared to the 1° fields. After preprocessing the data as described, the embedding transformation $f$ is applied. The diffusion model is trained with the preprocessed $f(ERA5)$ as a condition and the original 0.25° ERA5 data as a target.
Before we apply the embedding transformation $g$ we first pre-process the 1° GFDL data by applying quantile delta mapping (QDM \cite{cannon2015bias}) with 500 quantiles. The bilinear upsampling is then used to increase the resolution to 0.25$\times$0.25° (256$\times$256 pixels). The preprocessed data are used as input to the embedding transformation $g$. The corresponding output serves as the condition during the inference process of the diffusion model

\subsection{Embedding framework} 

Our framework introduces transformations $f$ \& $g$ that map OBS and ESM data to a shared embedding space $f: \mathbf{V^{obs}} \rightarrow \mathbf{V^{emb}}$ and $g: \mathbf{V^{esm}} \rightarrow \mathbf{V^{emb}}$. The goal is to do bias correction and downscaling of ESM fields, i.e., to obtain samples from the conditional distribution $\omega = p(OBS|ESM)$. Training a conditional model to approximate this distribution directly is not possible because OBS and ESM are unpaired. Therefore, we will train the model without the ESM data, only using OBS data and utilize a trick to enable transfer learning and inference on the ESM data. We apply transformations on ESM and OBS such that the resulting datasets are similarly distributed and therefore allow for generalization. The arrows in the diagram of Figure \ref{fig:main_fig} show that we can represent the mapping that achieves the bias correction and downscaling as $\omega = f^{-1} \circ g$. Our idea is to approximate $f^{-1}$ with a neural network $f^{-1} \approx \epsilon $. We chose a conditional diffusion model (DM), denoted by the conditional distribution $p(OBS|f(OBS))$, to approximate $f^{-1} = DM: \mathbf{V^{emb}} \rightarrow \mathbf{V^{obs}}$. The diffusion model (Fig. \ref{fig:main_fig}C) is only trained on pairs $(OBS,f(OBS))$. The shared embedding space allows us to evaluate the trained model on ESM embeddings $p(OBS|g(ESM))$, as all embeddings are identically distributed.

\subsubsection{Constructing the embedding space} 

The goal of $f$ and $g$ is to map OBS and ESM to a shared embedding space, where $f(OBS)$ and $g(ESM)$ are identically distributed (Fig. \ref{fig:main_fig}). To achieve this, both embedded datasets need to be unbiased towards each other. OBS and ESM are biased towards each other in terms of statistical biases between distributions and biases between small-scale patterns visible in the spatial power spectral density (PSD) (fig.~\ref*{fig:era5_bc_eval}A). 

As mentioned earlier, the input for the embedding transformation $f$ is 0.25° ERA5 data, which is first preprocessed, then downsampled and upsampled. The input to the embedding transformation $g$ is the preprocessed and upsampled 0.25° GFDL data. By first downsampling ERA5 to 1° and then upsampling it to 0.25° we ensure that the fields match the information content of the original 1° GFDL fields.

To remove small-scale pattern bias, we apply a noising procedure analogous to the forward diffusion process as part of $f$ and $g$. Gaussian noise contains all frequencies in equal measure and the Fourier transform of Gaussian noise is itself Gaussian noise, so its power must be equal across all frequencies in expectation. The power spectrum of pure Gaussian noise corresponds to a horizontal line in the spectrum of Fig. \ref{fig:noise_level}A, reflecting the fact that it contains all frequencies in equal amounts. Adding noise to an image results in a hinge shape in the PSD of the noisy images (Fig. \ref{fig:noise_level}B, \ref{fig:noise_level}C and \ref{fig:noise_level}D). Increasing the variance of the noise increases its power and, as a result, its PSD will shift upward. Adding noise hence acts as a low-pass filter, while the variance of the added noise determines the cut-off frequency. Increasing variance leads to higher cut-off points as the power of the noisy frequencies increases. Both ERA5 and GFDL data are noised up to the cutoff frequency, denoted by $s$. The scale $s$ is determined by the point where ERA5 and the ESM data (in our case GFDL) start to disagree in their spatial PSDs (Fig. \ref{fig:noise_level}), i.e., the intersection between the two. Adding noise in this way ensures that $f(ERA5)$ is unbiased compared to $g(GFDL)$ in the PSD by erasing all information beneath $s$. In our implementation, the transformations $f$ and $g$ utilize the same cosine scheduler as the forward diffusion process to add Gaussian noise to the data. ERA5 data undergoes 50 noise steps within $f$, while $g$ applies the same 50 noise steps to the GFDL data. We ensure that the observational and ESM data have aligned distributions by incorporating Quantile Mapping (QM) directly into the transformation $g$. It only needs to be included in $g$. The quantile-mapped and bilinearly downscaled data is then noised as described above, as part of the embedding transformation.
It is important to clarify that QM is not included because the diffusion model is unable to do bias correction. QM is only used as a tool in our framework to ensure that in the embedding space $f(ERA5)$ and $g(GFDL)$ are identically distributed, such that $g(GFDL)$ can be used for the inference of the diffusion model.

\subsubsection{Determining the noising scale}

The choice of the spatial scale $s$ influences up to which scale we correct the spatial PSD. We note that the PSD shows spectral distributions normalized to 1; therefore, we can still observe slight changes above $s$ when small-scale patterns are corrected. The point $s$ is a hyperparameter chosen before training and purely depends on the datasets ESM and OBS and can be adjusted to the specific needs in a given context and task.

In the extreme case, where $s$ is maximal, the conditional images will contain pure noise (Fig. \ref{fig:noise_level}A). In this case, the diffusion model is equivalent to an unconditional model. As an unconditional model, the diffusion model will correct all biases at all spatial scales, however, at the expense of completely losing any paring between the condition and the output. We chose $s$ to be at the intersection of the ERA5 and GFDL spectrum around 512 km (Fig. \ref{fig:noise_level}B). Thereby, we trust in the ESM's ability to model large-scale structures above the point $s$, which we do not want to correct with the diffusion model.

\subsection{Network architecture and training}

The general architecture of our diffusion model $DM$ consists of a Denoising Diffusion Probabilistic Model (DDPM) architecture \cite{ho2020denoising} conditioned on low resolution images. For details about diffusion models and conditional diffusion models, see SI Sec. S\ref*{secDM} and SI Sec. S\ref*{secCDM}. We employ current state-of-the-art techniques to facilitate faster convergence and find the following to be important for convergence and sample quality \cite{saharia2022photorealistic}: The memory efficient architecture, ``Efficient U-Net'', in combination with dynamic clipping and noise conditioning augmentation \cite{ho2022cascaded} turned out to be effective for our relatively small dataset. We adopt the Min-SNR \cite{hang2023efficient} formulation to weight the loss terms of different timesteps based on the clamped signal-to-noise ratios. The diffusion model architecture utilizes a cosine schedule for noising the target data and a linear schedule for the condition during noise condition augmentation with 100 steps each. The diffusion model is trained to do v-prediction. The U-Net follows the $64 \times 64 \rightarrow 256 \times 256$ Efficient U-Net architecture \cite{saharia2022photorealistic}. The diffusion model has approximately 730 million trainable parameters and is trained for 100 epochs using the ADAM optimizer \cite{kingma2014adam} with a batch size of 2 and a learning rate of $1e^{-4}$. Note that in the case of fig.~\ref*{fig:era5_bc_eval}, where the inference data is also embedded OBS data and there is no ESM data present, the model performs better when being trained and evaluated with 1000 denoising steps, instead of the 100 steps that we used in all our experiments that include ESM data. The model with 100 steps is superior in training and inference speed and also in correcting the histograms, when correcting ESM data. We also compared the effect of not adding noise (SI Sec. S\ref*{secNoise}) and the effect of not applying QM (SI Sec. S\ref*{secQM}) as shown in Figures \ref*{fig:climatology_SI}, \ref*{fig:no_noise_metrics}, \ref*{fig:cdd_cwd_special}, \ref*{fig:r95special}, \ref*{fig:bias_SI}, as well as different noise choices (SI Sec. S\ref*{secdiffNoise}, fig.~\ref*{fig:comp_diff_noise_lvls}) during both training and inference.

\clearpage

\section*{Acknowledgments}

\subsection*{Funding}
MA acknowledges funding from the Excellence Strategy of the Federal Government and the L\"ander through the TUM Innovation Network EarthCare. \\
SB, and NB acknowledge funding by ClimTip. This is ClimTip contribution \#21; the ClimTip project has received funding from the European Union's Horizon Europe research and innovation program under grant agreement No. 101137601.\\
PH, SB, and NB acknowledge funding by the Volkswagen Foundation.\\
BP acknowledges funding by the National Key R\&D Program of China (2021YFA0718000).\\
YH acknowledges the Alexander von Humboldt Foundation for the Humboldt Research Fellowship.\\

\subsection*{Data and Materials Availability}

All data needed to evaluate the conclusions in the paper are present in the paper and/or the Supplementary Materials.

The ERA5 reanalysis data is available for download at the Copernicus Climate Change Service (\url{https://cds.climate.copernicus.eu/cdsapp#!/dataset/reanalysis-era5-single-levels?tab=overview}).
The CMIP6 GFDL-ESM4 is available at \url{https://esgf-data.dkrz.de/search/cmip6-dkrz/}. \\

The code is available on GitHub (\url{https://github.com/aim56009/ESM_cdifffusion_downscaling_bc.git}). \\

\section*{List of Supplementary Materials}
Supplementary Text \\
Figures S1 to S28
\clearpage

\nocite{van2017neural, gneiting2007strictly, haynes2023creating}

\clearpage

\appendix

\setcounter{figure}{0}
\setcounter{table}{0}
\setcounter{equation}{0}
\renewcommand{\thefigure}{S\arabic{figure}}
\renewcommand{\thetable}{S\arabic{table}}
\renewcommand{\theequation}{S\arabic{equation}}

\section{Supplementary Text}

\subsection{Unconditional Diffusion Models}\label{secDM}

Diffusion Models can be separated into two parts, a forward and a backward diffusion process. The forward diffusion process is a probabilistic model $ q(\mathbf{x}_t|\mathbf{x}_{t-1})$ that produces a noisy version of a given image $\mathbf{x}_t$ in $t$ noising steps. The model is chosen to be a Gaussian model: $q(\mathbf{x}_t|\mathbf{x}_{t-1}) = \mathcal{N}(\mu(\mathbf{x}_{t-1}),\beta_{t} \mathbf{I})$, where $\beta_t$ controls the amount of noise that is added in each step. In other studies, the model is often chosen to be of the form $q(\mathbf{x}_t|\mathbf{x}_{t-1}) = \mathcal{N}(\sqrt{1-\beta_{t} } \mathbf{x}_{t-1},\beta_{t} \mathbf{I})$ \citep{ho2020denoising}. 
In practice, we use the reparametrization trick to sample from a Gaussian distribution by $\mathcal{N}(\mu, \sigma) = \mu + \sigma \epsilon $ where $\epsilon\sim \mathcal{N}(0, 1)$. 
Thus, a noisy version of $\mathbf{x}_0$ can be obtained as $\mathbf{x}_t = \sqrt{1-\beta_{t} } \mathbf{x}_{t-1} + \sqrt{\beta_{t}} \epsilon$ after the $t$ noising steps.The noise scheduler $\beta_{t}$ is chosen to add small amounts of noise in the beginning and larger amounts later, to preserve a reasonable amount of information throughout the process. 

The backward process $q(\mathbf{x}_{t-1}|\mathbf{x}_t)$ models how to restore the previous version $q(\mathbf{x}_{t-1})$ of a given image $\mathbf{x}_t$ at a certain noise step $t$. This process is also modelled by a Gaussian $q(\mathbf{x}_{t-1}|\mathbf{x}_t) = \mathcal{N}(\mu(\mathbf{x}_{t}),\sigma{(\mathbf{x}_t)} )$. The problem is that $\mu(\mathbf{x}_{t})$ is not known.

Using Bayes' theorem, the model can be rewritten as a product of Gaussians: $q(\mathbf{x}_{t-1}|\mathbf{x}_t) = \frac{q(\mathbf{x}_t|\mathbf{x}_{t-1}) q(\mathbf{x}_{t-1})}{q(\mathbf{x}_t)}$.
Each term is a Gaussian distribution, and their product is also a Gaussian distribution. Computing the product and taking the mean of that expression is a valid way to model the backward process. However, in practice, the distribution $q(\mathbf{x}_{t-1})$ is unknown, so we cannot explicitly compute $q(\mathbf{x}_{t-1}|\mathbf{x}_t)$.

Predicting the state before a noising operation $\mathbf{x}_{t-1}$ can be done by conditioning on the noisy image $\mathbf{x}_t$ and the noise free image $\mathbf{x}_0$
\begin{equation}
q(\mathbf{x}_{t-1}|\mathbf{x}_t,\mathbf{x}_0) = \frac{q(\mathbf{x}_t|\mathbf{x}_{t-1},\mathbf{x}_0) q(\mathbf{x}_{t-1}|\mathbf{x}_0)}{q(\mathbf{x}_t|\mathbf{x}_0)} 
\end{equation}

The terms on the right-hand side are Gaussian and can be explicitly computed. The resulting Gaussian has a mean term that depends on $\mathbf{x}_t$ and $x_0$, while the variance is a constant depending on the time step $t$. 

\begin{align}
q(\mathbf{x}_{t-1} \mid \mathbf{x}_t, \mathbf{x}_0) & = \mathcal{N}(\mathbf{x}_{t-1}; \tilde{\boldsymbol{\mu}}(\mathbf{x}_t, \mathbf{x}_0), \sigma_t^2 \mathbf{I}) \label{eq:raw_bw}\\
{\sigma}_t^2 & = \frac{1 - \bar{\alpha}_{t-1}}{1 - \bar{\alpha}_t} \cdot \beta_t \\
\tilde{\boldsymbol{\mu}}_t(\mathbf{x}_t, \mathbf{x}_0) & = \frac{\sqrt{\bar{\alpha}_{t-1}} \beta_t}{1 - \bar{\alpha}_t} \mathbf{x}_0 + \frac{\sqrt{\alpha_t}(1 - \bar{\alpha}_{t-1})}{1 - \bar{\alpha}_t} \mathbf{x}_t \label{eq:mu_t}
\end{align}

with $\alpha_t=1-\beta_t, \quad \bar{\alpha}_t=\prod_{s=0}^t \alpha_s$.

The following equation describes how $\mathbf{x}_0$ is connected to $\mathbf{x}_t$, when applying the forward diffusion model $T$ times: 

\begin{equation}\label{eq:fw_noise}
\begin{aligned}
\mathbf{x}_t & = \sqrt{1-\beta_t} \mathbf{x}_{t-1} + \sqrt{\beta_t} \epsilon_{t-1} \\
& = \sqrt{\alpha_t} \mathbf{x}_{t-2} + \sqrt{1-\alpha_t} \epsilon_{t-2} \\
& = \ldots \\
& = \sqrt{\bar{\alpha}_t} \mathbf{x}_0 + \sqrt{1-\bar{\alpha}_t} \boldsymbol{\epsilon}
\end{aligned}
\end{equation}

where $\epsilon, \ldots, \epsilon_{t-2}, \epsilon_{t-1} \sim \mathcal{N}(\mathbf{0}, \mathbf{I})$. Solving for $x_0$ yields:

\begin{equation}\label{eq:x_0}
\mathbf{x}_0=\frac{1}{\sqrt{\bar{\alpha}_t}}\left(\mathbf{x}_t-\sqrt{1-\bar{\alpha}_t} \epsilon\right).
\end{equation}

Combining Eq. \ref{eq:x_0} with Eq. \ref{eq:mu_t} leads to:
\[
\tilde{\boldsymbol{\mu}}_t\left(\mathbf{x}_t\right)=\frac{1}{\sqrt{\alpha_t}}\left(\mathbf{x}_t-\frac{\beta_t}{\sqrt{1-\bar{\alpha}_t}} \boldsymbol{\epsilon}\right).
\]

The backward process (Eq. \ref{eq:raw_bw}) is then modelled as:
\begin{equation}
q(\mathbf{x}_{t-1}|\mathbf{x}_t) = \mathcal{N}\left(\frac{1}{\sqrt{\alpha_t}}\left(\mathbf{x}_t - \frac{\beta_t}{\sqrt{\bar\alpha_t}} \epsilon\right), {\sigma}_t^2 \mathbf{I} \right),
\end{equation}

so given a noisy image in step t, this model will predict a less noisy version of that image $\mathbf{x}_{t-1}$. The only unknown in this equation is $\epsilon$. The idea is to parameterize $\epsilon$ with a neural network ${\epsilon}_\theta$. 
The objective of the network is then to estimate the noise that was added to a (noisy) image $\mathbf{x}_{t-1}$ at each time step $t$:

\begin{equation}\label{eq:final_mu}
\tilde{\boldsymbol{\mu}_\theta}\left(\mathbf{x}_t, t\right)=\frac{1}{\sqrt{\alpha_t}}\left(\mathbf{x}_t-\frac{\beta_t}{\sqrt{1-\bar{\alpha}_t}} \boldsymbol{\epsilon}_\theta\left(\mathbf{x}_t, t\right)\right)
\end{equation}

Using the reparametrization trick and inserting Eq. \ref{eq:final_mu} into Eq. \ref{eq:raw_bw}, the backward diffusion process (also called de-noising process) denotes as: 

\begin{equation}
q(\mathbf{x}_{t-1}|\mathbf{x}_t) = \mathcal{N}\left(\frac{1}{\sqrt{\alpha_t}}\left(\mathbf{x}_t-\frac{\beta_t}{\sqrt{1-\bar{\alpha}_t}} \boldsymbol{\epsilon}_\theta\left(\mathbf{x}_t, t\right)\right), {\sigma}_t^2 \mathbf{I} \right).
\end{equation}

Following the reparametrization trick, every iteration of the backward process takes the form: 

\begin{equation}
\mathbf{x}_{t-1} \leftarrow \frac{1}{\sqrt{\alpha_t}}\left(\mathbf{x}_t-\frac{\beta_t}{\sqrt{1-\bar{\alpha}_t}} {\epsilon}_\theta\left(\mathbf{x}_t, t\right)\right)+\sigma_t\boldsymbol{\epsilon}_t
\end{equation}

with $\boldsymbol{\epsilon}_t \sim \mathcal{N}(\mathbf{0}, \mathbf{I})$.

The neural network in the backward diffusion process can be learned through the following algorithm proposed by \citet{ho2020denoising}: 
\[
\begin{array}{l}
\hline
\textbf{Algorithm 1: }\text{Training} \\
\hline
1: \textbf{repeat} \\
\begin{array}{l}
2: \quad \mathbf{x}_0 \sim q(\mathbf{x}_0) \\
3: \quad t \sim \text{Uniform}(\{1, \ldots, T\}) \\
4: \quad \boldsymbol{\epsilon} \sim \mathcal{N}(\mathbf{0}, \mathbf{I}) \\
5: \quad \text{Take gradient descent step on} \\
       \quad \quad \nabla_\theta\left\|\boldsymbol{\epsilon} - \boldsymbol{\epsilon}_\theta\left(\sqrt{\bar{\alpha}_t} \mathbf{x}_0 + \sqrt{1-\bar{\alpha}_t} \boldsymbol{\epsilon}, t\right)\right\|^2 \\
6: \quad \textbf{until } \text{converged} \\
\end{array} \\
\hline
\end{array}
\]

Model inference can be achieved with the following algorithm \citet{ho2020denoising}: 
\[
\begin{array}{l}
\hline
\textbf{Algorithm 2: }\text{Sampling} \\
\hline
\begin{array}{l}
1: \mathbf{x}_T \sim \mathcal{N}(\mathbf{0}, \mathbf{I}) \\
2: \textbf{for } t=T, \ldots, 1 \textbf{ do} \\
3: \quad \mathbf{z} \sim \mathcal{N}(\mathbf{0}, \mathbf{I}) \text{ if } t>1, \text{ else } \mathbf{z}=\mathbf{0} \\
4: \quad \mathbf{x}_{t-1} = \frac{1}{\sqrt{\alpha_t}}\left(\mathbf{x}_t - \frac{1-\alpha_t}{\sqrt{1-\bar{\alpha}_t}} \boldsymbol{\epsilon}_\theta(\mathbf{x}_t, t)\right) + \sigma_t \mathbf{z} \\
5: \textbf{end for} \\
6: \textbf{return } \mathbf{x}_0
\end{array} \\
\hline
\end{array}
\]

The inference corresponds to generating a noise-free image, given a noisy input image. 

\subsection{Conditional Diffusion Models}\label{secCDM}

The goal of a conditional diffusion model is to learn an approximation of the distribution $p(\mathbf{x}|c)$, where $c$ is some conditional information. The method learns to model the reverse diffusion process as $p_{\theta}(\mathbf{x}_{t-1} | \mathbf{x}_t, c)$. Starting with a pure noisy image, it gets denoised in $T$ steps. The forward diffusion process is identical to the unconditional case. The difference to the unconditional case is that the model has knowledge about the condition during the backward process. Theoretically, the model could also be conditioned during the forward process. The backward process looks as follows: 

\begin{equation}
\mathbf{x}_{t-1} \leftarrow \frac{1}{\sqrt{\alpha_t}}\left(\mathbf{x}_t-\frac{\beta_t}{\sqrt{1-\bar{\alpha}_t}} {\epsilon}_\theta\left(\mathbf{x}_t, c, t\right)\right)+\sigma_t\boldsymbol{\epsilon}_t.
\end{equation}

The condition is integrated by concatenating the condition and the noisy image in the color channel of the images. The network takes a two-channel image as input and produces a one-channel image as output in each backward step.

\section{Comparative Analysis of Diffusion Model Variants}
\renewcommand{\thefigure}{S\arabic{figure}} 

\subsection{No noise}\label{secNoise}

In order to show the importance of noising the biased small scales of the ESM condition, we conducted an experiment where no noise was applied to the condition during both training and inference. The climatology (fig. \ref{fig:climatology_SI}D) of the noiseless model is very similar to our proposed model (fig. \ref{fig:climatology_SI}E), with a mean absolute bias of 0.29 mm d$^{-1}$. The difference between the two methods becomes evident when examining the spatial PSD (fig. \ref{fig:no_noise_metrics}A), where our method is far superior in correcting the small-scale biases. The histogram (fig. \ref{fig:no_noise_metrics}B) further highlights the superiority of our approach (noise level n=50), particularly in the range from 50-150 mm d$^{-1}$. Furthermore, the latitudinal and longitudinal profiles (fig. \ref{fig:no_noise_metrics}C and fig. \ref{fig:no_noise_metrics}D) reveal that the output of the noise-free model is overall less accurate and exhibits too much variability compared to our proposed approach. We conclude that the model's poorer performance stems from its overreliance on biased information from the GFDL conditions during inference. During training, the model was conditioned on low-resolution ERA5 fields, which are nearly unbiased compared to high-resolution ERA5. As a result, the model only had to make minimal corrections during training and thus will do the same at inference. The good performance of our method is not limited to daily precipitation statistics;
we also find that it performs better in representing the ERA5 statistics of consecutive dry and consecutive wet days (fig. \ref{fig:cdd_cwd_special}C and fig. \ref{fig:cdd_cwd_special}F). For the precipitation extremes, the R95p for not noising the data appears slightly worse than with our model (fig. \ref{fig:r95special}A and fig. \ref{fig:r95special}C).

\subsection{Different noise levels}\label{secdiffNoise}

We tested the effect of different noise levels (added to the conditions) by analyzing their impact on the downscaling results, maintaining the same noise level during both training and inference. The results for the noiseless case (n=0) were discussed in the previous paragraph, while our originally proposed approach uses a noise level of n=50. The dependence of climatology on the noise level appears small for the smaller noise levels n=0 (fig. \ref{fig:climatology_SI}D and fig. \ref{fig:bias_SI}D) and n=50 (fig. \ref{fig:climatology_SI}E and fig. \ref{fig:bias_SI}E), with a mean absolute bias of 0.32 mm d$^{-1}$. However, at a higher noise level (n=80), the climatology deviates slightly (fig. \ref{fig:climatology_SI}F) and the bias becomes more pronounced (fig. \ref{fig:bias_SI}F), with the mean absolute bias increasing to 0.36 mm d$^{-1}$. When looking at the PSD (fig. \ref{fig:comp_diff_noise_lvls}A) we find that our model (n=50) performs best. Also the histogram (fig. \ref{fig:comp_diff_noise_lvls}B) and latitudinal as well as longitudinal profiles (fig. \ref{fig:comp_diff_noise_lvls}C and fig. \ref{fig:comp_diff_noise_lvls}D) show that the choice of noise level n=50 is optimal. One could argue that, since high noise levels (n=80) remove more information, the model's reconstruction task is harder, because it needs to learn more dependencies and cannot rely on the condition as much. This shows that the conditional information is indeed valuable for the generation. Hence, as we train all models for the same amount of time, we would expect the diffusion model to initially perform worse when removing more information under the same amount of training time.

\subsection{No quantile mapping}\label{secQM}

Fig. \ref{fig:no_noise_metrics} (orange) shows that the diffusion model without quantile mapping (QM) struggles to fully correct the characteristic biases in the GFDL data. The climatology and mean average bias presented in fig. \ref{fig:climatology_SI}C and fig. \ref{fig:bias_SI}C highlight the critical role of QM as part of the embedding transformation. Without QM, the model reduces the mean absolute bias from 0.69 mm d$^{-1}$ to 0.59 mm d$^{-1}$, but falls short of the 0.32 mm d$^{-1}$ achieved when QM is applied. The spatial Pearson correlation between the temporally averaged fields of the DM bias-corrected GFDL data without quantile mapping is 0.89, which is still an improvement over 0.83 of the raw GFDL data. The PSD (fig. \ref{fig:no_noise_metrics}A) shows that the diffusion model is still perfectly able to correct the spectrum of the embedded GFDL data without applying quantile mapping. The reason why the method without QM struggles with the histogram (fig. \ref{fig:no_noise_metrics}B) and latitudinal and longitudinal profiles (fig. \ref{fig:no_noise_metrics}C and fig. \ref{fig:no_noise_metrics}D) is that without QM, the training and inference distribution of the condition are very different from the distribution of the training condition. The transformation learned by the model during training does not generalize to the out-of-distribution condition when QM is excluded. The PSD is almost the same as before, as applying QM has almost no effect on the spatial variability of individual fields. The application of QM also helps to represent the ERA5 statistics of consecutive dry and consecutive wet days (fig. \ref{fig:cdd_cwd_special}B and fig. \ref{fig:cdd_cwd_special}E). Extreme precipitation, as represented by the R95p metric in fig. \ref{fig:r95special}A with QM and fig. \ref{fig:r95special}B without QM, also benefits from the application of QM. 

Overall, we find that QM is an essential part of our method, by ensuring that the embedded GFDL data is distributed like the embedded ERA5 data. In other words, it is important because of our unpaired data setting. Alone, however, it cannot correct the spatial PSD (Fig. 5A) and is therefore not useful for downscaling. The histogram (Fig. 5B) is also corrected a lot better with our diffusion model. Applying our DM also improves the CDD (fig. \ref{fig:cdd_cwd_special}A and fig. \ref{fig:cdd_cwd_special}B) and CWD (fig. \ref{fig:cdd_cwd_special}D and fig. \ref{fig:cdd_cwd_special}E) statistics compared to the benchmark without QM.

\section{Comparative Analysis to VQ-VAE model} \label{VAE}

We have included an additional comparison with an established generative model, the VQ-VAE. This model is primarily designed for representation learning and compression, with its training objective focused on mapping data to a latent space and reconstructing it. To adapt the VQ-VAE for our downscaling task, we employ a two-step training process inspired by the original VQ-VAE work \citep{van2017neural}.

Step 1: We train the VQ-VAE to compress and reconstruct high-resolution (HR) ERA5 images.
Step 2: Using the encoder of the trained model, we construct a dataset of latent representations for each HR field. A conditional PixelCNN is then trained to autoregressively model the prior distribution $p(z \mid c)$ where $c$ are samples from the embedded ERA5 distribution. This allows us to sample latents $z$ conditioned on the embedded ERA5 fields, which are subsequently decoded by the VQ-VAE decoder to generate HR ERA5 fields.

A notable limitation of the VQ-VAE is its inability to generate high-frequency information, as shown by the power spectral density compared to the HR ERA5 ground truth (fig. \ref{fig:vqvae_eval}A). Furthermore, the histogram reveals that the VQ-VAE performs significantly worse in capturing the target distribution for high precipitation values (fig. \ref{fig:vqvae_eval}B). The model also struggles to consistently match the latitudinal and longitudinal means with the ground truth (fig. \ref{fig:vqvae_eval}C, fig. \ref{fig:vqvae_eval}D).

Overall, the VQ-VAE lacks the fine detail in small-scale variability and falls short in overall accuracy compared to our diffusion model approach.

\section{Ensemble uncertainty evaluation}

\renewcommand{\thefigure}{S\arabic{figure}} 

For the following evaluations, we generate a 50 member DM ensemble by conditioning our model 50 times on the same low-resolution ERA5 year. This results in a 50-member ensemble of one-year trajectories. The resulting ground truth will be the corresponding high-resolution ERA5 year.

\subsection{Ensemble mean evaluation} 
\renewcommand{\thefigure}{S\arabic{figure}} 
\label{ensembleSI}

To evaluate the accuracy of the DM ensemble in reproducing precipitation patterns, we compare the spatially averaged daily precipitation of a 50-member ensemble of downscaled high-resolution fields, obtained by conditioning our DM on low-resolution (i.e. upscaled) ERA5 fields, to the corresponding ERA5 high-resolution ground truth, at 0.25° resolution over one year. Figure \ref{fig:ensemble_mean} illustrates that the ensemble mean generated by the DM closely aligns with the ERA5 high-resolution ground truth throughout the annual cycle. This demonstrates the ability of the diffusion model ensemble to -- on average -- capture the temporal variability of precipitation while maintaining well-calibrated ensemble members.

\subsection{Continuous Ranked Probability Score}

We evaluate the probabilistic downscaling performance of our model using the Continuous Ranked Probability Score (CRPS), which extends the concept of the mean absolute error (MAE) to probabilistic forecasts (for details and definition, see \citep{gneiting2007strictly}). 
In our case, each downscaling realization corresponds to a forecast realization in tasks like weather forecasting. To compute the CRPS, we compare 50 downscaled outputs from our diffusion model to the corresponding high-resolution ERA5 ground truth. The 50-member ensemble is generated by running the diffusion model 50 times, always conditioned on the same low-resolution ERA5 year. As a baseline, we use a deterministic, bilinearly upsampled version of the low-resolution ERA5 reference year and a 50-member VQ-VAE ensemble generated the same way as the DM. The corresponding high-resolution ERA5 year serves as the ground-truth reference. This results in a CRPS for every day of the year and every spatial location. Comparing the spatially and temporally averaged CRPS values (0.90 mm d$^{-1}$ vs 0.76 mm d$^{-1}$) and the maximum CRPS values (14.31 mm d$^{-1}$ vs 26.66 mm d$^{-1}$), our DM significantly outperforms the baseline of bilinear upsampling, while also slightly surpassing the VQ-VAE in terms of both the mean CRPS value (0.80 mm d$^{-1}$) and the maximum CRPS value (15.79 mm d$^{-1}$) (lower CRPS is better).

In fig. \ref{fig:crps}A we compare the CRPS time series over 1 year, where we averaged the spatial dimensions. Our model is consistently below the bilinearly upsampled baseline and performs on par with the VQ-VAE. Next, the annual mean CRPS is computed, and then the difference is taken between the bilinearly upsampled ERA5 and the diffusion model ensemble (fig. \ref{fig:crps}B), as well as between the bilinearly upsampled ERA5 and the VQ-VAE ensemble (fig. \ref{fig:crps}C). The results show that both models outperform the bilinear baseline over the continent. Positive differences (blue) indicate a higher (worse) CRPS value for the upsampling baseline. The performance of the DM and VQ-VAE is comparable, except in the southern part of the Andes, where our DM consistently outperforms the baseline.

\subsection{Spread-skill plot}

We evaluated the statistical consistency of our model using a spread-skill plot, which evaluates the relationship between the predicted root mean squared spread (RMSS) and the root-mean-squared error (RMSE) of the ensemble mean. The spread-skill plot relates the predicted model spread to the actual model error. We follow the implementation of \citet{haynes2023creating}, for more details see their work. The x-axis represents the average standard deviation of the DM ensemble distribution, while the y-axis shows the RMSE of the model's mean prediction. Each point on the plot corresponds to a bin of predicted spread values. The model uncertainty estimate is biased if spread values are above the 1:1 line (under-dispersive, over-confident) and if they are under the 1:1 line (over-dispersive / under-confident). A perfect calibration is a spread–skill of 1:1, along the diagonal line.

In our comparison, we use 50 ensemble members generated by conditioning once the DM and once our VQ-VAE model on one year of low-resolution ERA5 data. Both our DM and also our VQ-VAE show very good calibration overall. However, the diffusion model is superior for very large spreads, where the VQ-VAE is overconfident (fig. \ref{fig:spread_skill_plot}). As desired, our DM matches the 1:1 spread-skill line, indicating that its uncertainty is well calibrated.

\subsection{Temporal consistency}

Our framework preserves the temporal structure inherited from the ESM and, to some extent, even corrects existing biases. We validated this over a longer validation period between 1995 and 2014. The raw GFDL data shows excessively high temporal autocorrelation, our DM corrects this, closely matching the ERA5 reference (fig. \ref*{fig:AC}). Similarly, the DM produces more realistic statistics for consecutive dry and wet days (CDD/CWD). This improvement is consistent across seasons, with the model substantially reducing GFDL's biases in spell durations (fig. \ref*{fig:seasonal_scatter}). Finally, we confirmed that all primary evaluation metrics are robust over the longer validation period, where the DM yields consistent results (fig. \ref*{fig:extended_metrics}) compared to Figure \ref*{fig:gfdl_bc_combined_bias_map}. Our method is in general not suited to produce fields that are time consistent on the small spatial scales, as it was trained on single snapshots during training. Achieving this would require training a video model on sequences of frames.

\section{Quantitative Performance Summary}

To provide a coarse quantitative comparison of the model performance, we calculate the Mean Absolute Bias, Pearson Correlation ($r$), and Root Mean Square Error (RMSE) between the temporally averaged climatologies of the reference data (ERA5) and the various model outputs. While these global metrics do not capture fine-scale spatial realism, they serve as a sanity check for overall model fidelity.

\renewcommand{\thetable}{S\arabic{table}}
\setcounter{table}{0} 

\begin{table}[ht]
\centering
\caption{\textbf{Quantitative comparison of model climatologies.} Metrics are calculated between the 0.25$^\circ$ ERA5 reference climatology and the respective model climatologies over the validation period (2011--2014). Lower values are better for Bias and RMSE; higher values are better for Pearson Correlation.}
\label{tab:metrics}
\begin{tabular}{@{}lccc@{}}
\toprule
\textbf{Model} & \textbf{Mean Absolute Bias [mm d$^{-1}$]} & \textbf{Pearson $r$} & \textbf{RMSE [mm d$^{-1}$]} \\ \midrule
GFDL & 0.69 & 0.83 & 1.19 \\
QM-Corrected GFDL & 0.26 & 0.98 & 0.37 \\
Diffusion Model (Ours) & 0.32 & 0.98 & 0.44 \\ \bottomrule
\end{tabular}
\end{table}

\clearpage

\clearpage
\renewcommand{\thefigure}{S\arabic{figure}}

\begin{figure*}
    \centering
    \includegraphics[width=0.7\textwidth]{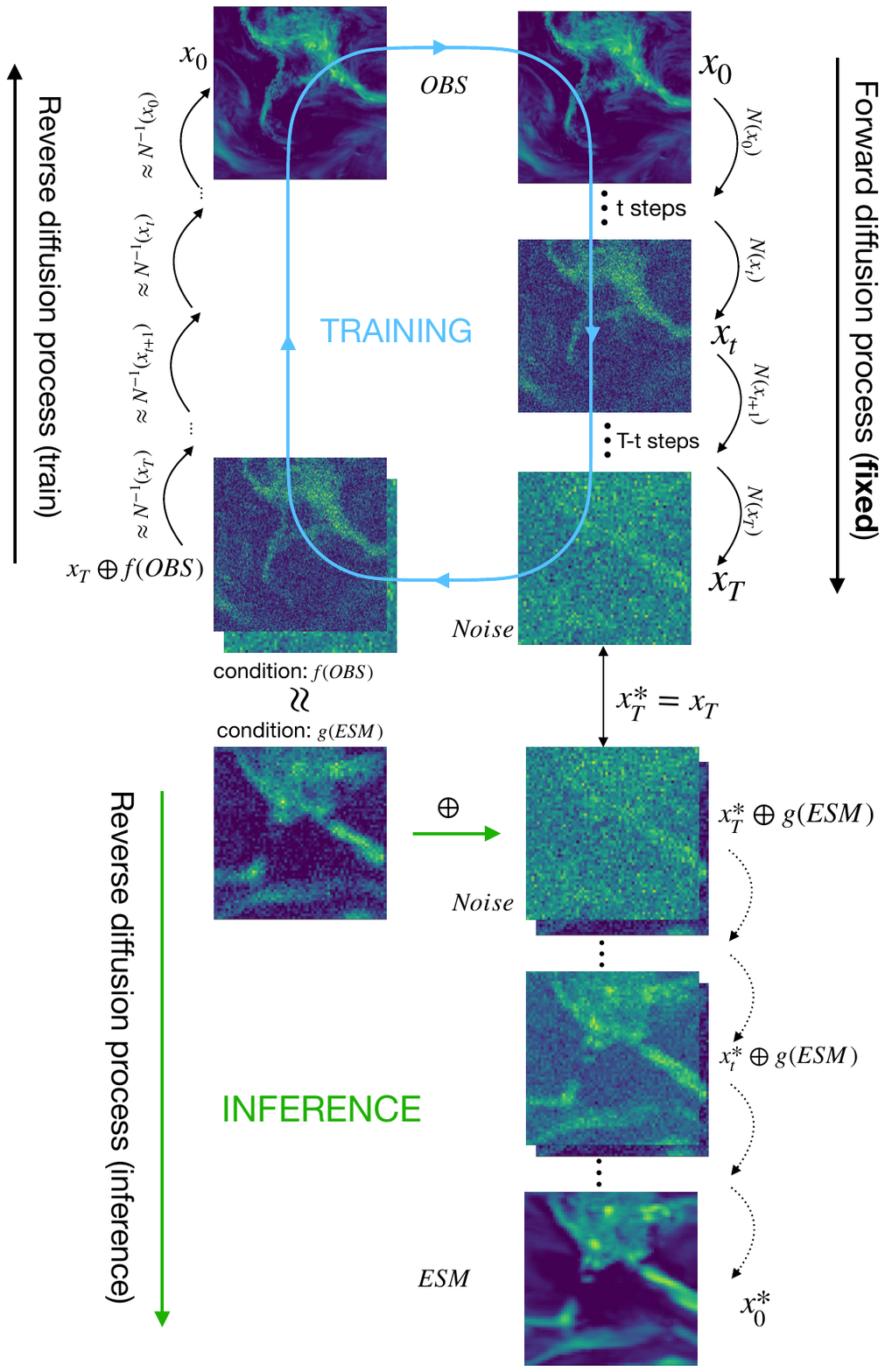}
    \caption{\textbf{Detailed visualisation of the training (blue) and inference process (green) of the conditional diffusion model introduced in our study}. During the forward process (top to center), Gaussian noise $N(x_t)$ is added to an input image $x_0$ over $t$ steps, following Eq.\ref{eq:fw_noise}, until $x_t$ contains just noise. During the backward process (center to bottom) we concatenate a conditional image $f(OBS)$ with the noisy image $x_t$ and train a neural network to remove the noise. The inference process uses the trained model. We also concatenate a purely noisy image $x_t*$ with the condition $g(ESM)$ and remove the noise from $x_t*$ with our trained model. The resulting image $x_0*$ follows the same distribution as $x_0$, indicating that the bias correction and downscaling is achieved. The arrows between the training and inference part indicate that the image distribution at which the training ends is the same as the distribution the inference starts with. }
    \phantomsection\label{fig:diffusion_model}
\end{figure*}

\clearpage

\begin{figure*}
    \centering
    \includegraphics[width=\textwidth]{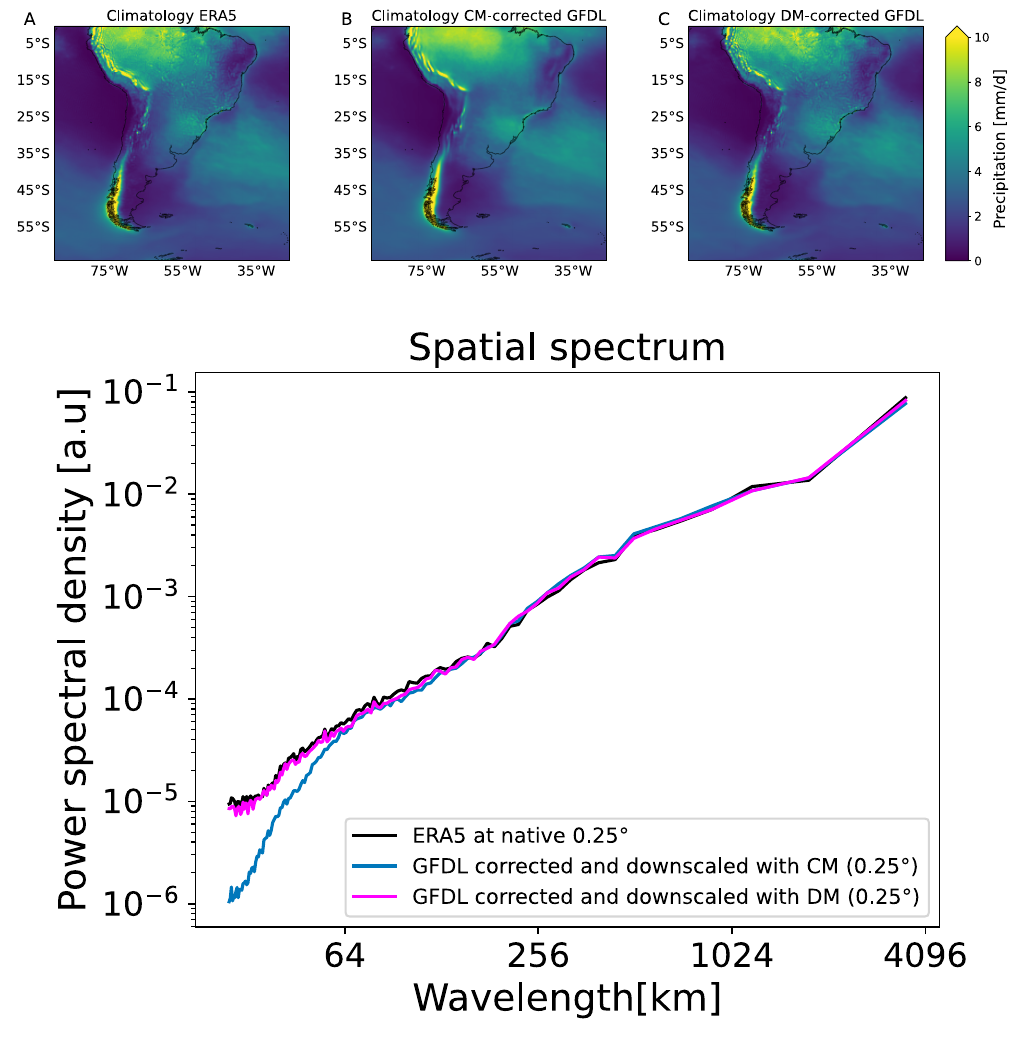}
    \caption{\textbf{Comparison of climatology to Consistency Model}.
    Comparing the 0.25° climatologies of ERA5 \textbf{(A)}, the consistency model \textbf{(B)} and our diffusion model correction \textbf{(C)}, we can see that the consistency model struggles to learn the target distribution accurately, leading to substantial blurring compared to ERA5 and our DM. The lack of detail in the small spatial scales is also apparent in the spatial PSD \textbf{(D)}.
    }
    \phantomsection\label{fig:CM_comp}
\end{figure*}

\begin{figure*}[!htb]
    \centering
    \includegraphics[width=1.1\textwidth]{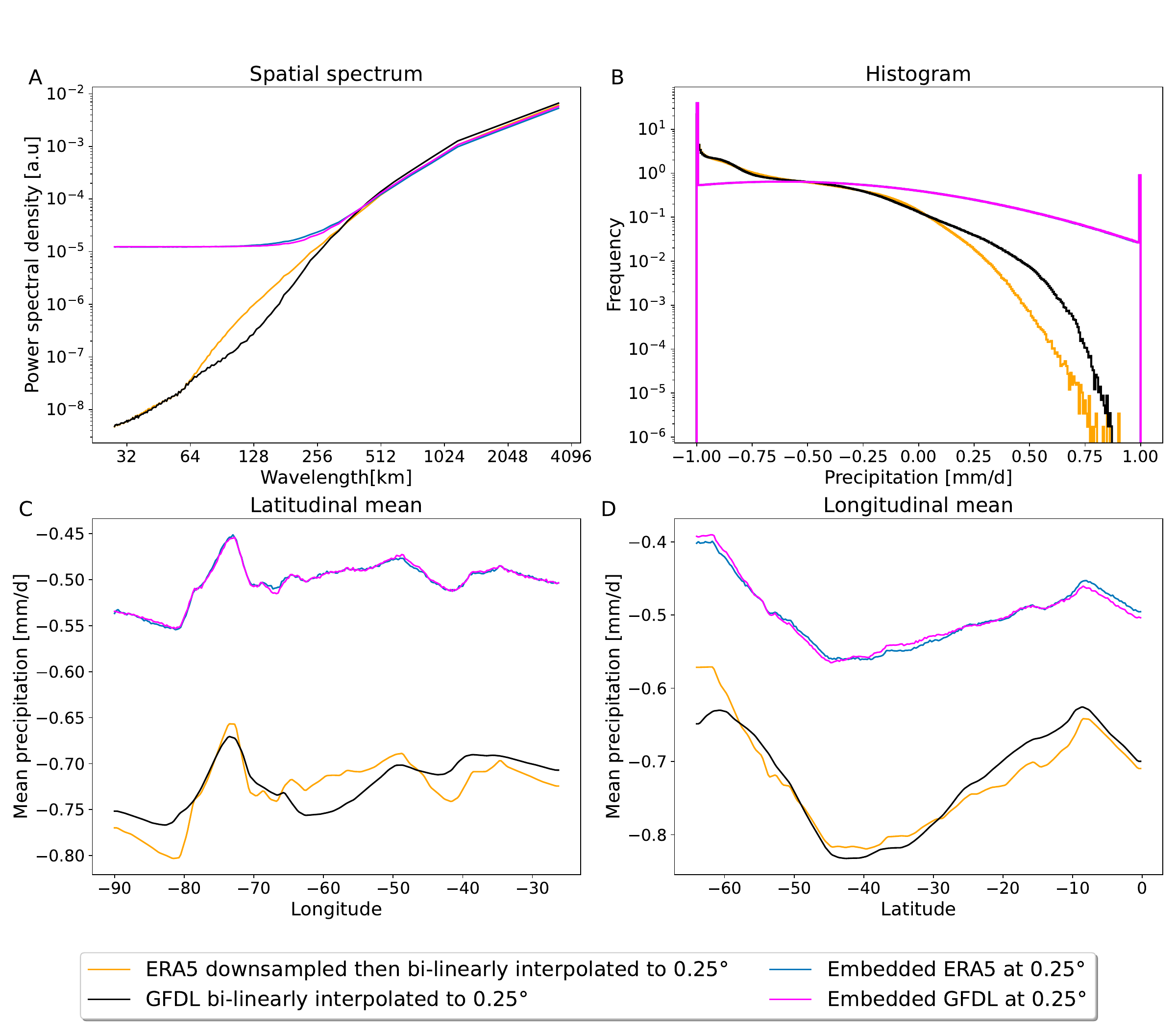}
    \caption{\textbf{We demonstrate the effect of the embedding transformations}. We apply the transformations to the first downsampled, then bilinearly upsampled 0.25° low-resolution ERA5 data and to the bilinearly upsampled GFDL data at 0.25°. The data shown is after our pre-processing transformation, hence the different precipitation units. Both embedded ERA5 and embedded GFDL fields are also at 0.25°. The PSD \textbf{(A)} of the embedded datasets align very well. The horizontal lines in the PSD plot are a sign of the added noise that acts as a low-pass filter on small scales. The histogram \textbf{(B)} as well as latitudinal / longitudinal profiles \textbf{(C)} / \textbf{(D)} show a strong alignment between embedded ERA5 and embedded GFDL data. Note that we clip the data at -1 and 1 after noising, as this is the maximal range of our pre-processed data. 
    }
    \phantomsection\label{fig:eval_emb}
\end{figure*}

\clearpage

\begin{figure*}[!htb]
    \centering
    \includegraphics[width=\textwidth]{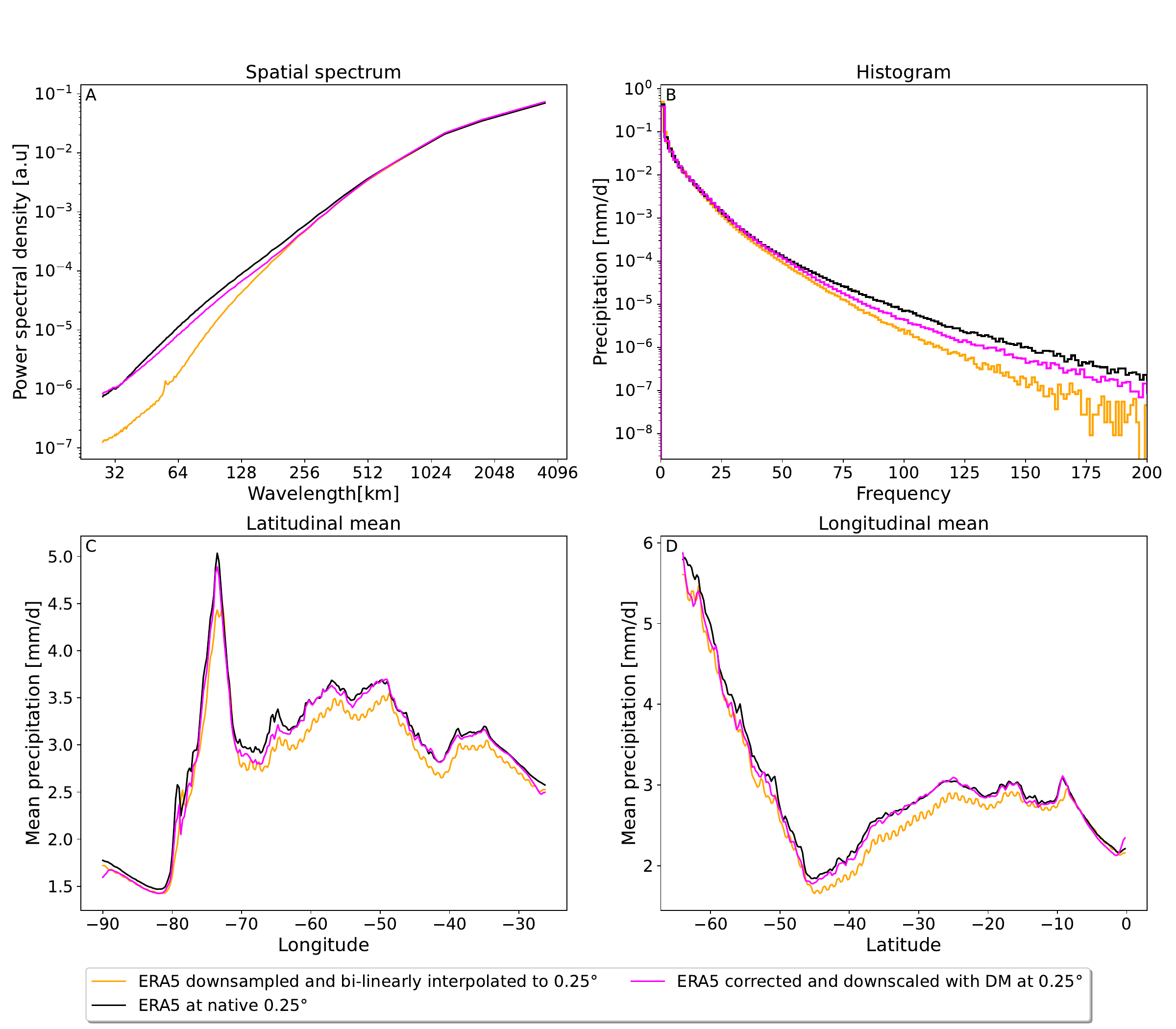}
    \caption{\textbf{Evaluation of the diffusion model's performance to reconstruct ERA5 at 0.25° resolution}. The reconstruction starts from the embedded ERA5 data, obtained by downsampling to 1° via choosing only every fourth grid cell, and then bilinearly interpolating back to 0.25°. In this case, the embedded ERA5 data mimics the ESM data in order to train the diffusion model (see Methods and main text). 
    \textbf{(A)} Mean spatial power spectral densities (PSDs). \textbf{(B)} Histogram indicating the precipitation frequencies. 
    \textbf{(C)} Latitude profile, given by the averaged longitudes over the validation period. \textbf{(D)} Longitude profile, given by the averaged latitudes over the validation period. Our diffusion model approximates the latitude and longitude profile of the original ERA5 reference data extremely well. The histogram also shows large improvements with slight deviations from the HR ERA5 reference data for extreme precipitation. The diffusion model manages to correct the small-scale spatial details and follows the target distribution closely.}
    \phantomsection\label{fig:era5_bc_eval}
\end{figure*}

\clearpage

\begin{figure*}
    \centering
    \includegraphics[width=\textwidth]{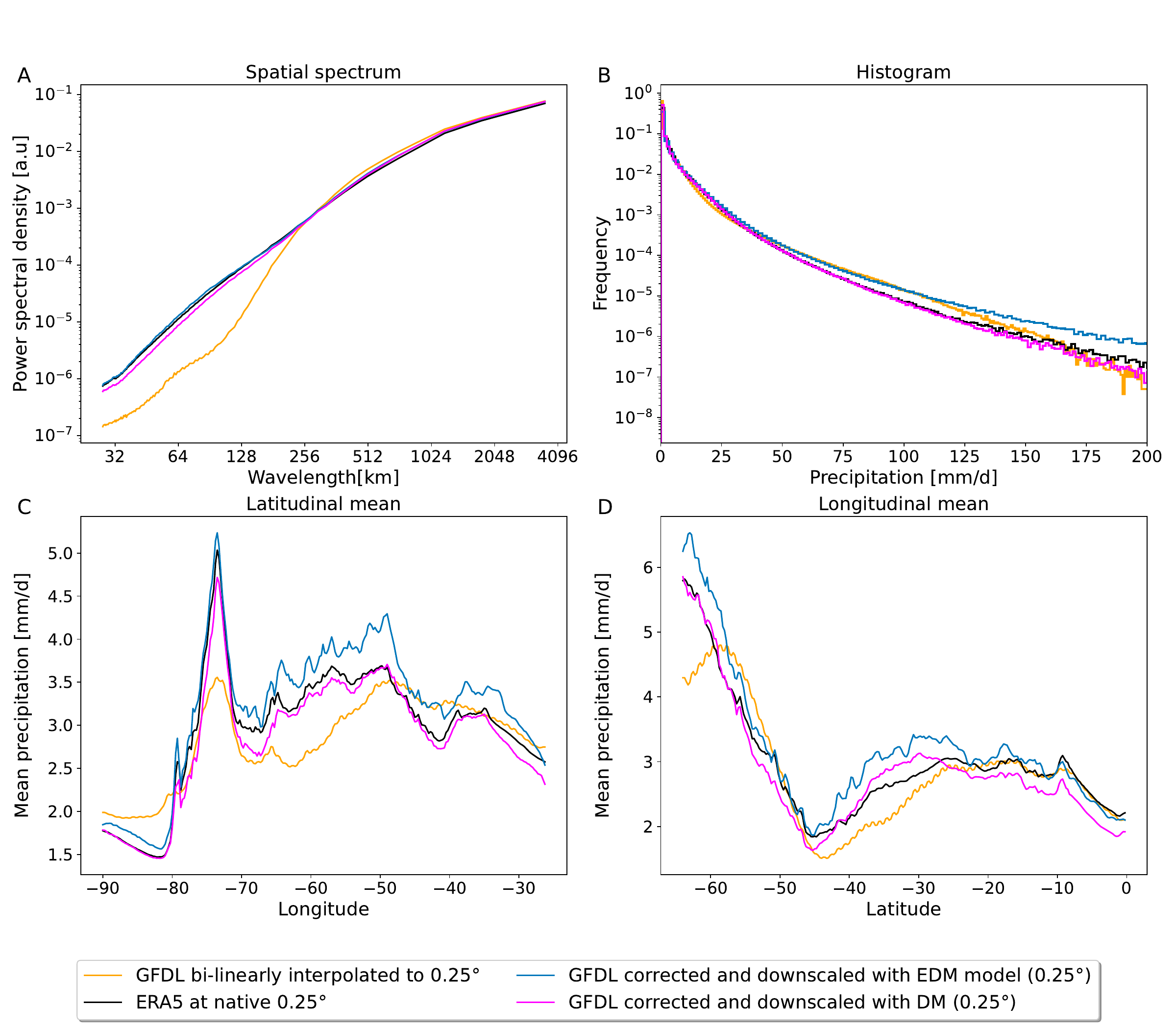}
    \caption{\textbf{Comparing the downscaling and bias correction performance of our DM to an EDM-diffusion model}. Comparison of GFDL (bilinearly upsampled to 0.25°) (orange) and ERA5 at 0.25° (black) to diffusion model-corrected GFDL fields with our proposed method at 0.25° (magenta) and diffusion model-corrected GFDL fields with the EDM diffusion model at 0.25° (blue). The Power spectral density (PSD) plot \textbf{(A)} shows that both diffusion models correct the small-scale spatial details very well. The spectrum aligns well with the high-resolution ERA5 target data. The EDM model is inferior in correcting the histogram \textbf{(B)} as well as the latitude \textbf{(C)} and longitude \textbf{(D)} profiles compared to our proposed method.}
    \phantomsection\label{fig:edm}
\end{figure*}

\clearpage

\begin{figure*}
    \centering
    \includegraphics[width=\textwidth]{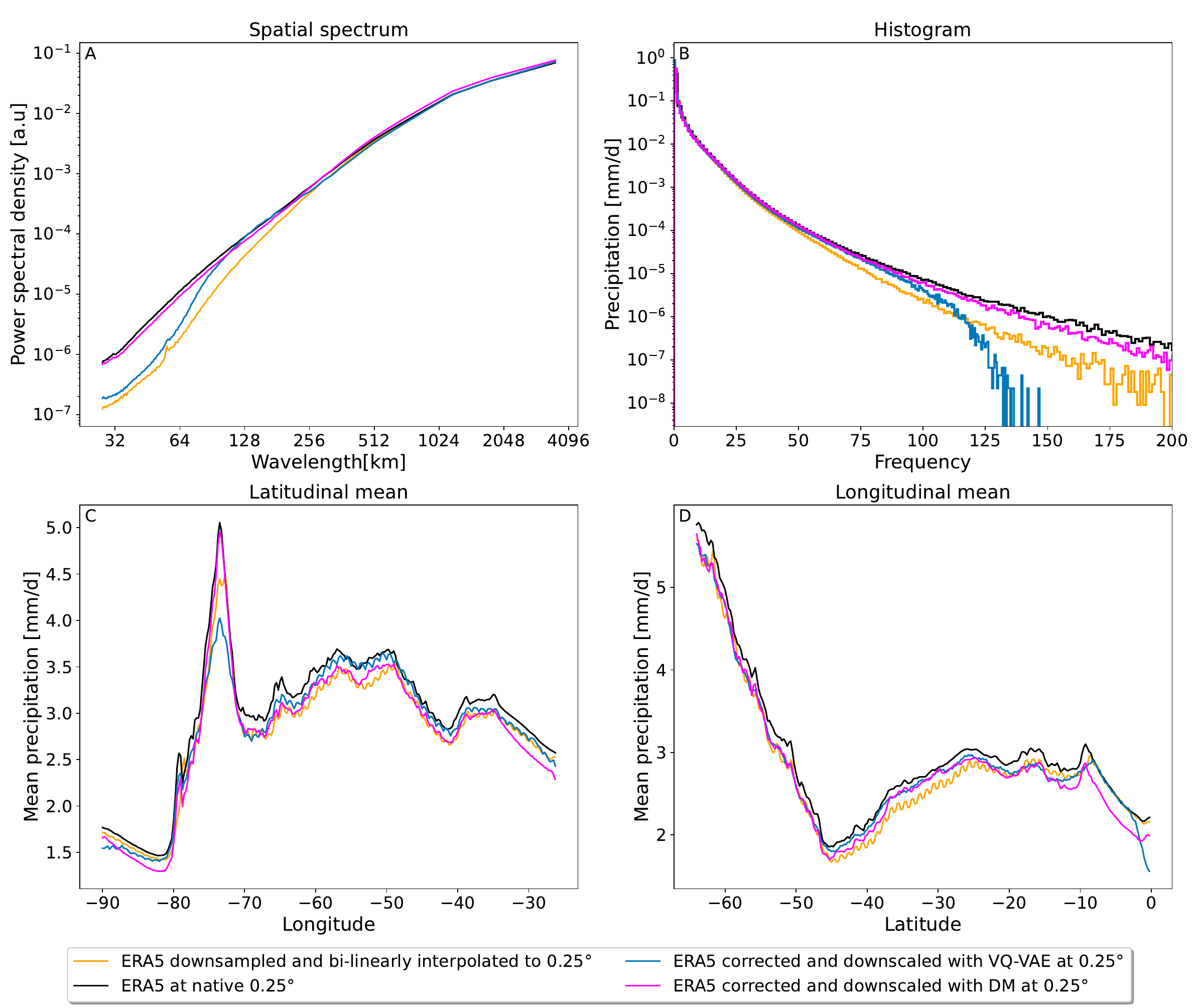}
    \caption{\textbf{Comparing the diffusion model’s downscaling and bias correction performance to a VQ-VAE}. We compare the VQ-VAE model (blue) to 0.25° ERA5 (black), ERA5 bilinearly upsampled to 0.25° (orange) and the diffusion model-corrected ERA5 fields with our proposed method at 0.25° (magenta). \textbf{(A)} Power spectral density (PSD) illustrates the inability of the VQ-VAE to generate high-frequency information, diverging from the HR ERA5 ground truth. \textbf{(B)} The histogram highlights the VQ-VAE's poor performance in capturing the target distribution for high precipitation values. The VQ-VAE also struggles to consistently match the latitudinal \textbf{(C)} and longitudinal \textbf{(D)} means with the ERA5 ground truth.}
    \phantomsection\label{fig:vqvae_eval}
\end{figure*}
\clearpage

\begin{figure*}
    \centering
    \includegraphics[width=\textwidth]{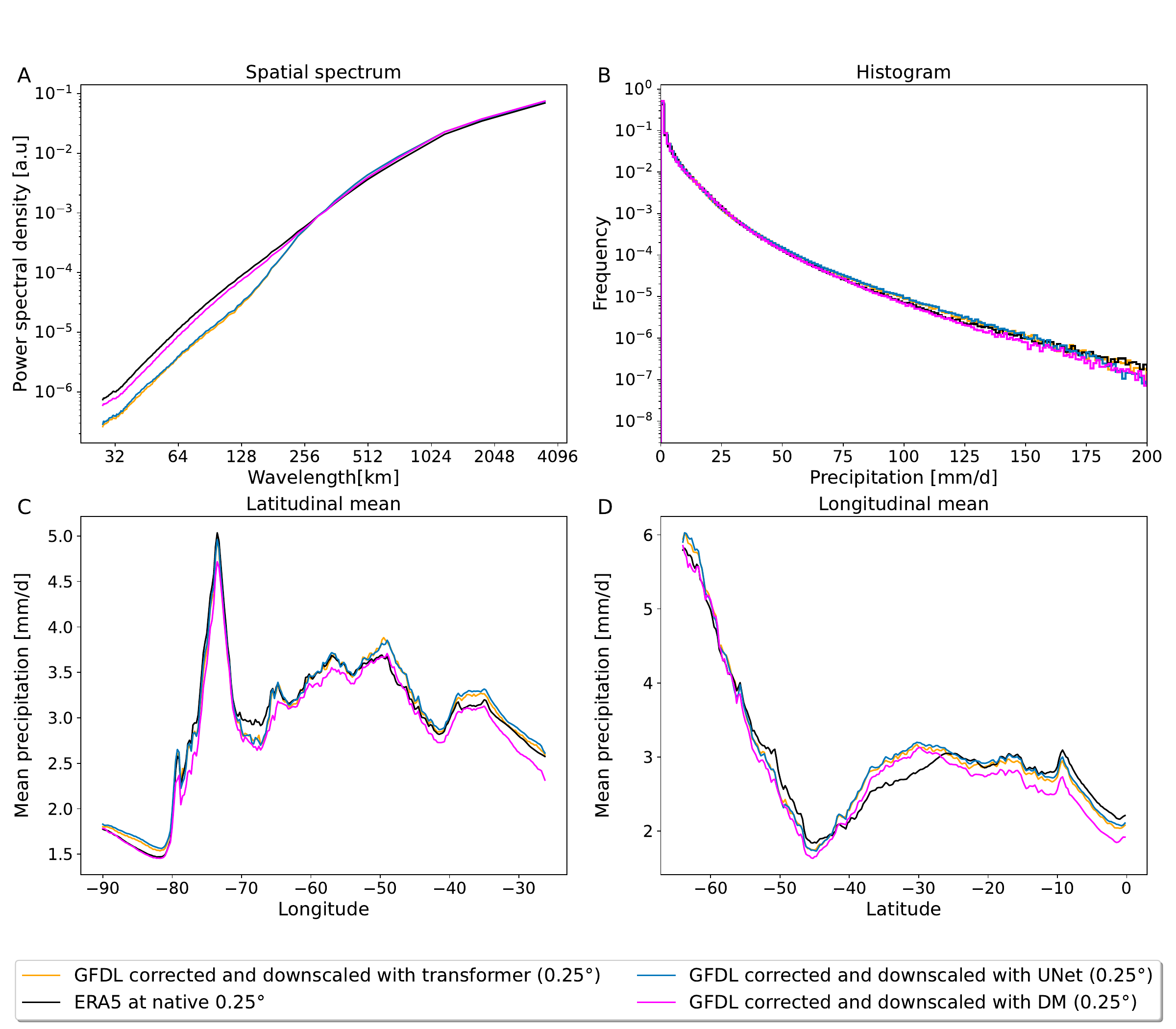}
   \caption{\textbf{Comparison of different machine learning backbones.} Comparison of precipitation fields from GFDL data corrected and downscaled using three different models: our Diffusion Model (magenta), a UNet (blue), and a Transformer (orange). The results are compared against ERA5 reanalysis data (black) as the reference. All datasets are at a 0.25° resolution.\textbf{(A)} The Power Spectral Density (PSD) plot shows that the diffusion model significantly outperforms the UNet and Transformer benchmarks in generating realistic small-scale spatial patterns that align with the ERA5 reference. 
   \textbf{(B)} The histogram of daily precipitation shows that all three methods are highly effective at correcting the distribution to match the ERA5 data.\textbf{(C, D)} The latitudinal and longitudinal mean profiles confirm that all three models perform similarly well in reproducing the large-scale mean precipitation.}
    \phantomsection\label{fig:compare_DM_transformer_CNN}
\end{figure*}

\clearpage

\begin{figure*}[htb]
    \centering
    \includegraphics[width=\textwidth]{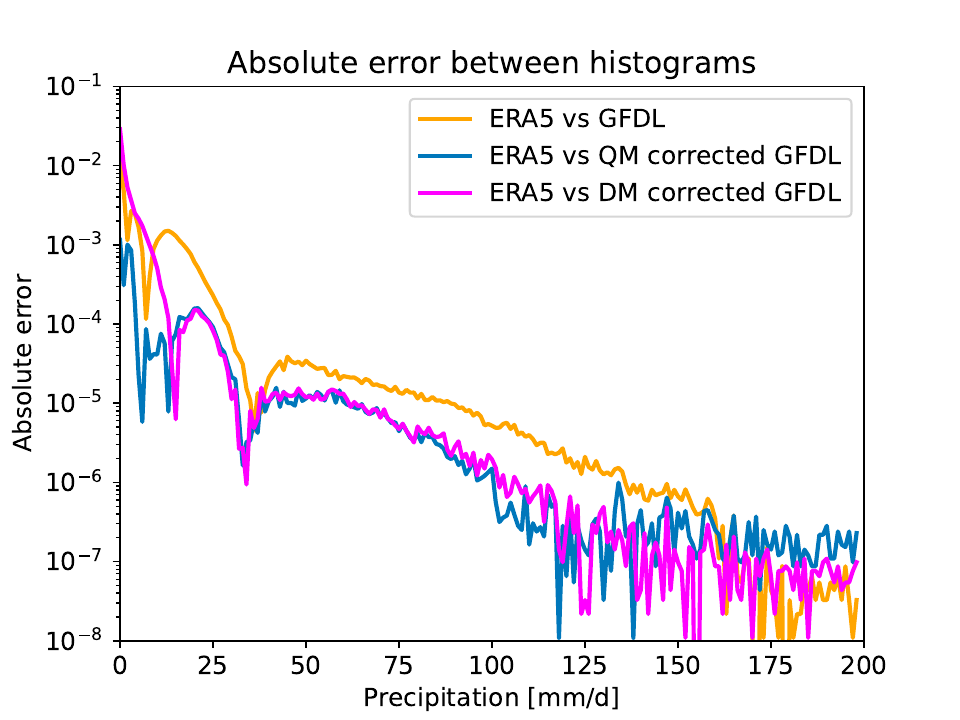}
    \caption{\textbf{Comparison of the absolute histogram errors}. The histogram shows the absolute errors between high-resolution ERA5 data (0.25°) and GFDL (bilinearly upsampled to 0.25°) (orange), ERA5 and DM-corrected GFDL (magenta) and ERA5 and QM-corrected GFDL data (blue). The dips around 10 mm d$^{-1}$ and 40 mm d$^{-1}$ correspond to points where the histograms of ERA5 and its comparisons intersect. For very large precipitation values, our diffusion model outperforms the QM baseline, while it has slightly larger absolute error for smaller precipitation values.
    }
    \phantomsection\label{fig:absolute_errors}
\end{figure*}

\clearpage

\begin{figure*}[!htb]
    \centering
    \includegraphics[width=\textwidth]{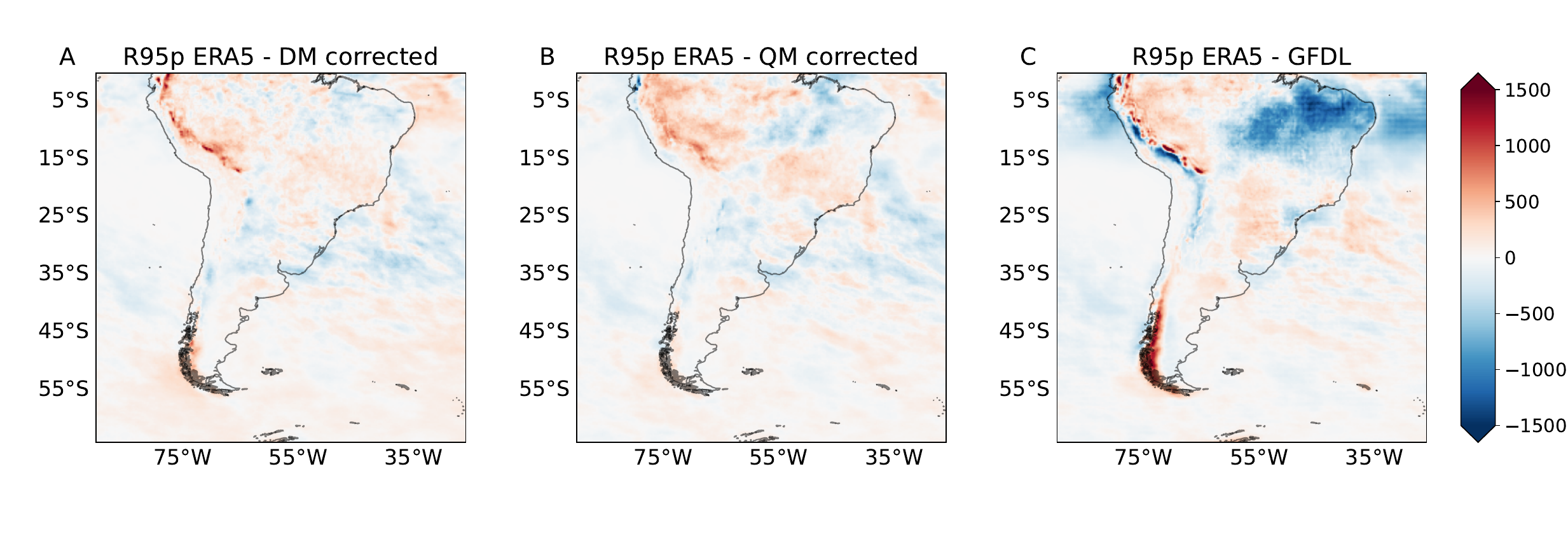}
    \caption{\textbf{ Performance evaluation of the diffusion model in representing extreme rainfall events}. The R95p metric represents the total annual precipitation from heavy rain days, calculated as the sum of daily precipitation on wet days (precipitation $>$ 1 mm d$^{-1}$) that exceed the 95th percentile of our reference period. We calculate the metric on the validation periods, giving the ERA5 training period as reference. We computed R95p for ERA5, GFDL, DM-corrected GFDL, and QM-corrected GFDL. For clarity, we plotted the difference between R95p values for ERA5 and DM-corrected GFDL \textbf{(A)}, ERA5 and QM-corrected GFDL \textbf{(B)}, and ERA5 and GFDL \textbf{(C)}. Our diffusion model effectively corrects the bias in extreme precipitation events, performing similarly to the quantile mapping correction. 
    }
    \phantomsection\label{fig:r95p}
\end{figure*}

\clearpage

\begin{figure*}
    \centering
    \includegraphics[width=\textwidth]{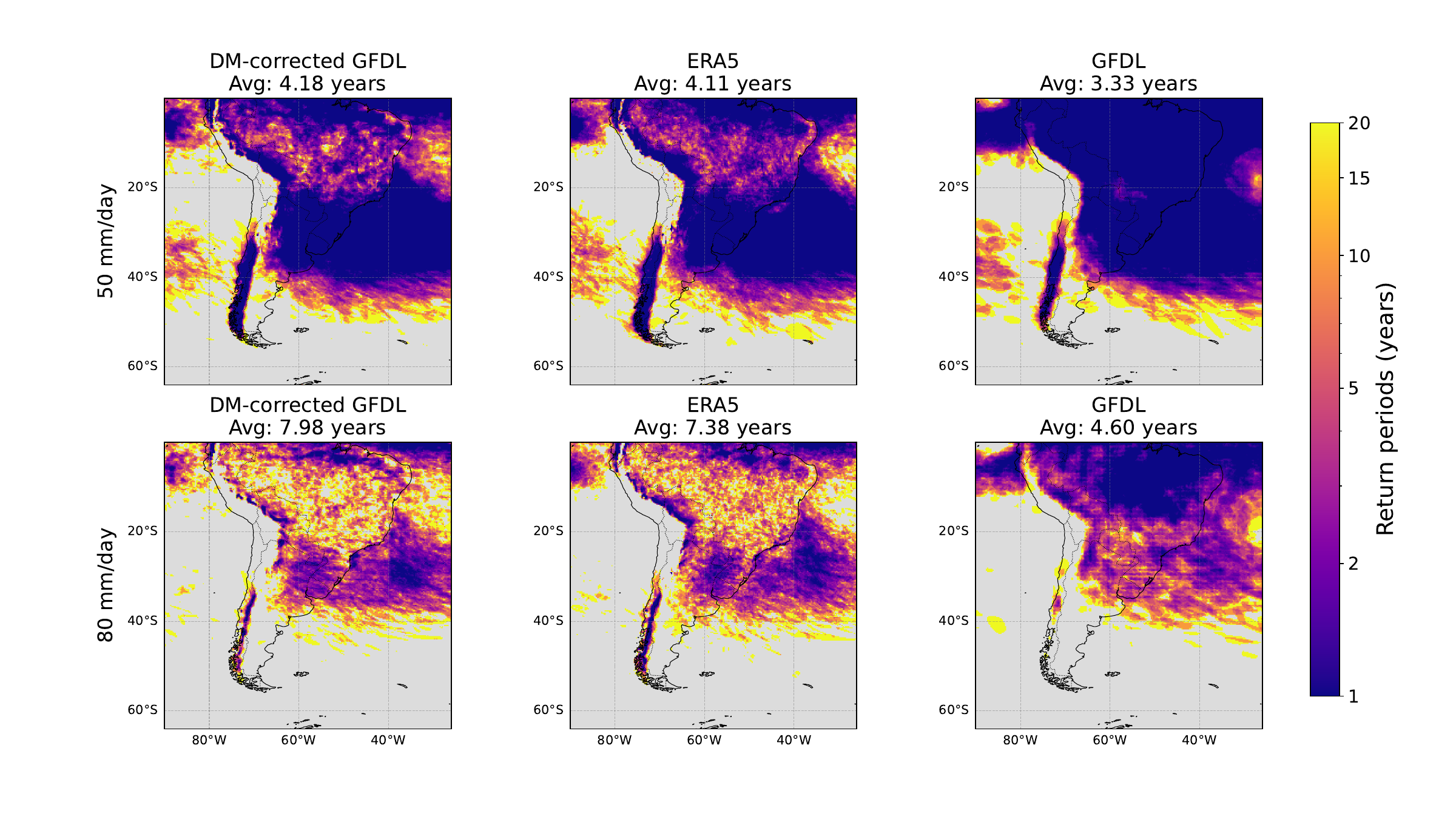}
   \caption{\textbf{Empirical return periods of extreme precipitation events.} The figure compares the spatial distribution of return periods (in years) for moderately extreme (50 mm d$^{-1}$, top row) and very extreme (80 mm d$^{-1}$, bottom row) events from our DM-corrected GFDL (left), the ERA5 reference (middle), and the raw GFDL model (right). The return period for each grid cell is calculated as the inverse of the mean annual frequency of exceedances. The raw GFDL model shows a significant wet bias, with unrealistically short return periods (spatially averaged values are shown above each plot) and diffuse spatial patterns. Our DM mitigates this bias, resulting in sharper, more localized patterns and average return periods that closely match the ERA5 reference for both thresholds.}
    \phantomsection\label{fig:return_intervals}
\end{figure*}

\begin{figure*}[!htb]
    \centering
    \includegraphics[width=\textwidth]{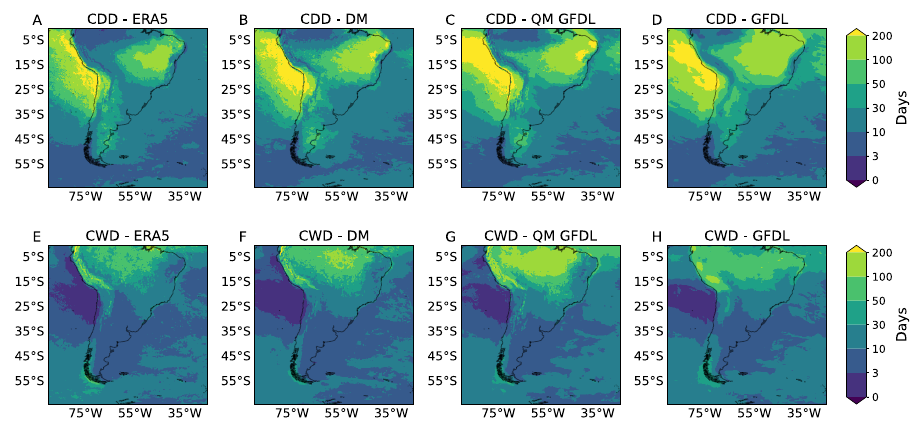}
    \caption{\textbf{Duration of consecutive dry and wet periods in the different datasets}. The first row shows a comparison between consecutive dry days (CDD) between ERA5 at 0.25° \textbf{(A)}, diffusion model (DM) correction GFDL fields at 0.25° \textbf{(B)}, GFDL first quantile mapped and then bilinearly upsampled to 0.25° \textbf{(C)} and the GFDL bilinearly upsampled to 0.25° \textbf{(D)}. The second row shows a comparison between consecutive wet days (CWD) between ERA5 at 0.25° \textbf{(E)}, diffusion model correction GFDL fields at 0.25° \textbf{(F)}, GFDL first quantile mapped and then bilinearly upsampled to 0.25° \textbf{(G)} and the GFDL bilinearly upsampled to 0.25° \textbf{(H)}. Overall our diffusion model is the most similar to ERA5 in both, the case of CDD (\textbf{(A)} and \textbf{(B)}) and the case of CWD (\textbf{(E)} and \textbf{(F)}).}
    \phantomsection\label{fig:dryday}
\end{figure*}

\begin{figure*}[!htb]
    \centering
    \includegraphics[width=1.1\textwidth]{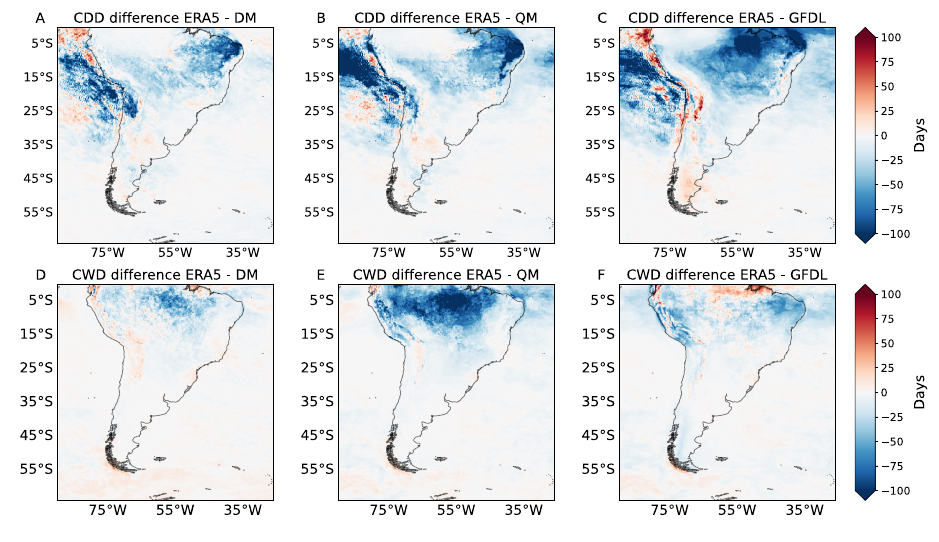}
    \caption{\textbf{Performance of different methods regarding consecutive dry and wet period statistics}. We compare the consecutive dry day (CDD) differences between 0.25° ERA5 and diffusion model (DM) correction GFDL fields at 0.25° \textbf{(A)}, as well as GFDL first quantile mapped and then bilinearly upsampled to 0.25° \textbf{(B)} and the GFDL bilinearly upsampled to 0.25° \textbf{(C)}. The second row shows a comparison between consecutive wet day (CWD) differences between ERA5 at 0.25° and diffusion model corrected GFDL fields at 0.25° \textbf{(D)}, as well as GFDL first quantile mapped and then bilinearly upsampled to 0.25° \textbf{(E)} and the GFDL bilinearly upsampled to 0.25° \textbf{(F)}. Our diffusion model exhibits substantially smaller errors than the benchmarks. 
    }
    \phantomsection\label{fig:dryday_diff}
\end{figure*}

\begin{figure*}
    \centering
    \includegraphics[width=1\textwidth]{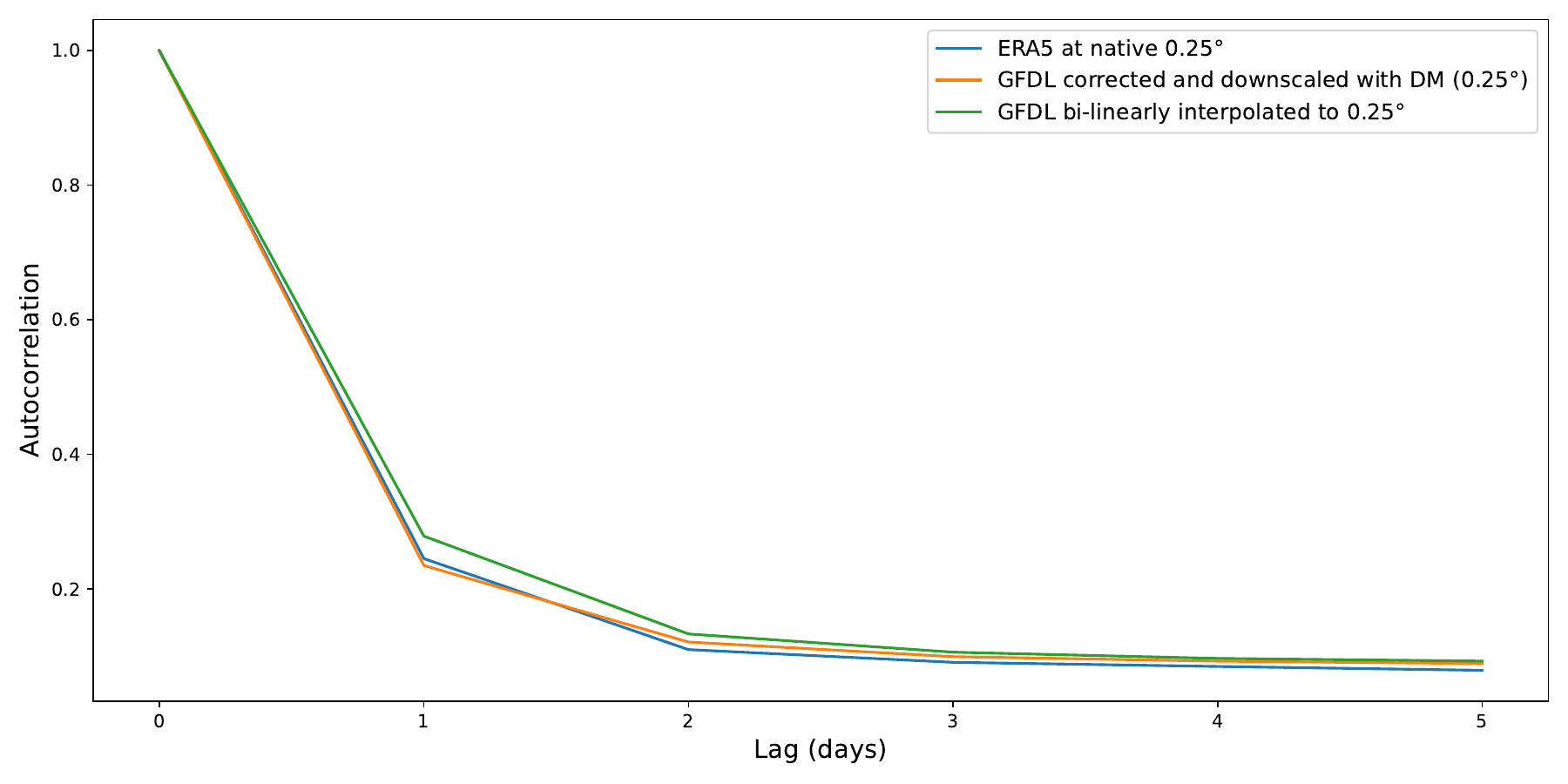}
   \caption{\textbf{Temporal autocorrelation of daily precipitation over the extended 1995-2014 period.} Spatially averaged temporal autocorrelation of daily precipitation for time lags up to 5 days. The plot compares the ERA5 reference (blue), the bilinearly interpolated GFDL data (green), and our DM-corrected GFDL data (orange). The GFDL data exhibits too high autocorrelation. The DM successfully corrects this temporal bias, producing a time series with an autocorrelation structure that closely matches ERA5.}
    \phantomsection\label{fig:AC}
\end{figure*}

\begin{figure*}
    \centering
    \includegraphics[width=1\textwidth]{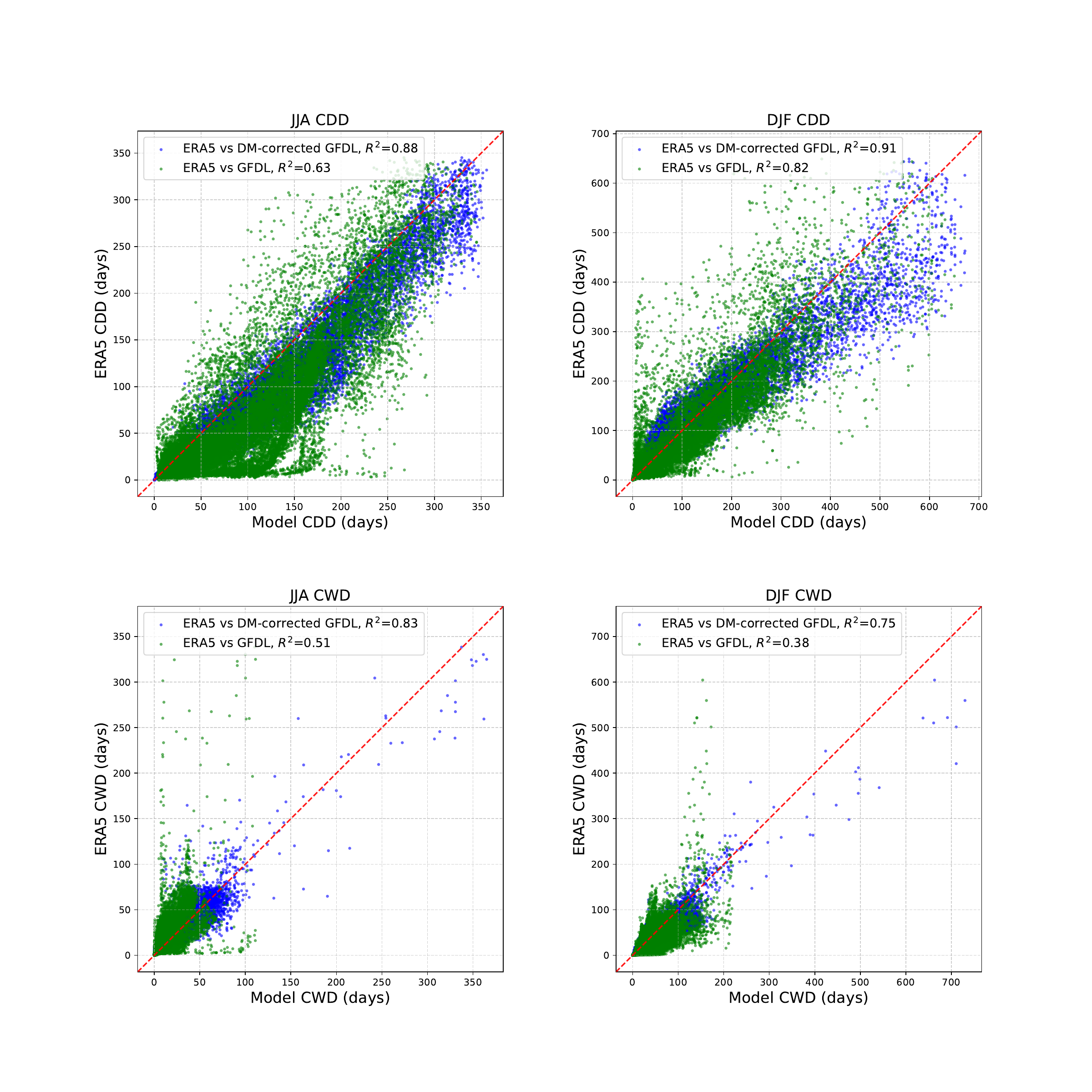}
   \caption{\textbf{Seasonal evaluation of consecutive dry and wet day spell durations.} Scatter plots comparing spell durations for Consecutive Dry Days (CDD, top row) and Consecutive Wet Days (CWD, bottom row) over the 1995-2014 period for three month periods JJA (left) and DJF (right). Each point represents a grid cell, comparing the raw GFDL model (green) and the DM-corrected GFDL (blue) against the ERA5 reference. The red dashed line indicates perfect agreement. The raw GFDL model exhibits significant biases, consistently underestimating the duration of long dry spells and overestimating the length of wet spells. Our DM correction substantially reduces these biases, aligning the spell durations much closer with the ERA5 reference, as shown by the tighter clustering of blue points around the 1:1 line. }
    \phantomsection\label{fig:seasonal_scatter}
\end{figure*}

\begin{figure*}
    \centering
    \includegraphics[width=1\textwidth]{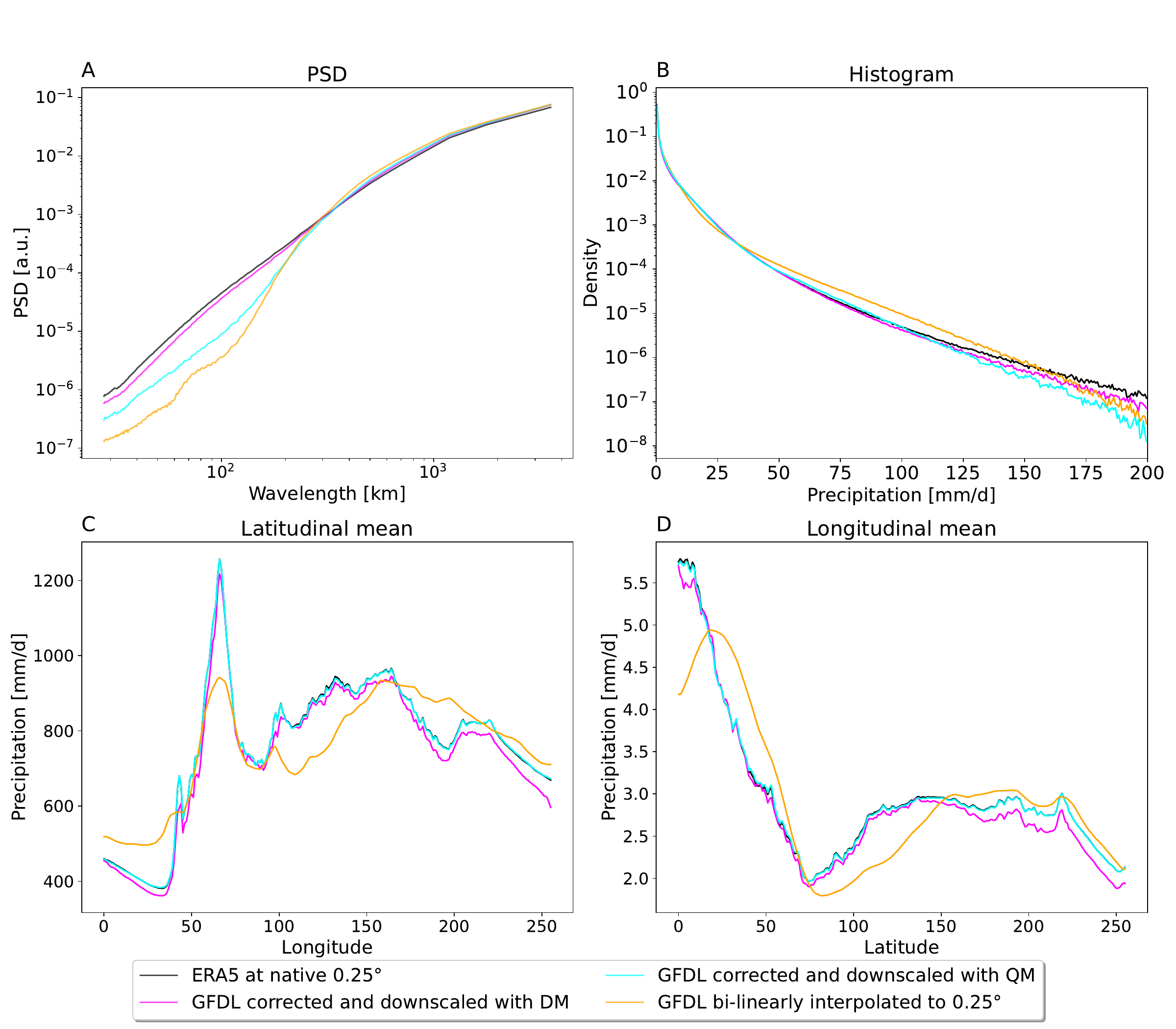}
   \caption{\textbf{Model performance evaluation over the extended 1995-2014 period.} This figure replicates the analysis from Fig. \ref*{fig:gfdl_bc_combined_bias_map} but for the longer evaluation period of 1995–2014 to demonstrate the robustness of the results. It compares our DM-corrected GFDL (magenta), the QM benchmark (cyan), and the bilinearly interpolated GFDL (orange) with the ERA5 reference (black), all at 0.25° resolution. The findings are consistent with those from the shorter 2011-2014 period. \textbf{(A)} The Power Spectral Density plot confirms the DM's superior ability to correct small-scale spatial patterns.\textbf{(B)} The histogram shows that the DM more accurately reproduces the frequency distribution of precipitation events than the benchmark. \textbf{(C, D)} In the mean profiles, both the DM and the benchmark offer comparable improvements over the raw GFDL data.}
    \phantomsection\label{fig:extended_metrics}
\end{figure*}

\clearpage

\begin{figure*}
    \centering
    \includegraphics[width=\textwidth]{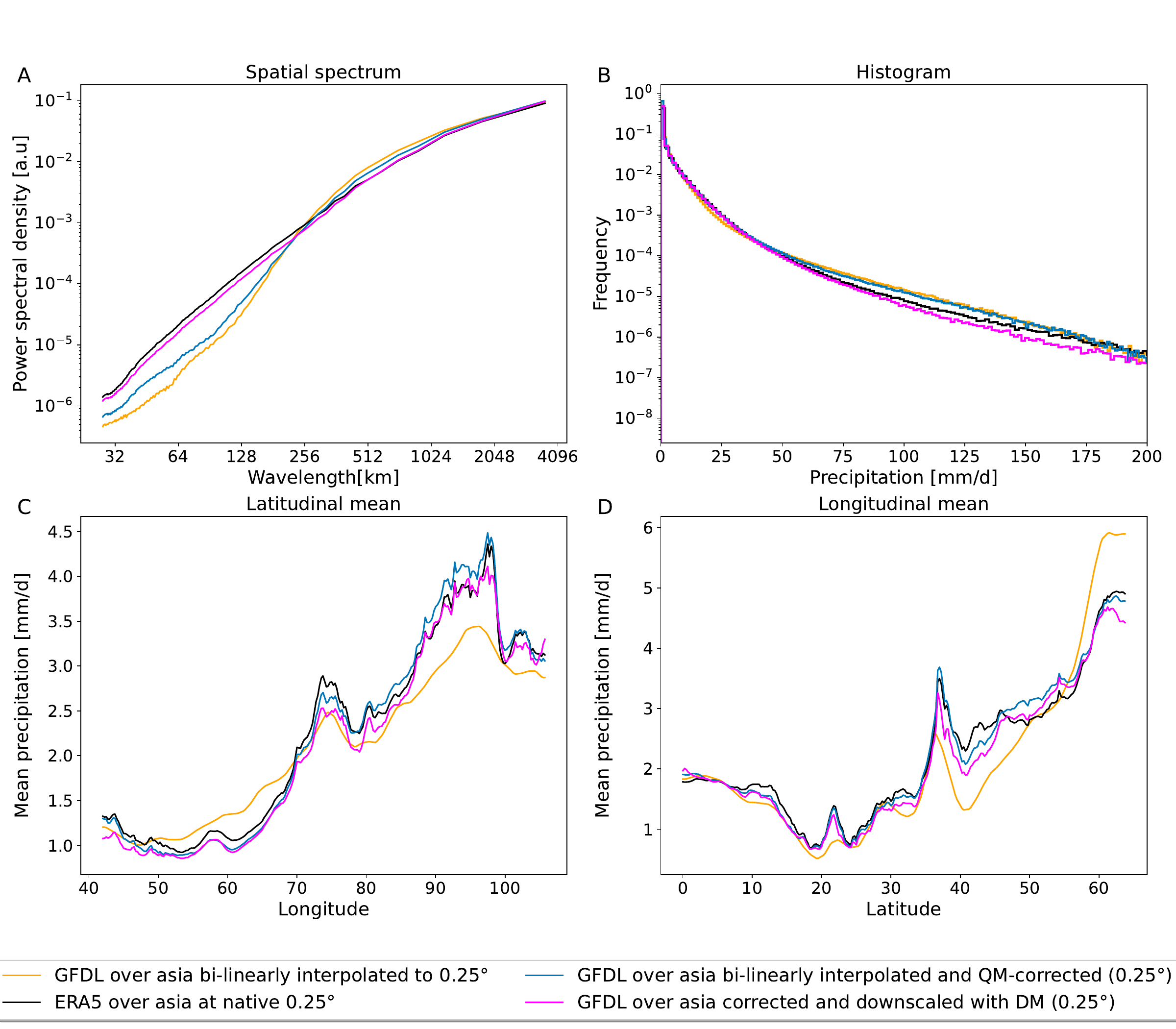}
    \caption{\textbf{Evaluation of the diffusion model's downscaling and bias correction performance over South Asia.} Comparison of precipitation fields from the DM-corrected GFDL model (magenta) with the original GFDL output (orange), the quantile mapping benchmark (blue), and ERA5 (black) as the reference. All datasets are at a 0.25° resolution. The original GFDL data and the data before QM were bilinearly upsampled.
    \textbf{(A)} The Power Spectral Density (PSD) plot shows the DM's superior correction of small-scale features.
    \textbf{(B)} The histogram of daily precipitation shows that both DM and QM corrections improve the distribution of moderate events (e.g., 10–40 mm d$^{-1}$) compared to the raw GFDL data.
    \textbf{(C, D)} The latitude and longitude mean profiles show that both correction methods offer similar improvements over the original GFDL data.}
    \phantomsection\label{fig:asia_metrics}
\end{figure*}
\clearpage

\begin{figure*}
    \centering
    \includegraphics[width=\textwidth]{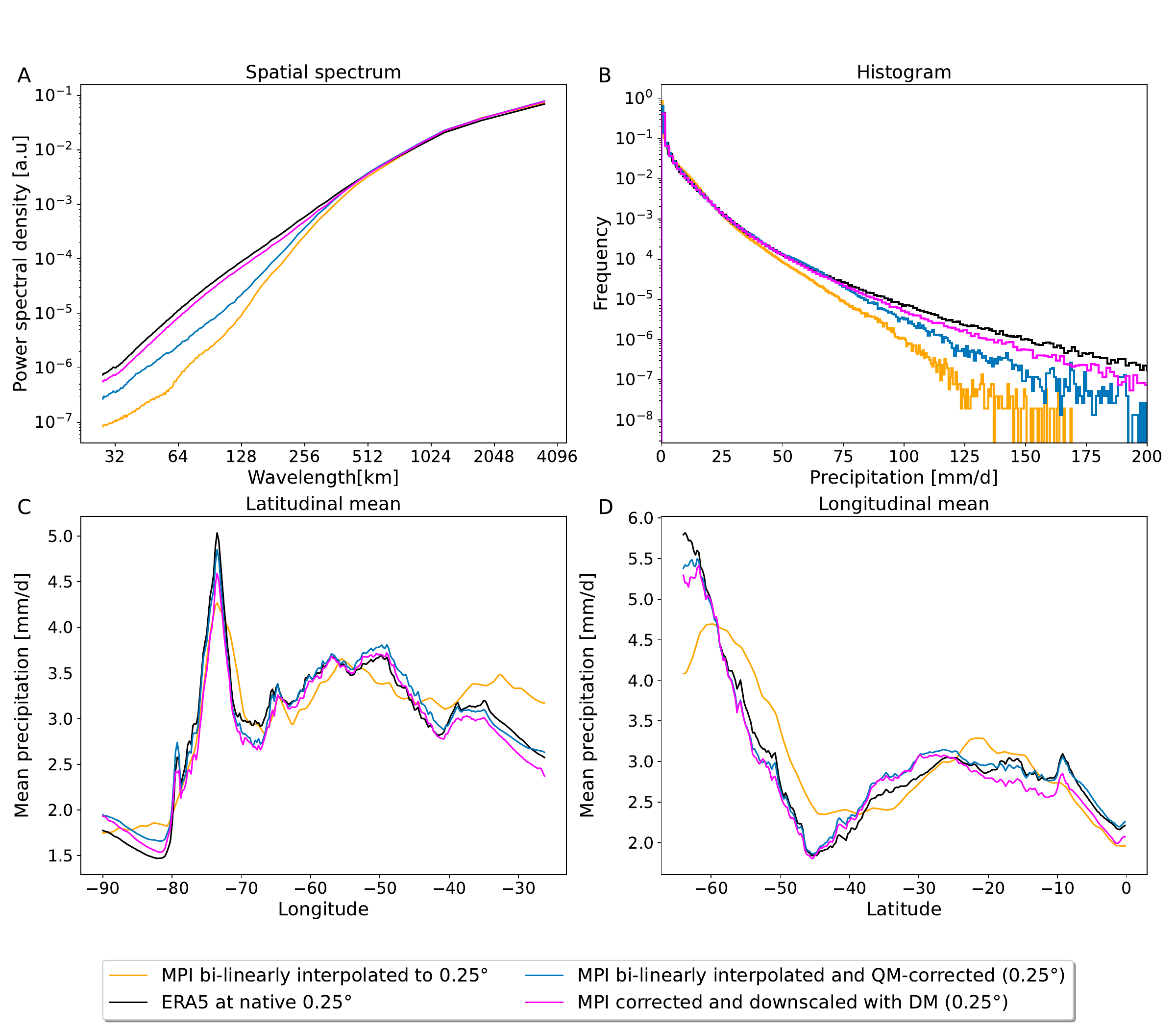}
    \caption{\textbf{Evaluation of the diffusion model's downscaling and bias correction performance for a different ESM.} Comparison of precipitation fields from the DM-corrected MPI model (magenta) with the original MPI output (orange), the quantile mapping benchmark MPI (blue), and ERA5 (black) as the reference. All datasets are at a 0.25° resolution. The original MPI data and the data before QM were bilinearly upsampled.
    \textbf{(A)} The Power Spectral Density plot shows that the DM almost perfectly restores spatial variability across all scales, significantly outperforming both the raw MPI and the benchmark at small scales.
    \textbf{(B)} The histogram of daily precipitation demonstrates the DM's superior ability to reproduce the frequency of extreme events, matching the ERA5 reference far better than the benchmark.
    \textbf{(C, D)} The latitudinal and longitudinal mean profiles show a similar correction of our DM and the benchmark over the raw MPI data.}
    \phantomsection\label{fig:MPI_metrics}
\end{figure*}

\clearpage

\begin{figure*}
    \centering
    \includegraphics[width=1.1\textwidth]{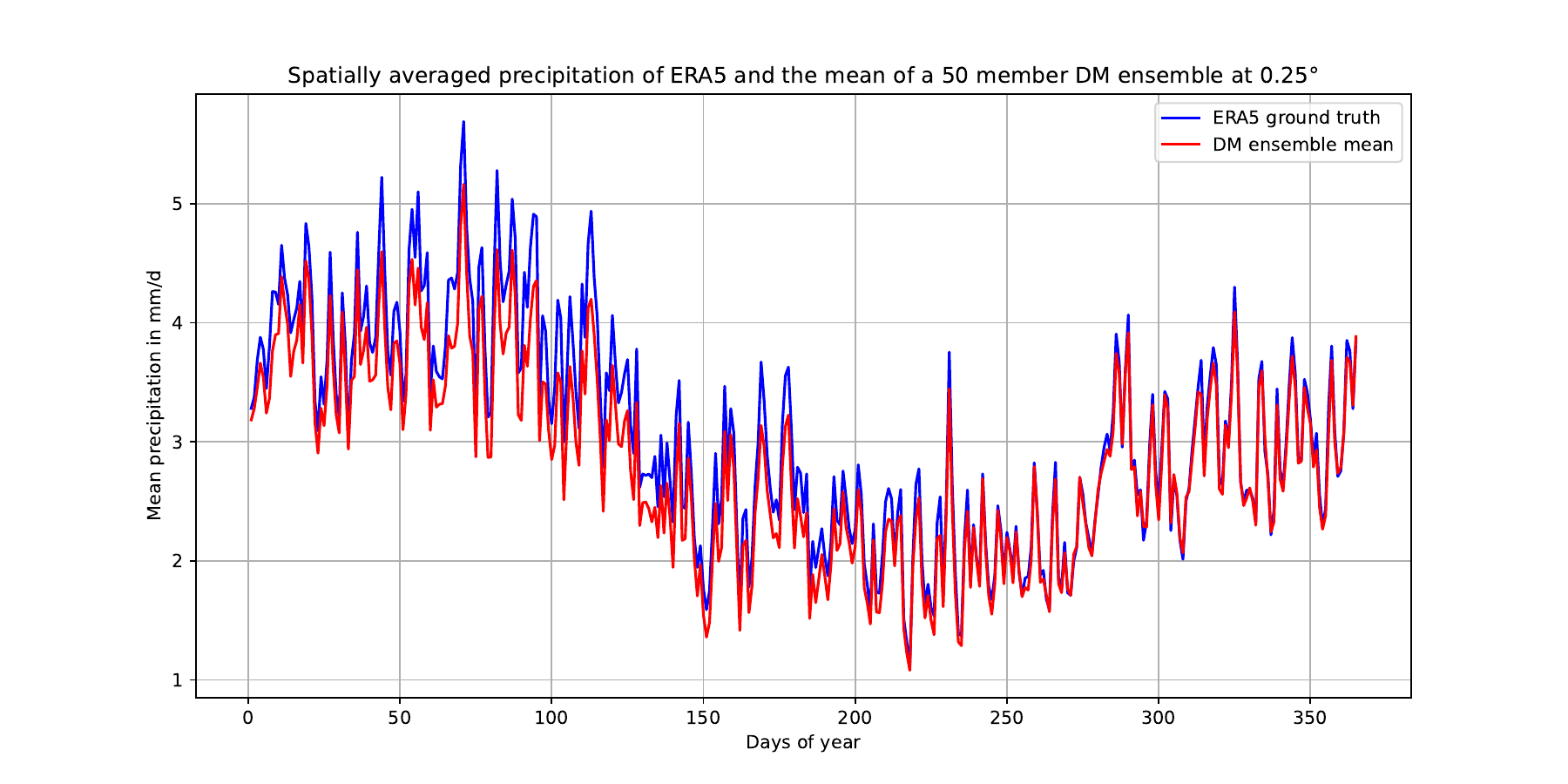}
    \caption{\textbf{Spatially averaged daily precipitation at 0.25° resolution over one year}. We compare the ERA5 ground truth (blue) with the mean of a 50-member ensemble of DM-downscaled fields (red). The close alignment demonstrates the DM ensemble's ability to capture temporal variability and maintain accurate ensemble calibration on average.}
    \phantomsection\label{fig:ensemble_mean}
\end{figure*}

\clearpage

\begin{figure*}
    \centering
    \includegraphics[width=0.85\textwidth]{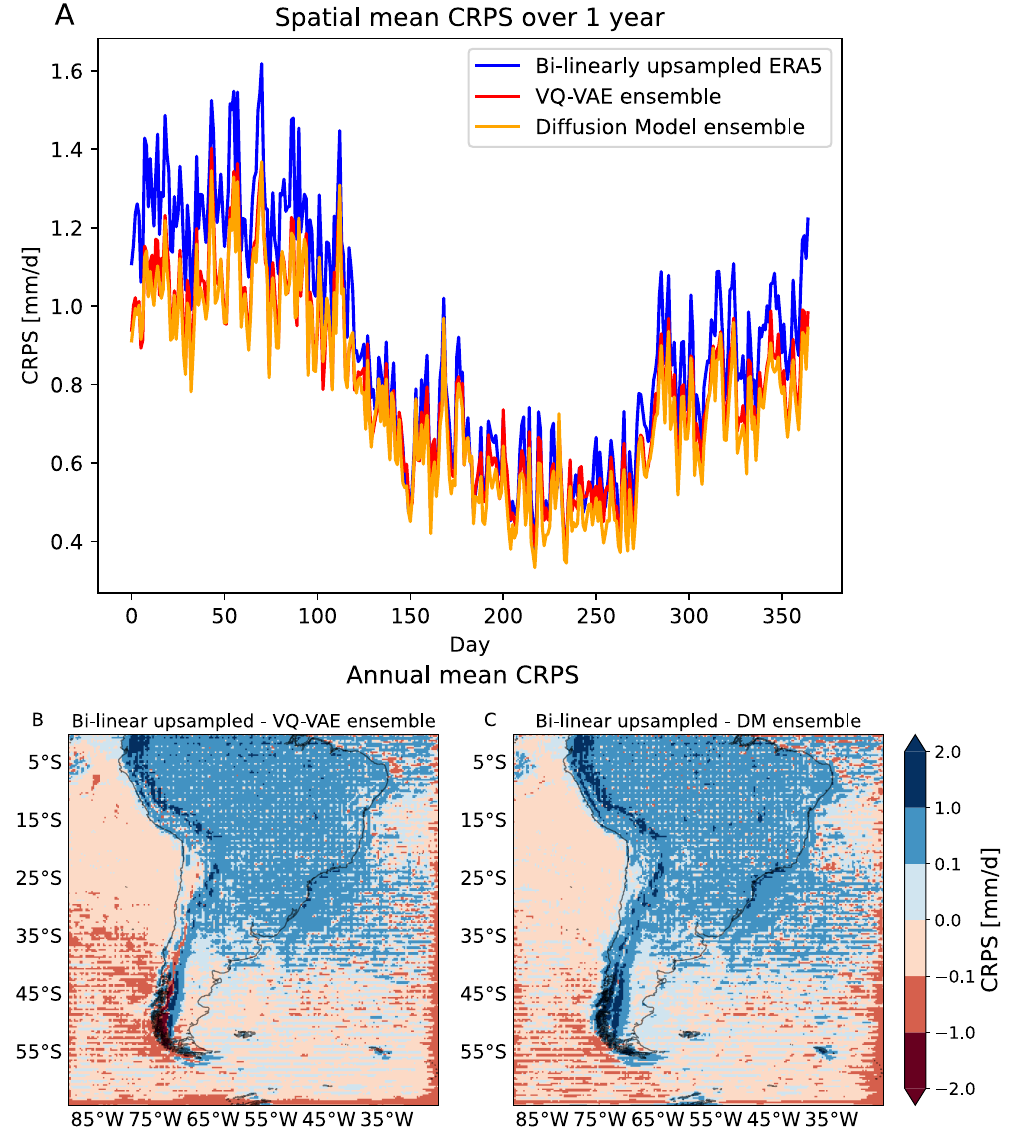}
    \caption{\textbf{Evaluating our DM using the Continuous Ranked Probability Score (CRPS)}. We compare a 50-member DM downscaling ensemble (orange), a 50-member VQ-VAE downscaling ensemble (red) and a deterministic baseline generated by bilinearly upsampling 1 year of low-resolution ERA5 fields (blue). \textbf{(A)} shows a time series of CRPS over 1 year averaged over the spatial domain. Our DM ensemble as well as the VQ-VAE ensemble are consistently under the baseline (lower is better). \textbf{(B)} and \textbf{(C)} display the annual mean CRPS difference between the bilinearly upsampled ERA5 and our DM ensemble, as well as the difference between the bilinearly upsampled ERA5 and the VQ-VAE ensemble. Both models consistently outperform the bilinear baseline, achieving lower mean CRPS values across the continent. The blue regions indicate that the baseline has higher (worse) CRPS than the ML models. While our DM and VQ-VAE perform similarly overall, in the southern part of the Andes, our DM achieves lower CRPS.}
    \phantomsection\label{fig:crps}
\end{figure*}

\begin{figure*}
    \centering
    \includegraphics[width=1.1\textwidth]{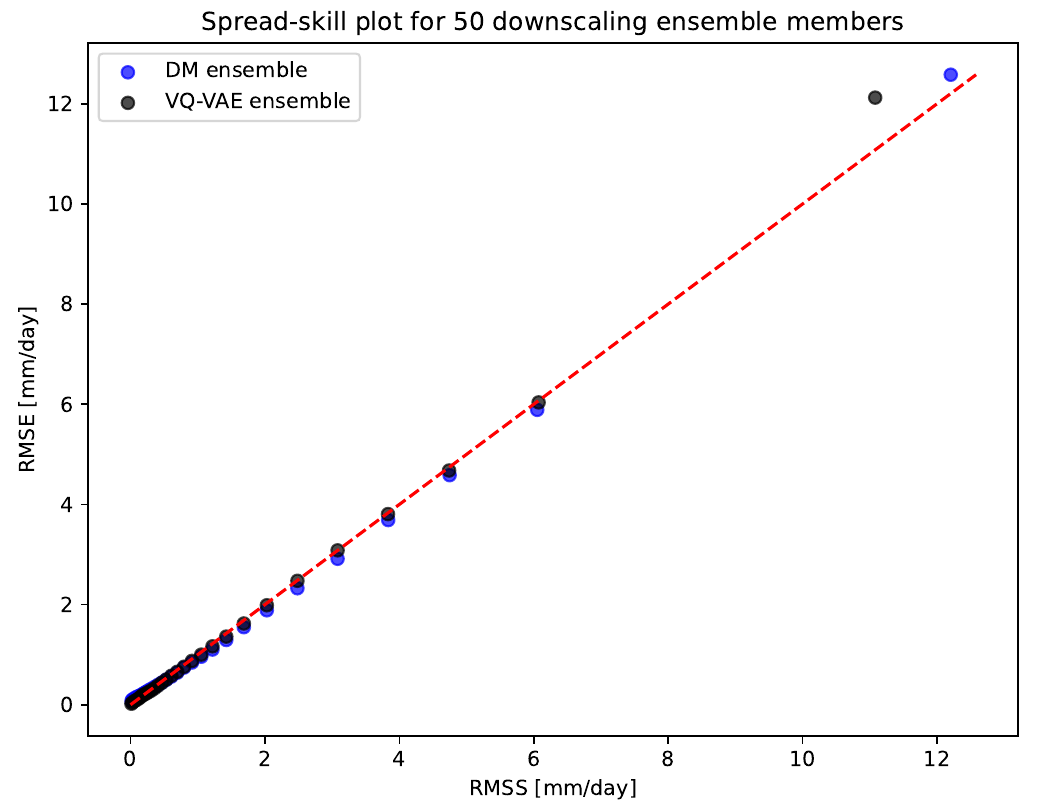}
    \caption{\textbf{Spread-skill plot for 50 downscaling ensemble members generated by conditioning on one year of low-resolution ERA5 data}. The spread measures the uncertainty in terms of root-mean-squared spread (RMSS), whereas the skill is indicated by the root-mean-squared error (RMSE). The plot compares the uncertainty calibration of the DM ensemble (blue) and the VQ-VAE ensemble (black). Both models demonstrate overall good calibration, with most points aligning along the 1:1 spread-skill line (red dashed line). However, for very large spreads, the DM exhibits superior calibration, while the VQ-VAE ensemble is overconfident. The DM's close alignment with the 1:1 line indicates well-calibrated uncertainty estimates.}
    \phantomsection\label{fig:spread_skill_plot}
\end{figure*}

\begin{figure*}
    \centering
    \includegraphics[width=0.85\textwidth]{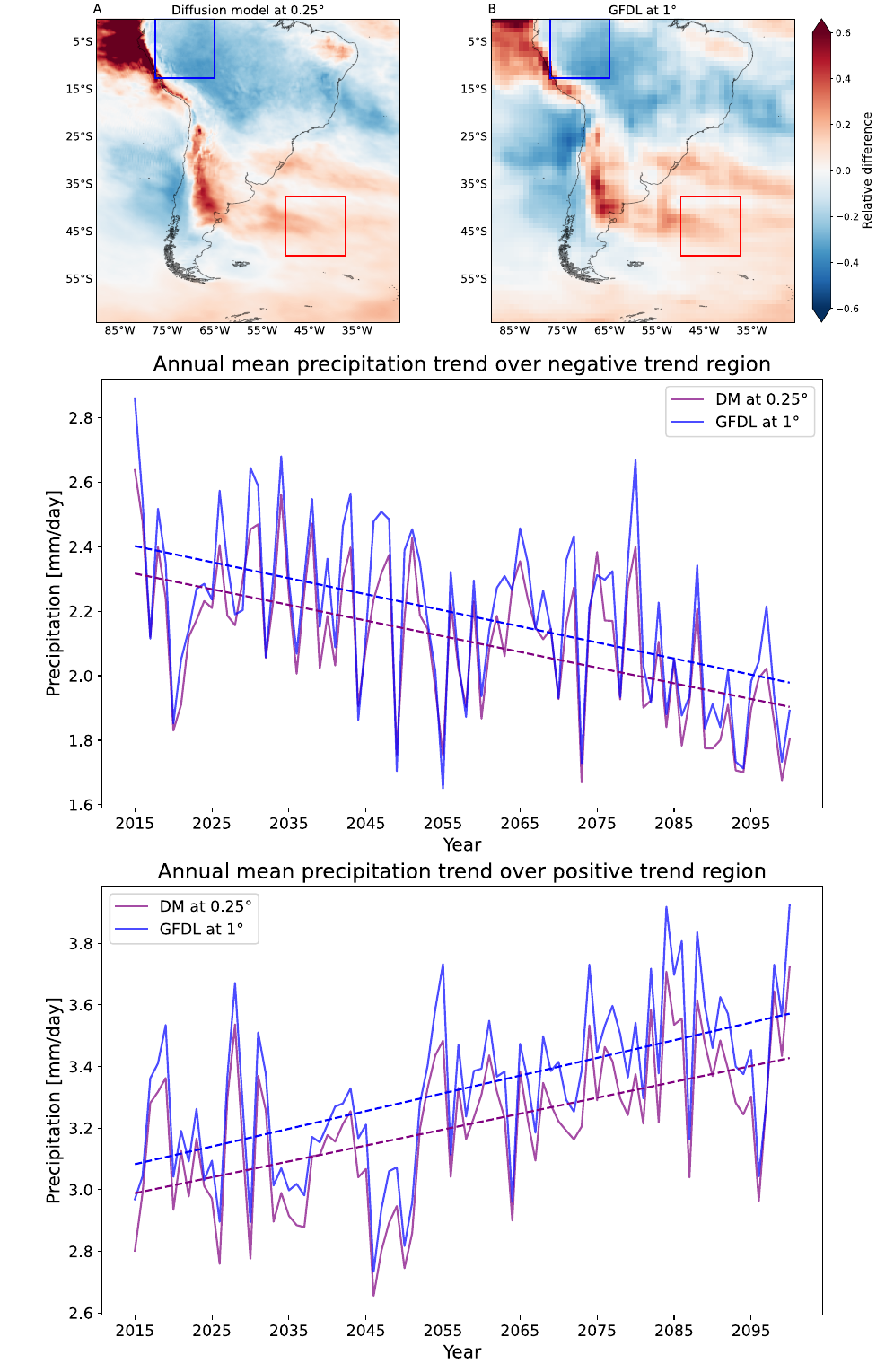}
    \caption{\textbf{Comparison of the trends for the SSP5-8.5 scenario over different sub-regions}. \textbf{(A)} shows the relative climate change signal between the late 21st century (2081-2100) under the DM-corrected GFDL SSP5-8.5 scenario and the historical DM-corrected GFDL period (1995–2014). \textbf{(B)} shows the same for relative climate change signal for GFDL without the correction. 
    We specifically chose one region with a strong negative trend (blue box) and one with a strong positive trend (red box). \textbf{(C)} The annually averaged trend over the region with negative trend between GFDL and DM-corrected GFDL is consistent. \textbf{(D)} The annually averaged trend over the region with positive trend also shows that the DM-corrected GFDL preserves the trend in GFDL. Note that the DM-corrected trend is bias corrected and hence a deviation in mean is to be expected.
    }
    \phantomsection\label{fig:trends_ssp_regional}
\end{figure*}

\begin{figure*}
    \centering
    \includegraphics[width=1\textwidth]{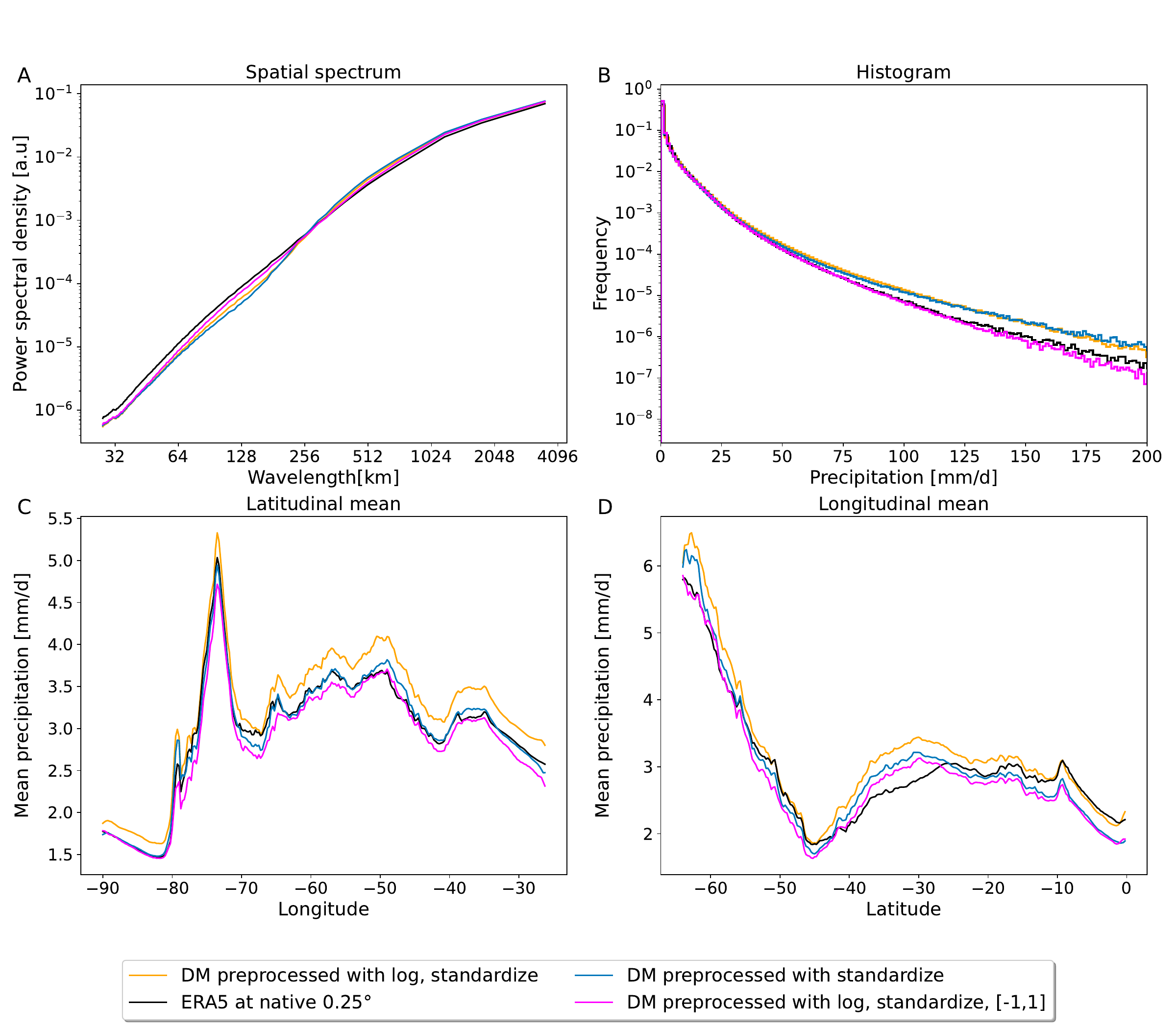}
   \caption{\textbf{Ablation study of data pre-processing strategies.} We compare the proposed pipeline (magenta: log-transform, standardization, and [-1,1] rescaling) against variants using only standardization (blue) and log-transform plus standardization (orange). The black line represents the ERA5 ground truth. (A) Spatial power spectral density. (B) Precipitation frequency histogram (C) Latitudinal and (D) Longitudinal mean profiles. Overall, the proposed method (magenta) aligns best with ERA5.
   }
    \phantomsection\label{fig:compare_preprocess}
\end{figure*}

\clearpage

\begin{figure*}
    \centering
    \includegraphics[width=\textwidth]{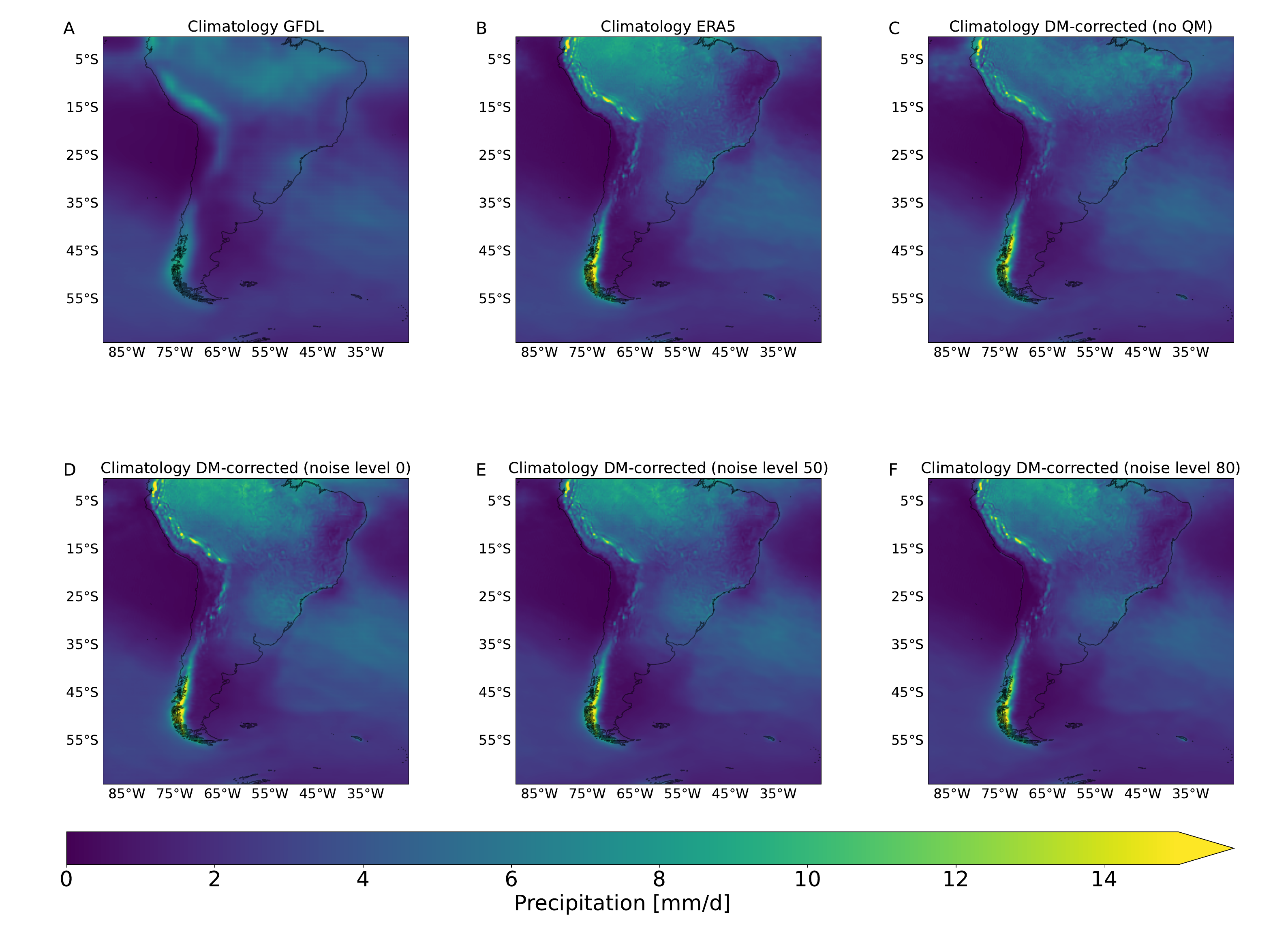}
    \caption{\textbf{Time-averaged precipitation for different diffusion model training and inference variants}. We show GFDL upsampled to 0.25° \textbf{(A)} and 0.25° ERA5 \textbf{(B)} for reference. The climatology of the diffusion model corrected GFDL data without applying QM \textbf{(C)} looks noticeably different from ERA5. Figures \textbf{(D)}-\textbf{(F)} all show the DM-corrected GFDL fields at 0.25° with different noise variations (n=0, n=50, n=80). Adding no noise during training and inference \textbf{(D)} and our proposed approach \textbf{(E)} are most similar to ERA5. Choosing a very large noise level (n=80) \textbf{(F)} still results in a similar climatology.}
    \phantomsection\label{fig:climatology_SI}
\end{figure*}
\clearpage

\begin{figure*}
    \centering
    \includegraphics[width=\textwidth]{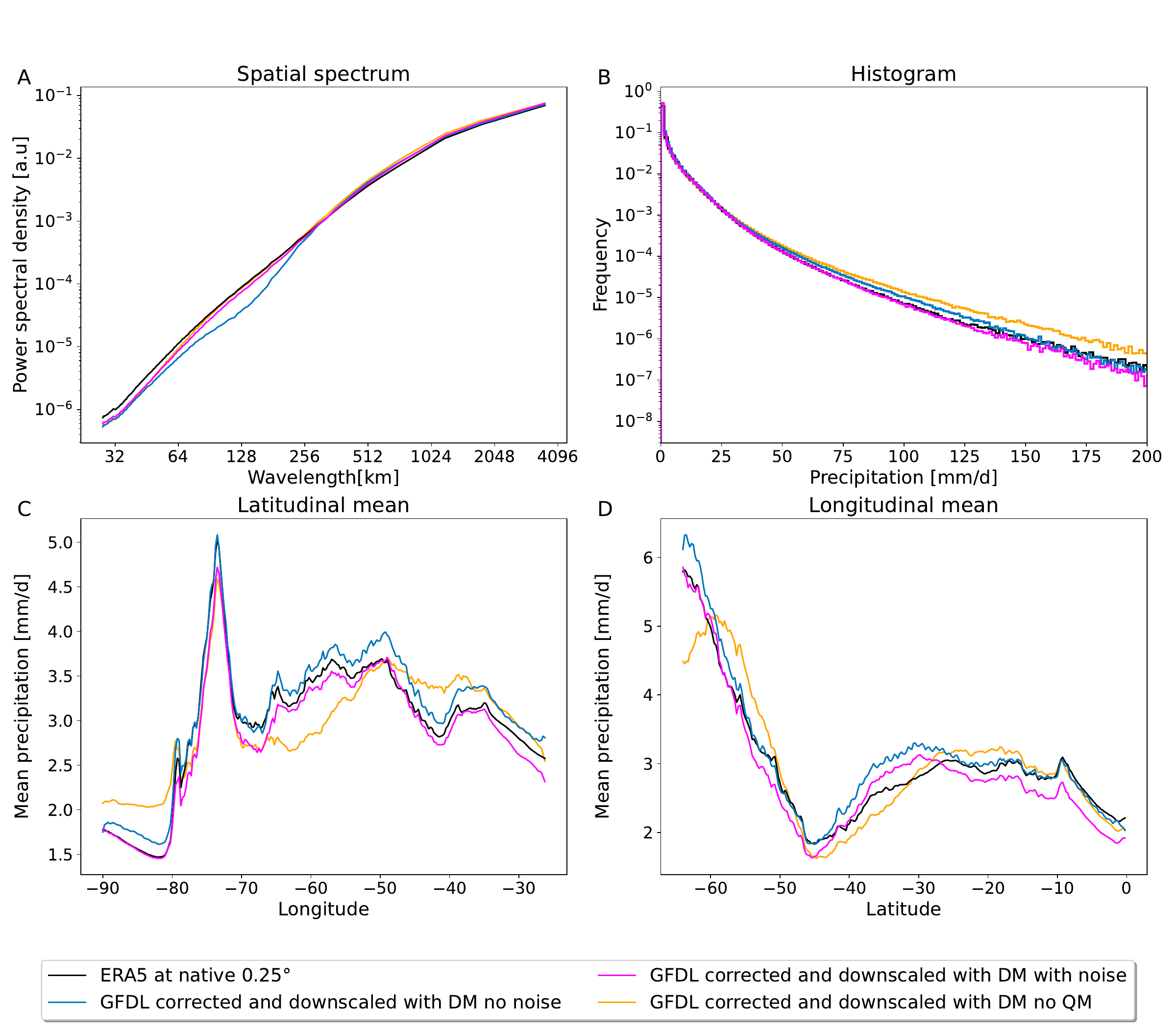}
    \caption{\textbf{Evaluating the role of QM and noising in the embedding transformation}. We compare the diffusion model-corrected GFDL fields at 0.25° without applying quantile mapping in the embedding transformation (orange) and 0.25° ERA5 (black) to the regularly noised diffusion model-corrected GFDL at 0.25° (magenta) and diffusion model-corrected GFDL fields without adding noise at 0.25° (blue). The spatial Power spectral density (PSD) plot \textbf{(A)} shows that the regularly noised diffusion models correct the small-scale spatial details far better than the noiseless variant. The histogram \textbf{(B)} for the version without noise is also worse for the range 50-150 mm d$^{-1}$. The model without quantile mapping has a completely shifted histogram compared to the ground truth. The latitude \textbf{(C)} and longitude \textbf{(D)} profiles show that the noisy version is less smooth and the version without quantile mapping is completely shifted compared to ERA5.}
    \phantomsection\label{fig:no_noise_metrics}
\end{figure*}
\clearpage

\begin{figure*}[!htb]
    \centering
    \includegraphics[width=1.1\textwidth]{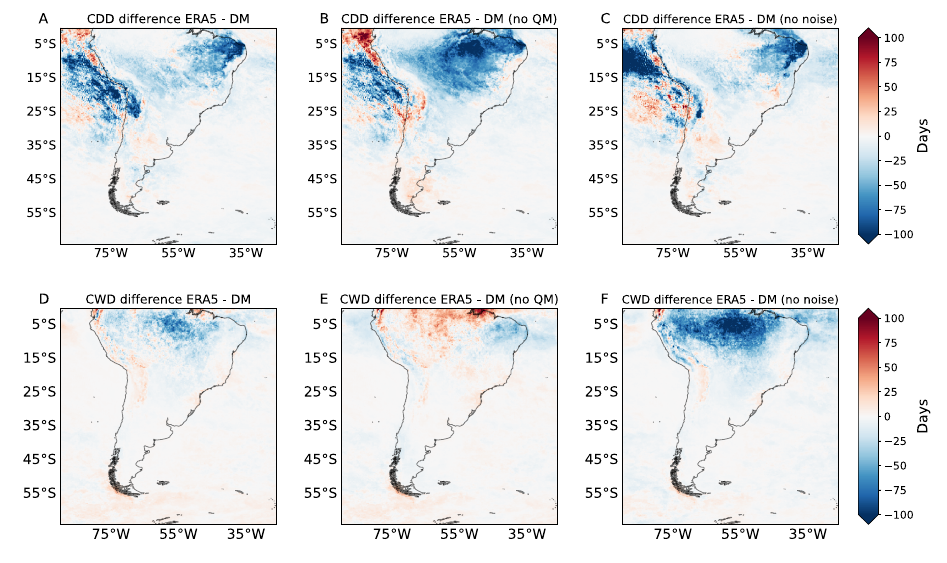}
    \caption{\textbf{Comparing consecutive dry day (CDD) and consecutive wet days (CWD) differences}. We compare the CDD/CWD between 0.25° ERA5 and diffusion model correction GFDL fields with our method at 0.25° \textbf{(A)}/\textbf{(D)}, as well as ERA5 and DM-corrected GFDL fields without using quantile mapping at 0.25° \textbf{(B)}/\textbf{(E)}, ERA5, and DM-corrected GFDL fields at 0.25° without adding noise to the condition neither during training nor inference \textbf{(C)}/\textbf{(F)}. Our diffusion model exhibits substantially smaller differences than the versions without QM and without adding noise. 
    }
    \phantomsection\label{fig:cdd_cwd_special}
\end{figure*}
\clearpage

\begin{figure*}[!htb]
    \centering
    \includegraphics[width=1.1\textwidth]{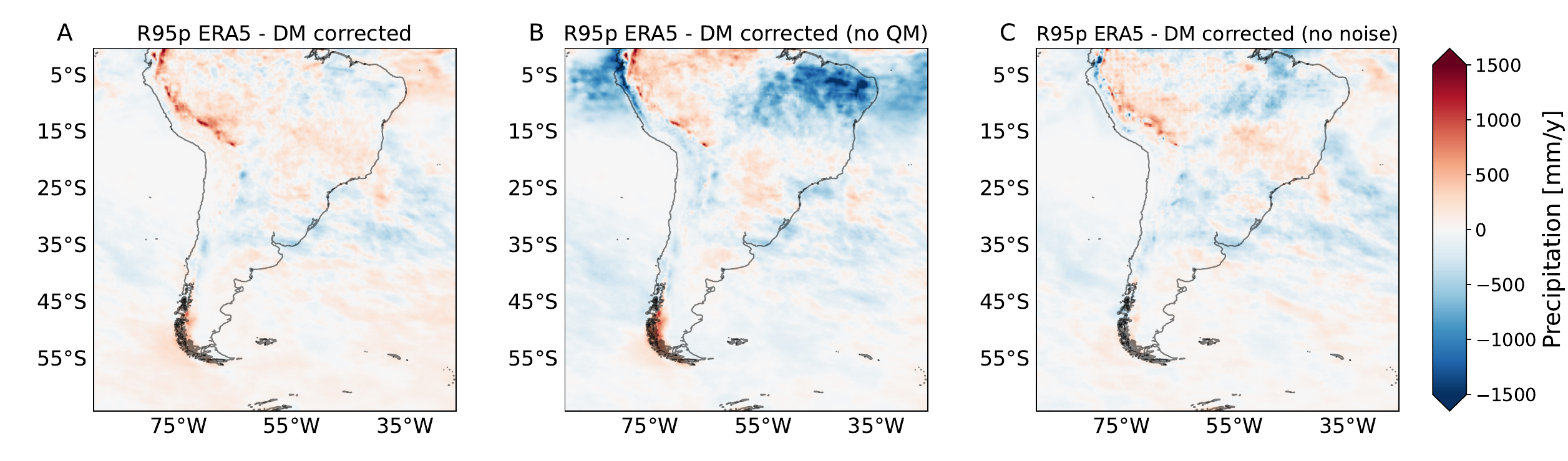}
    \caption{\textbf{We use the R95p metric to investigate the performance of different diffusion model variants on extreme events}. For clarity, we plotted the difference between R95p values for ERA5 and DM-corrected GFDL with our proposed model \textbf{(A)}, ERA5 and DM-corrected GFDL fields without applying quantile mapping \textbf{(B)}, and ERA5 and DM-corrected GFDL fields without adding noise to the condition neither during training nor inference \textbf{(C)}. Our proposed model is most effective in correcting extreme precipitation events.
    }
    \phantomsection\label{fig:r95special}
\end{figure*}
\clearpage

\begin{figure*}
    \centering
    \includegraphics[width=\textwidth]{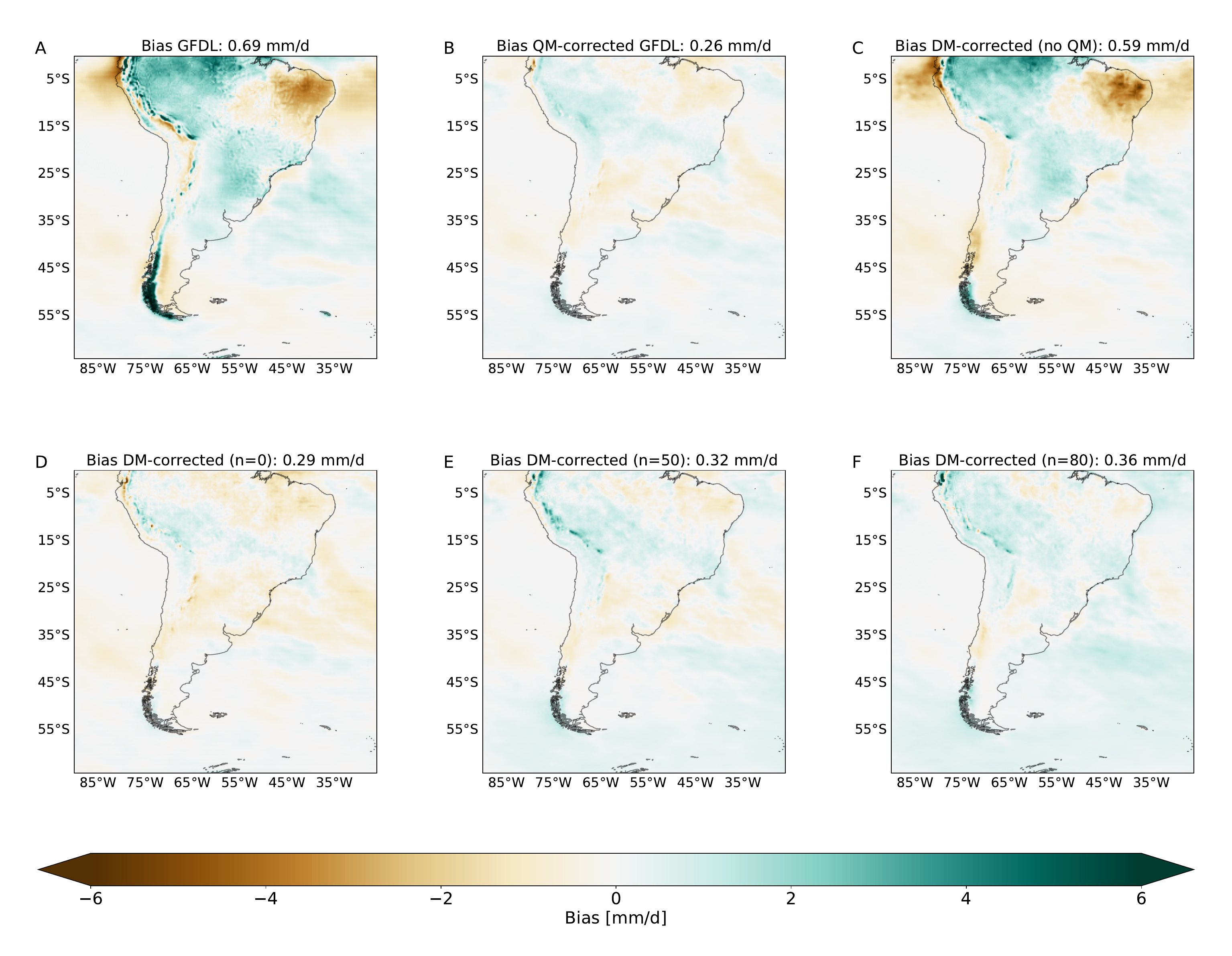}
    \caption{\textbf{Time-averaged precipitation differences for various diffusion model training and inference variants}. We show GFDL upsampled to 0.25° \textbf{(A)} and 0.25° QM-corrected GFDL \textbf{(B)} for comparison. The bias of the diffusion model corrected GFDL data without applying quantile mapping \textbf{(C)} looks noticeably different from ERA5. Figures \textbf{(D)}-\textbf{(F)} all show the DM-corrected GFDL fields at 0.25° with different noise variations (n=0, n=50, n=80). Adding no noise during training and inference \textbf{(D)} and our proposed approach \textbf{(E)} are most similar to ERA5. Choosing a very large noise level \textbf{(F)} still results in a similar climatology.}
    \phantomsection\label{fig:bias_SI}
\end{figure*}
\clearpage

\begin{figure*}
    \centering
    \includegraphics[width=\textwidth]{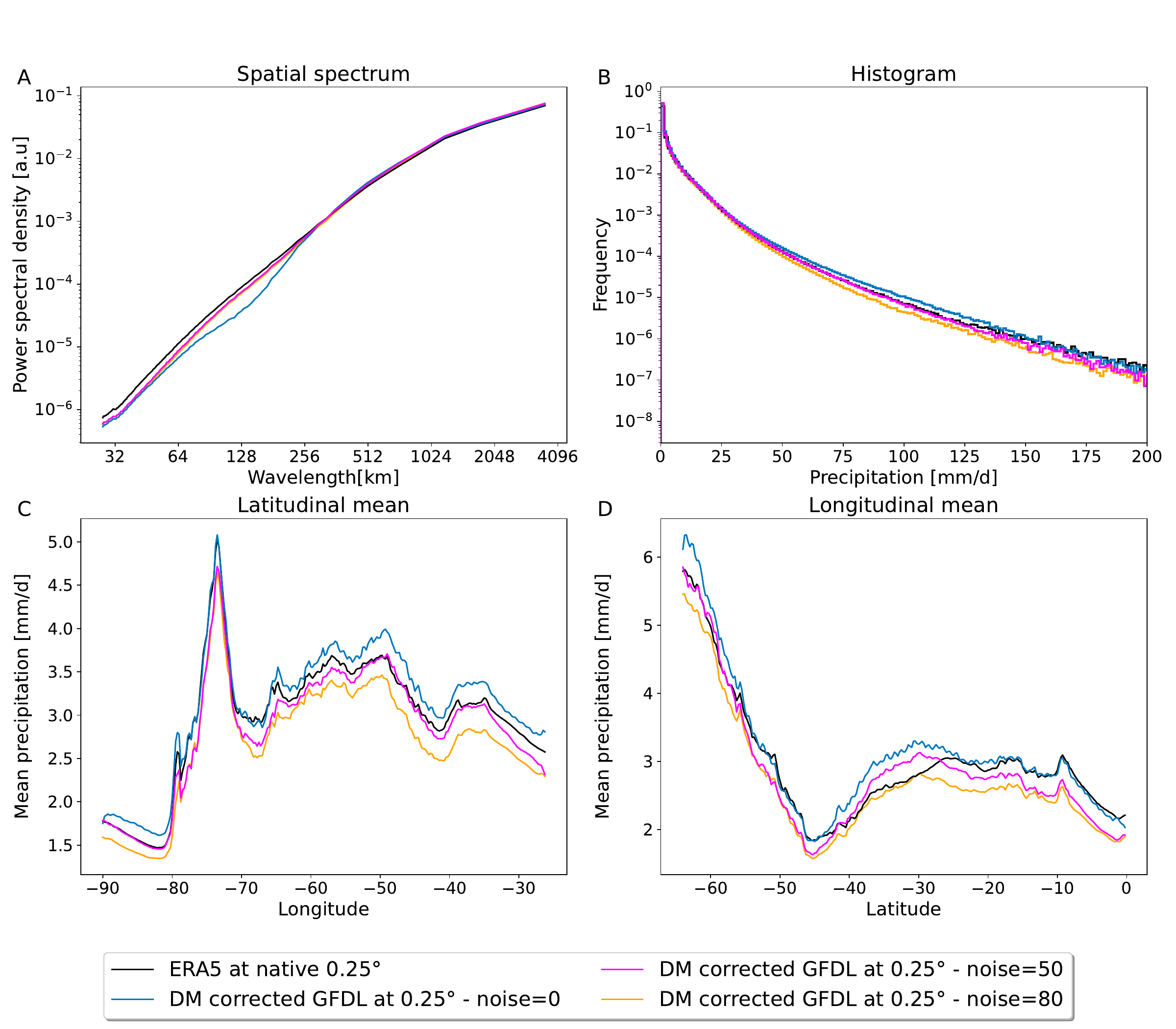}
    \caption{\textbf{Evaluating the role of different noising strengths in the embedding transformation}. Comparison of 0.25° ERA5 (orange), the diffusion model-corrected GFDL fields without adding noise (n=0) at 0.25° (black), the regularly noised (n=50) diffusion model-corrected GFDL at 0.25° (blue), and the diffusion model-corrected GFDL with a higher noise level (n=80) at 0.25° (magenta). The power spectral density plot \textbf{(A)} shows that the model with n=50 corrects the small-scale spatial details better than the n=80 version and a lot better than the n=0 version. The histogram \textbf{(B)} for both n=0 and n=50 are superior to n=80. The n=0 model falls behind the n=50 model in the range of 50-150 mm d$^{-1}$. The latitude \textbf{(C)} and longitude \textbf{(D)} profiles also show the superiority of the n=50 model over both alternatives, especially the n=80 model.}
    \phantomsection\label{fig:comp_diff_noise_lvls}
\end{figure*}


\begin{thebibliography}{45}
\providecommand{\natexlab}[1]{#1}
\providecommand{\url}[1]{\texttt{#1}}
\expandafter\ifx\csname urlstyle\endcsname\relax
  \providecommand{\doi}[1]{doi: #1}\else
  \providecommand{\doi}{doi: \begingroup \urlstyle{rm}\Url}\fi

\bibitem[Aich(2025)]{Aich2025_weights}
M.~Aich.
\newblock {Model weights for Conditional diffusion models for downscaling \&
  bias correction of ESM precipitation}, 2025.
\newblock URL \url{https://doi.org/10.5281/zenodo.18069119}.

\bibitem[Aich(2026)]{Aich2026_code}
M.~Aich.
\newblock {aim56009/ESM\_cdifffusion\_downscaling\_bc: GMD (Version v0)}, 2026.
\newblock URL \url{https://doi.org/10.5281/zenodo.18368891}.

\bibitem[Cannon et~al.(2015)Cannon, Sobie, and Murdock]{cannon2015bias}
Alex~J Cannon, Stephen~R Sobie, and Trevor~Q Murdock.
\newblock Bias correction of gcm precipitation by quantile mapping: how well do
  methods preserve changes in quantiles and extremes?
\newblock \emph{Journal of Climate}, 28\penalty0 (17):\penalty0 6938--6959,
  2015.

\bibitem[Cuturi(2013)]{cuturi2013sinkhorn}
Marco Cuturi.
\newblock Sinkhorn distances: Lightspeed computation of optimal transport.
\newblock \emph{Advances in neural information processing systems}, 26, 2013.

\bibitem[Doury et~al.(2023)Doury, Somot, Gadat, Ribes, and
  Corre]{doury2023regional}
Antoine Doury, Samuel Somot, Sebastien Gadat, Aur{\'e}lien Ribes, and Lola
  Corre.
\newblock Regional climate model emulator based on deep learning: Concept and
  first evaluation of a novel hybrid downscaling approach.
\newblock \emph{Climate Dynamics}, 60\penalty0 (5):\penalty0 1751--1779, 2023.

\bibitem[Doury et~al.(2024)Doury, Somot, and Gadat]{doury2024suitability}
Antoine Doury, Samuel Somot, and Sebastien Gadat.
\newblock On the suitability of a convolutional neural network based
  rcm-emulator for fine spatio-temporal precipitation.
\newblock \emph{Climate Dynamics}, 62\penalty0 (9):\penalty0 8587--8613, 2024.

\bibitem[Dunne et~al.(2020)Dunne, Horowitz, Adcroft, Ginoux, Held, John,
  Krasting, Malyshev, Naik, Paulot, Shevliakova, Stock, Zadeh, Balaji, Blanton,
  Dunne, Dupuis, Durachta, Dussin, Gauthier, Griffies, Guo, Hallberg, Harrison,
  He, Hurlin, McHugh, Menzel, Milly, Nikonov, Paynter, Ploshay, Radhakrishnan,
  Rand, Reichl, Robinson, Schwarzkopf, Sentman, Underwood, Vahlenkamp, Winton,
  Wittenberg, Wyman, Zeng, and Zhao]{dunne_gfdl_2020}
J.~P. Dunne, L.~W. Horowitz, A.~J. Adcroft, P.~Ginoux, I.~M. Held, J.~G. John,
  J.~P. Krasting, S.~Malyshev, V.~Naik, F.~Paulot, E.~Shevliakova, C.~A. Stock,
  N.~Zadeh, V.~Balaji, C.~Blanton, K.~A. Dunne, C.~Dupuis, J.~Durachta,
  R.~Dussin, P.~P.~G. Gauthier, S.~M. Griffies, H.~Guo, R.~W. Hallberg,
  M.~Harrison, J.~He, W.~Hurlin, C.~McHugh, R.~Menzel, P.~C.~D. Milly,
  S.~Nikonov, D.~J. Paynter, J.~Ploshay, A.~Radhakrishnan, K.~Rand, B.~G.
  Reichl, T.~Robinson, D.~M. Schwarzkopf, L.~T. Sentman, S.~Underwood,
  H.~Vahlenkamp, M.~Winton, A.~T. Wittenberg, B.~Wyman, Y.~Zeng, and M.~Zhao.
\newblock The {GFDL} {Earth} {System} {Model} {Version} 4.1 ({GFDL}-{ESM} 4.1):
  {Overall} {Coupled} {Model} {Description} and {Simulation} {Characteristics}.
\newblock \emph{Journal of Advances in Modeling Earth Systems}, 12\penalty0
  (11):\penalty0 e2019MS002015, 2020.
\newblock ISSN 1942-2466.
\newblock \doi{10.1029/2019MS002015}.
\newblock URL
  \url{https://onlinelibrary.wiley.com/doi/abs/10.1029/2019MS002015}.
\newblock \_eprint:
  https://onlinelibrary.wiley.com/doi/pdf/10.1029/2019MS002015.

\bibitem[Fran{\c{c}}ois et~al.(2021)Fran{\c{c}}ois, Thao, and
  Vrac]{franccois2021adjusting}
Bastien Fran{\c{c}}ois, Soulivanh Thao, and Mathieu Vrac.
\newblock Adjusting spatial dependence of climate model outputs with
  cycle-consistent adversarial networks.
\newblock \emph{Climate dynamics}, 57:\penalty0 3323--3353, 2021.

\bibitem[Gneiting and Raftery(2007)]{gneiting2007strictly}
Tilmann Gneiting and Adrian~E Raftery.
\newblock Strictly proper scoring rules, prediction, and estimation.
\newblock \emph{Journal of the American statistical Association}, 102\penalty0
  (477):\penalty0 359--378, 2007.

\bibitem[Goodfellow et~al.(2020)Goodfellow, Pouget-Abadie, Mirza, Xu,
  Warde-Farley, Ozair, Courville, and Bengio]{goodfellow2020generative}
Ian Goodfellow, Jean Pouget-Abadie, Mehdi Mirza, Bing Xu, David Warde-Farley,
  Sherjil Ozair, Aaron Courville, and Yoshua Bengio.
\newblock Generative adversarial networks.
\newblock \emph{Communications of the ACM}, 63\penalty0 (11):\penalty0
  139--144, 2020.

\bibitem[Gudmundsson et~al.(2012)Gudmundsson, Bremnes, Haugen, and
  Engen-Skaugen]{gudmundsson2012downscaling}
Lukas Gudmundsson, John~Bj{\o}rnar Bremnes, Jan~Erik Haugen, and Torill
  Engen-Skaugen.
\newblock Downscaling rcm precipitation to the station scale using statistical
  transformations--a comparison of methods.
\newblock \emph{Hydrology and Earth System Sciences}, 16\penalty0 (9):\penalty0
  3383--3390, 2012.

\bibitem[Gutjahr et~al.(2019)Gutjahr, Putrasahan, Lohmann, Jungclaus, von
  Storch, Br{\"u}ggemann, Haak, and St{\"o}ssel]{gutjahr2019max}
Oliver Gutjahr, Dian Putrasahan, Katja Lohmann, Johann~H Jungclaus, Jin-Song
  von Storch, Nils Br{\"u}ggemann, Helmuth Haak, and Achim St{\"o}ssel.
\newblock Max planck institute earth system model (mpi-esm1. 2) for the
  high-resolution model intercomparison project (highresmip).
\newblock \emph{Geoscientific Model Development}, 12\penalty0 (7):\penalty0
  3241--3281, 2019.

\bibitem[Gutmann et~al.(2014)Gutmann, Pruitt, Clark, Brekke, Arnold, Raff, and
  Rasmussen]{gutmann2014intercomparison}
Ethan Gutmann, Tom Pruitt, Martyn~P Clark, Levi Brekke, Jeffrey~R Arnold,
  David~A Raff, and Roy~M Rasmussen.
\newblock An intercomparison of statistical downscaling methods used for water
  resource assessments in the united states.
\newblock \emph{Water Resources Research}, 50\penalty0 (9):\penalty0
  7167--7186, 2014.

\bibitem[Hang et~al.(2023)Hang, Gu, Li, Bao, Chen, Hu, Geng, and
  Guo]{hang2023efficient}
Tiankai Hang, Shuyang Gu, Chen Li, Jianmin Bao, Dong Chen, Han Hu, Xin Geng,
  and Baining Guo.
\newblock Efficient diffusion training via min-snr weighting strategy.
\newblock In \emph{Proceedings of the IEEE/CVF International Conference on
  Computer Vision}, pages 7441--7451, 2023.

\bibitem[Haynes et~al.(2023)Haynes, Lagerquist, McGraw, Musgrave, and
  Ebert-Uphoff]{haynes2023creating}
Katherine Haynes, Ryan Lagerquist, Marie McGraw, Kate Musgrave, and Imme
  Ebert-Uphoff.
\newblock Creating and evaluating uncertainty estimates with neural networks
  for environmental-science applications.
\newblock \emph{Artificial Intelligence for the Earth Systems}, 2\penalty0
  (2):\penalty0 220061, 2023.

\bibitem[Hersbach et~al.(2020)Hersbach, Bell, Berrisford, Hirahara, Horányi,
  Muñoz-Sabater, Nicolas, Peubey, Radu, Schepers, Simmons, Soci, Abdalla,
  Abellan, Balsamo, Bechtold, Biavati, Bidlot, Bonavita, De~Chiara, Dahlgren,
  Dee, Diamantakis, Dragani, Flemming, Forbes, Fuentes, Geer, Haimberger,
  Healy, Hogan, Hólm, Janisková, Keeley, Laloyaux, Lopez, Lupu, Radnoti,
  de~Rosnay, Rozum, Vamborg, Villaume, and Thépaut]{hersbach_era5_2020}
Hans Hersbach, Bill Bell, Paul Berrisford, Shoji Hirahara, András Horányi,
  Joaquín Muñoz-Sabater, Julien Nicolas, Carole Peubey, Raluca Radu, Dinand
  Schepers, Adrian Simmons, Cornel Soci, Saleh Abdalla, Xavier Abellan,
  Gianpaolo Balsamo, Peter Bechtold, Gionata Biavati, Jean Bidlot, Massimo
  Bonavita, Giovanna De~Chiara, Per Dahlgren, Dick Dee, Michail Diamantakis,
  Rossana Dragani, Johannes Flemming, Richard Forbes, Manuel Fuentes, Alan
  Geer, Leo Haimberger, Sean Healy, Robin~J. Hogan, Elías Hólm, Marta
  Janisková, Sarah Keeley, Patrick Laloyaux, Philippe Lopez, Cristina Lupu,
  Gabor Radnoti, Patricia de~Rosnay, Iryna Rozum, Freja Vamborg, Sebastien
  Villaume, and Jean-Noël Thépaut.
\newblock The {ERA5} global reanalysis.
\newblock \emph{Quarterly Journal of the Royal Meteorological Society},
  146\penalty0 (730):\penalty0 1999--2049, 2020.
\newblock ISSN 1477-870X.
\newblock \doi{10.1002/qj.3803}.
\newblock URL \url{https://onlinelibrary.wiley.com/doi/abs/10.1002/qj.3803}.
\newblock \_eprint: https://onlinelibrary.wiley.com/doi/pdf/10.1002/qj.3803.

\bibitem[Hess et~al.(2022)Hess, Dr{\"u}ke, Petri, Strnad, and
  Boers]{hess2022physically}
Philipp Hess, Markus Dr{\"u}ke, Stefan Petri, Felix~M Strnad, and Niklas Boers.
\newblock Physically constrained generative adversarial networks for improving
  precipitation fields from earth system models.
\newblock \emph{Nature Machine Intelligence}, 4\penalty0 (10):\penalty0
  828--839, 2022.

\bibitem[Hess et~al.(2023)Hess, Lange, Sch{\"o}tz, and Boers]{hess2023deep}
Philipp Hess, Stefan Lange, Christof Sch{\"o}tz, and Niklas Boers.
\newblock Deep learning for bias-correcting cmip6-class earth system models.
\newblock \emph{Earth's Future}, 11\penalty0 (10):\penalty0 e2023EF004002,
  2023.

\bibitem[Hess et~al.(2025)Hess, Aich, Pan, and Boers]{hess2025fast}
Philipp Hess, Michael Aich, Baoxiang Pan, and Niklas Boers.
\newblock Fast, scale-adaptive and uncertainty-aware downscaling of earth
  system model fields with generative machine learning.
\newblock \emph{Nature Machine Intelligence}, pages 1--11, 2025.

\bibitem[Ho et~al.(2020)Ho, Jain, and Abbeel]{ho2020denoising}
Jonathan Ho, Ajay Jain, and Pieter Abbeel.
\newblock Denoising diffusion probabilistic models.
\newblock \emph{Advances in neural information processing systems},
  33:\penalty0 6840--6851, 2020.

\bibitem[Ho et~al.(2022)Ho, Saharia, Chan, Fleet, Norouzi, and
  Salimans]{ho2022cascaded}
Jonathan Ho, Chitwan Saharia, William Chan, David~J Fleet, Mohammad Norouzi,
  and Tim Salimans.
\newblock Cascaded diffusion models for high fidelity image generation.
\newblock \emph{Journal of Machine Learning Research}, 23\penalty0
  (47):\penalty0 1--33, 2022.

\bibitem[Hobeichi et~al.(2023)Hobeichi, Nishant, Shao, Abramowitz, Pitman,
  Sherwood, Bishop, and Green]{hobeichi2023using}
Sanaa Hobeichi, Nidhi Nishant, Yawen Shao, Gab Abramowitz, Andy Pitman, Steve
  Sherwood, Craig Bishop, and Samuel Green.
\newblock Using machine learning to cut the cost of dynamical downscaling.
\newblock \emph{Earth's Future}, 11\penalty0 (3):\penalty0 e2022EF003291, 2023.

\bibitem[{IPCC}(2023)]{IPCC_AR6_SYR_SPM_2023}
{IPCC}.
\newblock \emph{Climate Change 2023: Synthesis Report. Contribution of Working
  Groups I, II and III to the Sixth Assessment Report of the Intergovernmental
  Panel on Climate Change}.
\newblock Intergovernmental Panel on Climate Change (IPCC), Geneva,
  Switzerland, 2023.
\newblock ISBN 978-92-9169-164-7.
\newblock \doi{10.59327/IPCC/AR6-9789291691647.001}.
\newblock URL \url{https://www.ipcc.ch/report/ar6/syr/}.
\newblock Summary for Policymakers.

\bibitem[Karras et~al.(2022)Karras, Aittala, Aila, and
  Laine]{karras2022elucidating}
Tero Karras, Miika Aittala, Timo Aila, and Samuli Laine.
\newblock Elucidating the design space of diffusion-based generative models.
\newblock \emph{Advances in neural information processing systems},
  35:\penalty0 26565--26577, 2022.

\bibitem[Kingma and Ba(2015)]{kingma2014adam}
Diederik Kingma and Jimmy Ba.
\newblock Adam: A method for stochastic optimization.
\newblock In \emph{International Conference on Learning Representations
  (ICLR)}, San Diego, CA, USA, 2015.

\bibitem[Li et~al.(2022)Li, Pan, Xia, and Duan]{li2022convolutional}
Wentao Li, Baoxiang Pan, Jiangjiang Xia, and Qingyun Duan.
\newblock Convolutional neural network-based statistical post-processing of
  ensemble precipitation forecasts.
\newblock \emph{Journal of hydrology}, 605:\penalty0 127301, 2022.

\bibitem[Luhman and Luhman(2021)]{luhman_knowledge_2021}
Eric Luhman and Troy Luhman.
\newblock Knowledge {Distillation} in {Iterative} {Generative} {Models} for
  {Improved} {Sampling} {Speed}, January 2021.
\newblock URL \url{http://arxiv.org/abs/2101.02388}.
\newblock arXiv:2101.02388 [cs].

\bibitem[Miao et~al.(2019)Miao, Pan, Wang, Hsu, and
  Sorooshian]{miao2019improving}
Qinghua Miao, Baoxiang Pan, Hao Wang, Kuolin Hsu, and Soroosh Sorooshian.
\newblock Improving monsoon precipitation prediction using combined
  convolutional and long short term memory neural network.
\newblock \emph{Water}, 11\penalty0 (5):\penalty0 977, 2019.

\bibitem[Pan et~al.(2019)Pan, Hsu, AghaKouchak, and
  Sorooshian]{pan2019improving}
Baoxiang Pan, Kuolin Hsu, Amir AghaKouchak, and Soroosh Sorooshian.
\newblock Improving precipitation estimation using convolutional neural
  network.
\newblock \emph{Water Resources Research}, 55\penalty0 (3):\penalty0
  2301--2321, 2019.

\bibitem[Pan et~al.(2021)Pan, Anderson, Goncalves, Lucas, Bonfils, Lee, Tian,
  and Ma]{pan2021learning}
Baoxiang Pan, Gemma~J Anderson, Andr{\'e} Goncalves, Donald~D Lucas,
  C{\'e}line~JW Bonfils, Jiwoo Lee, Yang Tian, and Hsi-Yen Ma.
\newblock Learning to correct climate projection biases.
\newblock \emph{Journal of Advances in Modeling Earth Systems}, 13\penalty0
  (10):\penalty0 e2021MS002509, 2021.

\bibitem[Rampal et~al.(2022)Rampal, Gibson, Sood, Stuart, Fauchereau,
  Brandolino, Noll, and Meyers]{rampal2022high}
Neelesh Rampal, Peter~B Gibson, Abha Sood, Stephen Stuart, Nicolas~C
  Fauchereau, Chris Brandolino, Ben Noll, and Tristan Meyers.
\newblock High-resolution downscaling with interpretable deep learning:
  Rainfall extremes over new zealand.
\newblock \emph{Weather and Climate Extremes}, 38:\penalty0 100525, 2022.

\bibitem[Rampal et~al.(2024)Rampal, Hobeichi, Gibson, Ba{\~n}o-Medina,
  Abramowitz, Beucler, Gonz{\'a}lez-Abad, Chapman, Harder, and
  Guti{\'e}rrez]{rampal2024downscaling}
Neelesh Rampal, Sanaa Hobeichi, Peter~B Gibson, Jorge Ba{\~n}o-Medina, Gab
  Abramowitz, Tom Beucler, Jose Gonz{\'a}lez-Abad, William Chapman, Paula
  Harder, and Jos{\'e}~Manuel Guti{\'e}rrez.
\newblock Enhancing regional climate downscaling through advances in machine
  learning.
\newblock \emph{Artificial Intelligence for the Earth Systems}, 3\penalty0
  (2):\penalty0 230066, 2024.

\bibitem[Rampal et~al.(2025)Rampal, Gibson, Sherwood, Abramowitz, and
  Hobeichi]{rampal2025reliable}
Neelesh Rampal, Peter~B Gibson, Steven Sherwood, Gab Abramowitz, and Sanaa
  Hobeichi.
\newblock A reliable generative adversarial network approach for climate
  downscaling and weather generation.
\newblock \emph{Journal of Advances in Modeling Earth Systems}, 17\penalty0
  (1):\penalty0 e2024MS004668, 2025.

\bibitem[Rombach et~al.(2022)Rombach, Blattmann, Lorenz, Esser, and
  Ommer]{rombach2022high}
Robin Rombach, Andreas Blattmann, Dominik Lorenz, Patrick Esser, and Bj{\"o}rn
  Ommer.
\newblock High-resolution image synthesis with latent diffusion models.
\newblock In \emph{Proceedings of the IEEE/CVF conference on computer vision
  and pattern recognition}, pages 10684--10695, 2022.

\bibitem[Saharia et~al.(2022{\natexlab{a}})Saharia, Chan, Chang, Lee, Ho,
  Salimans, Fleet, and Norouzi]{saharia2022palette}
Chitwan Saharia, William Chan, Huiwen Chang, Chris Lee, Jonathan Ho, Tim
  Salimans, David Fleet, and Mohammad Norouzi.
\newblock Palette: Image-to-image diffusion models.
\newblock In \emph{ACM SIGGRAPH 2022 Conference Proceedings}, pages 1--10,
  2022{\natexlab{a}}.

\bibitem[Saharia et~al.(2022{\natexlab{b}})Saharia, Chan, Saxena, Li, Whang,
  Denton, Ghasemipour, Gontijo~Lopes, Karagol~Ayan, Salimans, Ho, Fleet, and
  Norouzi]{saharia2022photorealistic}
Chitwan Saharia, William Chan, Saurabh Saxena, Lala Li, Jay Whang, Emily~L
  Denton, Kamyar Ghasemipour, Raphael Gontijo~Lopes, Burcu Karagol~Ayan, Tim
  Salimans, Jonathan Ho, David Fleet, and Mohammad Norouzi.
\newblock Photorealistic text-to-image diffusion models with deep language
  understanding.
\newblock \emph{Advances in neural information processing systems},
  35:\penalty0 36479--36494, 2022{\natexlab{b}}.

\bibitem[Saharia et~al.(2022{\natexlab{c}})Saharia, Ho, Chan, Salimans, Fleet,
  and Norouzi]{saharia2022image}
Chitwan Saharia, Jonathan Ho, William Chan, Tim Salimans, David~J Fleet, and
  Mohammad Norouzi.
\newblock Image super-resolution via iterative refinement.
\newblock \emph{IEEE Transactions on Pattern Analysis and Machine
  Intelligence}, 45\penalty0 (4):\penalty0 4713--4726, 2022{\natexlab{c}}.

\bibitem[Tong et~al.(2021)Tong, Gao, Han, Xu, Xu, and Giorgi]{tong2021bias}
Yao Tong, Xuejie Gao, Zhenyu Han, Yaqi Xu, Ying Xu, and Filippo Giorgi.
\newblock Bias correction of temperature and precipitation over china for rcm
  simulations using the qm and qdm methods.
\newblock \emph{Climate Dynamics}, 57:\penalty0 1425--1443, 2021.

\bibitem[Van Den~Oord et~al.(2017)Van Den~Oord, Vinyals, and
  Kavukcuoglu]{van2017neural}
Aaron Van Den~Oord, Oriol Vinyals, and Koray Kavukcuoglu.
\newblock Neural discrete representation learning.
\newblock \emph{Advances in neural information processing systems}, 30, 2017.

\bibitem[van~der Meer et~al.(2023)van~der Meer, de~Roda~Husman, and
  Lhermitte]{van2023deep}
Marijn van~der Meer, Sophie de~Roda~Husman, and Stef Lhermitte.
\newblock Deep learning regional climate model emulators: A comparison of two
  downscaling training frameworks.
\newblock \emph{Journal of Advances in Modeling Earth Systems}, 15\penalty0
  (6):\penalty0 e2022MS003593, 2023.

\bibitem[Vandal et~al.(2017)Vandal, Kodra, Ganguly, Michaelis, Nemani, and
  Ganguly]{vandal2017deepsd}
Thomas Vandal, Evan Kodra, Sangram Ganguly, Andrew Michaelis, Ramakrishna
  Nemani, and Auroop~R Ganguly.
\newblock Deepsd: Generating high resolution climate change projections through
  single image super-resolution.
\newblock In \emph{Proceedings of the 23rd acm sigkdd international conference
  on knowledge discovery and data mining}, pages 1663--1672, 2017.

\bibitem[Wan et~al.(2024)Wan, Baptista, Boral, Chen, Anderson, Sha, and
  Zepeda-N{\'u}{\~n}ez]{wan2024debias}
Zhong~Yi Wan, Ricardo Baptista, Anudhyan Boral, Yi-Fan Chen, John Anderson, Fei
  Sha, and Leonardo Zepeda-N{\'u}{\~n}ez.
\newblock Debias coarsely, sample conditionally: Statistical downscaling
  through optimal transport and probabilistic diffusion models.
\newblock \emph{Advances in Neural Information Processing Systems}, 36, 2024.

\bibitem[Wang et~al.(2004)Wang, Bovik, Sheikh, and Simoncelli]{SSIM}
Zhou Wang, Alan~C Bovik, Hamid~R Sheikh, and Eero~P Simoncelli.
\newblock Image quality assessment: from error visibility to structural
  similarity.
\newblock \emph{IEEE transactions on image processing}, 13\penalty0
  (4):\penalty0 600--612, 2004.

\bibitem[Zelinka et~al.(2020)Zelinka, Myers, McCoy, Po-Chedley, Caldwell,
  Ceppi, Klein, and Taylor]{zelinka2020causes}
Mark~D Zelinka, Timothy~A Myers, Daniel~T McCoy, Stephen Po-Chedley, Peter~M
  Caldwell, Paulo Ceppi, Stephen~A Klein, and Karl~E Taylor.
\newblock Causes of higher climate sensitivity in cmip6 models.
\newblock \emph{Geophysical Research Letters}, 47\penalty0 (1):\penalty0
  e2019GL085782, 2020.

\bibitem[Zhu et~al.(2017)Zhu, Park, Isola, and Efros]{Zhu_2017_ICCV}
Jun-Yan Zhu, Taesung Park, Phillip Isola, and Alexei~A. Efros.
\newblock Unpaired image-to-image translation using cycle-consistent
  adversarial networks.
\newblock In \emph{Proceedings of the IEEE International Conference on Computer
  Vision (ICCV)}, Oct 2017.

\end{thebibliography}
\end{document}